\PassOptionsToPackage{unicode}{hyperref}
\PassOptionsToPackage{hyphens}{url}
\PassOptionsToPackage{dvipsnames,svgnames,x11names}{xcolor}
\documentclass[
  10pt,
  a4paper,
]{article}
\usepackage{xcolor}
\usepackage[margin=1in]{geometry}
\usepackage{amsmath,amssymb}
\usepackage{iftex}
\ifPDFTeX
  \usepackage[T1]{fontenc}
  \usepackage[utf8]{inputenc}
  \usepackage{textcomp} % provide euro and other symbols
\else % if luatex or xetex
  \usepackage{unicode-math} % this also loads fontspec
  \defaultfontfeatures{Scale=MatchLowercase}
  \defaultfontfeatures[\rmfamily]{Ligatures=TeX,Scale=1}
\fi
\usepackage{lmodern}
\ifPDFTeX\else
\fi
\IfFileExists{upquote.sty}{\usepackage{upquote}}{}
\IfFileExists{microtype.sty}{% use microtype if available
  \usepackage[]{microtype}
  \UseMicrotypeSet[protrusion]{basicmath} % disable protrusion for tt fonts
}{}
\makeatletter
\@ifundefined{KOMAClassName}{% if non-KOMA class
  \IfFileExists{parskip.sty}{%
    \usepackage{parskip}
  }{% else
    \setlength{\parindent}{0pt}
    \setlength{\parskip}{6pt plus 2pt minus 1pt}}
}{% if KOMA class
  \KOMAoptions{parskip=half}}
\makeatother
\usepackage{longtable,booktabs,array}
\usepackage{caption}
\usepackage{calc} % for calculating minipage widths
\usepackage{etoolbox}
\makeatletter
\patchcmd\longtable{\par}{\if@noskipsec\mbox{}\fi\par}{}{}
\makeatother
\providecommand{\tightlist}{%
  \setlength{\itemsep}{0pt}\setlength{\parskip}{0pt}}
\usepackage{newunicodechar}
\usepackage{pmboxdraw}
\usepackage{accsupp}
\usepackage{pdfrender}
\newcommand{\pfvbox}[1]{\pfvU{#1}{\rlap{\textpdfrender{TextRenderingMode=Invisible}{-}}\pmboxdrawuni{#1}}}
\usepackage{amssymb}
\usepackage{graphicx}
\newlength{\pfvcell}
\newcommand{\pfvtt}[1]{\settowidth{\pfvcell}{\ttfamily M}\makebox[\pfvcell][c]{#1}}
\newcommand{\pfvU}[2]{\BeginAccSupp{method=hex,unicode,ActualText=#1}#2\EndAccSupp{}}
\newcommand{\pfvsup}[2]{\pfvU{#1}{\textsuperscript{#2}}}
\newcommand{\pfvsub}[2]{\pfvU{#1}{\textsubscript{#2}}}
\newcommand{\pfvm}[2]{\pfvU{#1}{\ensuremath{#2}}}
\makeatletter
\newcommand{\pfvsym}[2]{\ifdim\fontdimen3\font=\z@\pfvU{#1}{\pfvtt{\ensuremath{#2}}}\else\pfvm{#1}{#2}\fi}
\makeatother
\newunicodechar{±}{\pfvsym{00B1}{\pm}}
\newunicodechar{·}{\pfvsym{00B7}{\cdot}}
\newunicodechar{²}{\pfvsup{00B2}{2}}
\newunicodechar{³}{\pfvsup{00B3}{3}}
\newunicodechar{¹}{\pfvsup{00B9}{1}}
\newunicodechar{⁰}{\pfvsup{2070}{0}}
\newunicodechar{⁴}{\pfvsup{2074}{4}}
\newunicodechar{⁵}{\pfvsup{2075}{5}}
\newunicodechar{⁶}{\pfvsup{2076}{6}}
\newunicodechar{⁷}{\pfvsup{2077}{7}}
\newunicodechar{⁸}{\pfvsup{2078}{8}}
\newunicodechar{⁹}{\pfvsup{2079}{9}}
\newunicodechar{⁻}{\pfvsup{207B}{\ensuremath{-}}}
\newunicodechar{₀}{\pfvsub{2080}{0}}
\newunicodechar{₁}{\pfvsub{2081}{1}}
\newunicodechar{ș}{\pfvU{0219}{\textcommabelow{s}}}
\newunicodechar{Δ}{\pfvsym{0394}{\Delta}}
\newunicodechar{‖}{\pfvsym{2016}{\|}}
\newunicodechar{′}{\pfvsym{2032}{{}^{\prime}}}
\newunicodechar{→}{\pfvsym{2192}{\rightarrow}}
\newunicodechar{∀}{\pfvsym{2200}{\forall}}
\newunicodechar{∃}{\pfvsym{2203}{\exists}}
\newunicodechar{−}{\pfvsym{2212}{-}}
\newunicodechar{√}{\pfvsym{221A}{\surd}}
\newunicodechar{∧}{\pfvsym{2227}{\wedge}}
\newunicodechar{≠}{\pfvsym{2260}{\neq}}
\newunicodechar{≤}{\pfvsym{2264}{\leq}}
\newunicodechar{⌊}{\pfvsym{230A}{\lfloor}}
\newunicodechar{⌋}{\pfvsym{230B}{\rfloor}}
\newunicodechar{►}{\pfvsym{25BA}{\blacktriangleright}}
\newunicodechar{◄}{\pfvsym{25C4}{\blacktriangleleft}}
\newcommand{\pfvellipsis}{\pfvU{2026}{\ldots}}
\newunicodechar{…}{\ifdim\fontdimen3\font=0pt \pfvU{2026}{\pfvtt{\fontsize{4}{4}\selectfont\ldots}}\else\pfvU{2026}{\ldots}\fi}
\newunicodechar{─}{\pfvbox{2500}}
\newunicodechar{│}{\pfvbox{2502}}
\newunicodechar{└}{\pfvbox{2514}}
\newunicodechar{┘}{\pfvbox{2518}}
\usepackage{fvextra}
\usepackage{etoolbox}
\AtBeginEnvironment{longtable}{\footnotesize}
\usepackage{bookmark}
\IfFileExists{xurl.sty}{\usepackage{xurl}}{} % add URL line breaks if available
\makeatletter
\@ifundefined{xmpquote}{}{}
\makeatother
\hypersetup{
  pdftitle={Verifying deployed bytecode against executable EVM semantics{,} with exact gas and no added axioms},
  pdfauthor={Seulkee Baek},
  colorlinks=true,
  linkcolor={black},
  filecolor={Maroon},
  citecolor={Blue},
  urlcolor={black},
  pdfcreator={LaTeX via pandoc}}

\title{Verifying
deployed
bytecode
against
executable
EVM
semantics,
with exact
gas and no
added
axioms}
\author{Seulkee
Baek}
\date{}

\begin{document}
\maketitle
\begin{abstract}
Much of
Ethereum's
most
consequential
code is
immutable
bytecode:
old output
of
unverified
compilers,
and
protocol
logic run
as
contract
code. We
show that
such
bytecode
can be
verified
as it is,
without
trusting
what
produced
it,
without
simplifying
the
machine it
runs on,
and with
no axioms
beyond
Lean's
own.
Blanc, a
Lean 4
verification
stack,
reads
fixed
runtime
bytes via
kernel-checked
certificates
and states
every
theorem
against
Jaune, an
executable
semantics
of the EVM
and block
processing
that
passes all
34,205
blockchain-test
cases of
Ethereum's
fixture
release
v20.0.2
that
involve
only forks
that have
governed
mainnet
since May
2025.
Eight
artifacts,
deployed
from 2017
to 2026,
come from
seven
compiler
versions
and an
assembler.
For the
EIP-7002
contract,
we prove
that the
end-of-block
call to
it, whose
failure
would
invalidate
the block,
succeeds
whatever
its
storage
holds, and
that each
block's
withdrawal
requests
are
exactly
the oldest
pending
submissions;
mainnet
met these
theorems'
starting
premises
when
Prague
began. For
a Uniswap
V2 pair,
we prove
that after
any valid
history
whose
hashed
slots stay
apart,
from any
checkpoint
representing
its
Solidity
model's
state, the
storage
represents
that model
replayed
over
exactly
the
committed
calls, a
balanced
liquidity
ledger
stays
balanced,
and, under
a premise
on token
answers,
geometric-mean
reserves
per share
fall only
through
the
protocol
fee. Exact
gas makes
liveness
provable:
a tracked
WETH9
holder
without
code can
withdraw
by a
transaction
whose gas
has a
closed
form,
matching
all 3,748
comparable
mainnet
receipts
of its
shape in
30,000
blocks. On
Curve pool
bytes
compiled
with the
lock bug
exploited
in 2023,
Jaune
executes
such a
reentrancy
from
deployment;
for bytes
from a
fixed
compiler,
we prove
the lock
excludes
it. AI
agents
wrote the
proofs in
fifteen
days under
one
author's
direction,
atop an
existing
stack.
\end{abstract}

\emph{Statements
about
Blanc's
theorems,
and every
figure
drawn from
its Git
history,
refer to
Blanc
commit
\texttt{5048a51f}
on its
main
branch and
to the
Jaune
revision
it pins,
\texttt{2737c8eb}.
Unless a
section
says
otherwise,
on-chain
figures
were read
at block
26,141,229
(7 October
2026,
14:53 UTC)
from the
public
providers
named in
Appendix
B. We say
that a
result
uses} no
added
axioms
\emph{when
everything
it depends
on reaches
only the
three
standard
axioms of
Lean's
core
library,
\texttt{propext},
\texttt{Classical.\allowbreak{}choice}
and
\texttt{Quot.sound},
with no
\texttt{sorry},
no native
evaluation
and no
foreign
function
anywhere
in its
proof.}

\subsection{1
Introduction}\label{introduction}

On 30 July
2023,
attackers
drained
several
Curve
liquidity
pools of
assets
worth tens
of
millions
of dollars
{[}LlamaRisk{]}.
The pools'
source
marked
every
function
that moves
liquidity
with the
same
reentrancy
key,
\texttt{@nonreentrant\penalty200{}(\allowbreak{}'lock')},
which asks
the
compiler
for one
lock
shared by
all of
them. The
affected
Vyper
versions,
0.2.15,
0.2.16 and
0.3.0,
instead
reserved a
separate
lock slot
for every
decorated
function
{[}VyperPM{]}.
An
attacker
could
receive
ether from
\texttt{remove\_\allowbreak{}liquidity}
while it
held its
lock, call
\texttt{add\_liquidity},
find that
function's
lock free,
and mint
pool
tokens
against a
pool that
was only
half
updated. A
proof
about the
source
that took
the
compiler
at its
word would
have
certified
the guard;
the defect
could be
seen only
in the
bytes.

The bytes
are what
users
ultimately
rely on:
every node
executes
them,
nobody can
patch
them, and
many are
old.
WETH9, the
canonical
wrapped
ether,
holds
about 2.16
million
ether in
3,124
bytes that
solc
0.4.19
produced
in
December
2017. The
Beacon
deposit
contract,
6,358
bytes
produced
by solc
0.6.11, is
the only
way onto
the beacon
chain:
every
validator
on mainnet
was
created by
one of the
2,565,727
deposits
made
through
it. One
11,293-byte
runtime,
produced
by solc
0.5.16 in
2020, runs
at every
one of the
524,109
trading
pairs that
Uniswap
V2's
factory
has
created
(Appendix
B).\footnote{Balances,
  the
  deposit
  contract's
  \texttt{get\_\allowbreak{}deposit\_\allowbreak{}count(\allowbreak{})},
  and the
  factory's
  \texttt{allPairsLeng\penalty200{}th(\allowbreak{})},
  at block
  26,141,229
  (7
  October
  2026),
  read
  from
  three
  separately
  operated
  public
  providers,
  which
  agreed
  (Appendix
  B).} And
since the
Prague
upgrade,
every
withdrawal
request
triggered
from the
execution
layer
passes
through a
504-byte
contract
written in
assembly,
and a
block
whose
end-of-block
call to
that
contract
fails is
invalid. A
proof
about the
source of
any of
these
reaches
the bytes
only
through a
link
between
source and
bytes:
trust in
the
compiler,
a verified
compiler,
or a
translation
checked
after the
fact, by
translation
validation
{[}Pnueli98{]}
or an
equivalence
check of
the bytes
against a
specification
{[}Act{]}.
None of
the tools
that
produced
these
bytes is a
verified
compiler;
a proof
about the
bytes
themselves
needs no
such link.

We ask six
questions
of any
verification
result
about such
code.

\begin{enumerate}
\def\labelenumi{\arabic{enumi}.}
\tightlist
\item
  \textbf{Is
  it about
  the code
  that
  runs?} A
  theorem
  about
  source
  code, a
  model or
  a
  recompilation
  reaches
  the
  deployed
  contract
  only
  through
  the link
  between
  source
  and
  bytes,
  and the
  2023
  incident
  shows
  what
  trusting
  a
  compiler
  can
  cost.
\item
  \textbf{What
  does the
  proof
  rest
  on?} A
  proof
  checked
  by a
  small
  kernel
  rests on
  less
  than one
  that
  admits
  lemmas,
  evaluates
  by
  compiled
  code, or
  adds
  axioms
  such as
  the
  convenient
  and
  false
  axiom
  that
  Keccak-256
  is
  injective.
\item
  \textbf{Was
  the code
  written
  for the
  verifier?}
  Code
  written
  in a
  verifier's
  language,
  or
  emitted
  by a
  verifier's
  own
  compiler,
  is
  rarely
  the code
  that
  holds
  today's
  assets.
\item
  \textbf{Is
  the
  machine
  the real
  machine?}
  The
  semantics
  should
  be the
  whole
  EVM,
  with
  every
  opcode,
  precompile,
  transaction
  type,
  fork
  rule and
  gas
  cost,
  and
  executable,
  so that
  the
  official
  conformance
  suite
  can test
  it.
\item
  \textbf{Does
  the
  theorem
  matter?}
  It
  should
  tell
  users
  what
  they
  want to
  know:
  that
  balances
  are
  backed,
  a queue
  delivers
  what was
  put in,
  a lock
  excludes
  what it
  should.
\item
  \textbf{Does
  it say
  the
  contract
  works?}
  A
  contract
  that
  always
  reverts
  preserves
  every
  invariant
  its
  storage
  satisfies.
  Users
  also
  need
  liveness,
  the
  assurance
  that a
  transaction
  they
  send is
  admitted
  and
  succeeds,
  and on
  the EVM
  liveness
  is a
  statement
  about
  gas. Our
  liveness
  theorems
  execute
  a
  transaction
  at the
  state a
  history
  reaches,
  or at a
  given
  position
  of a
  given
  valid
  block;
  whether
  any
  block
  will
  include
  it is
  outside
  them.
\end{enumerate}

Each
question
has been
answered
before,
though
rarely
together.
Symbolic
tools such
as hevm,
halmos and
the
Certora
Prover
reach
bytecode
at a
scale, and
with a
degree of
automation,
that
interactive
proof does
not
approach,
each
approximating
Keccak-256
in a
documented
way
{[}hevm;
halmos;
Certora{]}.
Bytecode
of
contracts
in our
corpus, of
other
builds of
their
sources,
or of
earlier
versions
of them,
has been
verified
in K, with
gas
{[}KEVM-Deposit;
RV-DepositSol;
dapp-UniV2{]},
and
checked
with
Halmos,
loops
bounded
{[}Halmos-Pectra{]};
verified
compilers
cover the
code they
emit, not
code
already
deployed
{[}DeepSEA;
powdr;
Solidus;
Verity{]};
and EquiVM
proves
that local
builds of
more than
twenty
contracts
refine
whole-interface
specifications,
leaving
gas out
{[}EquiVM{]}.
To our
knowledge
no prior
result
answers
all six
questions
at once;
Section 6
describes
what each
system
does that
we do not.

This paper
shows that
the six
answers
can be had
together,
for
ordinary
compiler
output and
hand-written
assembly
that our
tools had
no part in
shaping.
Blanc
{[}Blanc{]}
is a
verification
stack for
the EVM in
Lean 4
{[}Lean4{]},
with its
own
contract
language
and
compiler;
the work
reported
here is a
recent
extension
that
\emph{lifts}
bytecode
Blanc did
not
produce.
An
untrusted
program
proposes a
structured
reading of
the bytes,
a
\emph{certificate},
which a
decision
procedure
evaluated
by Lean's
kernel
accepts or
rejects
against
the
literal
bytes.
Generic
theorems
carry
reasoning
about the
reading
back to
the bytes,
for
safety,
for
executions
of any
outcome,
and for
liveness
given
enough
gas. Every
theorem is
stated
against
Jaune
{[}Jaune{]},
an
executable
semantics
of the EVM
and of
block
processing,
also in
Lean,
which
passes
every
blockchain
test of
Ethereum's
fixture
release
\texttt{tests@v20.0.2}
that
involves
only the
forks that
have
governed
mainnet
since May
2025
(Section
3.1). No
declaration
in the
725,186-line
Blanc
library
reaches an
axiom
beyond
Lean's
three, and
Lean's
kernel,
replaying
the
compiled
library in
a separate
process,
accepts
every
declaration
(Section
3.11).

Here is
the kind
of
statement
this makes
possible.

\begin{quote}
\textbf{WETH9
exit.} Fix
a
checkpoint
at which
the WETH9
bytes are
installed
and the
contract's
storage
satisfies
its
footprint
invariant,
and take
\emph{any}
history of
valid
blocks
from
there,
under the
rules of
Prague,
Osaka,
BPO1 or
BPO2, in
which the
storage
keys
touched do
not
collide.
Let
\emph{E}
be an
account
without
code, so
neither a
contract
nor an
account
with an
EIP-7702
delegation,
whose
balance
key is
tracked
and whose
WETH
balance is
at least
\emph{w}
(an amount
in wei). A
signed
type-2
transaction
from
\emph{E}
calling
\texttt{withdraw(w)},
with zero
value and
no access
list, the
usual
chain,
nonce,
fee,
funding
and
block-room
conditions,
a fee
recipient
other than
\emph{E}
and the
contract,
and a gas
limit
between
its
intrinsic
cost plus
the
frame's
cost plus
811 and
2²⁴, is
admitted
by Jaune's
\texttt{processTrans\penalty200{}acti\penalty200{}on}
at the
state the
history
reaches.
In the
world it
returns,
\emph{E}'s
WETH
balance is
\emph{w}
lower, the
contract
has paid
\emph{w}
wei to
\emph{E},
no other
storage
has
changed,
and the
transaction
has used
exactly

\textbf{gasUsed
= 21,000 +
4·tokens(calldata)
+ (13,940
if
\emph{w} ≠
0,
otherwise
4,440) −
(4,800 if
\emph{w} ≠
0 and the
withdrawal
empties
the
balance)}

where
tokens(calldata)
counts
each zero
byte once
and each
nonzero
byte four
times, so
that the
calldata
costs 4
gas per
zero byte
and 16 per
nonzero
byte.
\end{quote}

This one
theorem
answers
all six
questions
(Section
2): it
concerns
the 3,124
bytes that
three
providers
return for
\texttt{0xC02a\pfvellipsis{}6Cc2},
2017
compiler
output
never
meant to
be
verified,
under
every fork
that has
governed
mainnet
since 7
May 2025;
its proof
rests only
on the
kernel and
Lean's
three
axioms;
and it
tells a
holder
that an
exit
transaction
is
admitted
at the
state any
such
history
reaches,
pays out,
and costs
exactly
the stated
gas. In a
window of
30,000
blocks,
every
mainnet
withdrawal
of this
shape
whose
starting
balance
the logs
determine,
3,748 of
3,780,
used
exactly
the
predicted
gas
(Section
2.4).

A seventh
question
is whether
a
theorem's
starting
assumption
holds of
the
contract
on mainnet
today. For
WETH9, the
deposit
contract,
3Crv and
the
Uniswap
pair,
deployed
long
before
Prague, we
answer it
only for a
modeled
fresh
deployment;
the live
contracts
need an
invariant
checked on
a complete
enumeration
of their
storage
(Section
7), which
we have
not done.
The
EIP-7002
contract,
below, is
the
exception,
and the
CircuitBreaker's
live
registry
passes its
finite
check,
evaluated
by a
script
rather
than by
the kernel
(Section
7).

WETH9 is
one of
eight
artifacts
(Table 2).
For the
protocol
itself,
the most
consequential
is the
EIP-7002
contract,
through
which
every
execution-layer
withdrawal
request
passes and
whose
failing
end-of-block
call would
invalidate
the block.
Whatever
its
storage
holds, as
long as
its
canonical
code is
installed,
the
client's
end-of-block
call to it
succeeds,
using at
most
210,000
gas, and
no history
removes
that code.
Each
block's
withdrawal
requests
are
exactly
the first
queued
entries,
at most
sixteen,
in order.
Each
request
records
exactly
one
committed
submission,
and no
submission
is
dropped,
duplicated,
reordered
or emitted
early.
These
theorems
need no
hash
premise
and no
entry
condition
on the
frames
that run
the
contract.
They cover
every fork
since the
contract's
use began
at
Prague's
activation,
and
mainnet's
state at
that
moment, as
providers
report it,
met their
starting
premises
(Section
5.1). The
other
results
replay
histories
exactly,
for the
deposit
contract's
accumulator,
3Crv's
ledger and
the
Uniswap
pair's
nested
calls,
with
share-value
bounds for
the pair
under a
premise on
the
tokens'
actual
answers.
They keep
the
CircuitBreaker's
registry
consistent
under
universal
hash
premises,
which a
finite
tier
replaces
by
decidable
checks for
a single
update.
And they
separate
the two
Curve pool
implementations,
executing
the
reentrancy
on the
vulnerable
bytes and
proving
that the
fixed
bytes'
lock
excludes
it.

This paper
makes the
following
contributions.

\begin{itemize}
\tightlist
\item
  \textbf{A
  corpus}
  of
  machine-checked
  theorems
  about
  eight
  artifacts
  deployed
  on
  Ethereum
  mainnet
  between
  2017 and
  2026,
  compiled
  by solc
  0.4.19,
  0.5.16,
  0.6.11
  and
  0.8.34
  and by
  Vyper
  0.2.4,
  0.2.15
  and
  0.3.7,
  or
  written
  in
  assembly,
  among
  them the
  protocol's
  own
  EIP-7002
  contract,
  whose
  end-of-block
  call is
  proved
  to
  succeed
  whatever
  its
  storage
  holds
  (Section
  4).
\item
  \textbf{A
  method}
  for
  bytecode
  produced
  by
  someone
  else's
  tools,
  with no
  compiler,
  no added
  axiom
  and no
  abstraction
  of the
  EVM in
  its
  trust
  base:
  kernel-checked
  certificates,
  histories
  replayed
  over
  exactly
  the
  calls
  whose
  effects
  survived,
  other
  contracts'
  answers
  as
  inputs,
  and hash
  premises
  over the
  keys a
  history
  touches
  or
  decidable
  by the
  kernel
  (Section
  3).
\item
  \textbf{Liveness
  from
  exact
  gas},
  from a
  single
  frame up
  to a
  signed
  transaction
  admitted
  after
  any
  valid
  history,
  with
  explicit
  premises
  about
  the
  callees
  a
  function
  depends
  on. The
  theorems
  prove
  admission
  and
  execution,
  not
  inclusion
  (Section
  3.8).
\item
  \textbf{Statements
  that can
  fail.}
  Kernel-checked
  counterexamples
  show
  that key
  premises
  cannot
  be
  dropped;
  one
  refutes
  a
  premise
  our own
  earlier
  statements
  made,
  and
  another
  the
  natural
  guarantee
  that
  every
  EIP-7002
  submission
  paid the
  EIP's
  fee
  (Sections
  3.10,
  4.2 and
  4.5).
\item
  \textbf{A
  case
  study in
  agent-written
  proof.}
  The lift
  was
  written
  by AI
  agents
  in
  fifteen
  days,
  under
  one
  author's
  direction,
  on top
  of an
  existing
  stack;
  the
  bottleneck
  was not
  writing
  proofs
  but
  knowing
  that
  their
  statements
  mean
  what
  they
  should
  (Section
  5.3).
\end{itemize}

Section 2
examines
the WETH9
theorem,
Section 3
the
semantics
and the
method,
Section 4
the corpus
and
Section 5
the
evaluation;
Sections 6
and 7
discuss
related
work and
limitations.

\subsection{2
One
theorem,
six
answers}\label{one-theorem-six-answers}

\subsubsection{2.1
The
statement}\label{the-statement}

Figure 1
shows the
theorem in
Lean,
abridged.
Its
conclusion
is a fact
about
\texttt{processTrans\penalty200{}acti\penalty200{}on},
Jaune's
function
for
executing
one
transaction
inside a
block,
which
block
processing
calls for
every
transaction
and the
conformance
fixtures
therefore
exercise.

\begin{figure}[!htbp]
\begin{minipage}{\linewidth}
\begin{Verbatim}[fontsize=\footnotesize]
-- Blanc/Lift/Weth9/LiveTx.lean, abridged. Omitted: implicit variables; the chain-id,
-- fee-field, nonce-bound and non-precompile premises; three of the nine conclusions.
-- `bal` abbreviates (future.state.getStor ca).get (balSlot E).
def withdrawIntrinsicGas (wad : B256) : Nat := 21000 + 4 * calldataTokens (withdrawCalldata wad)
def withdrawFrameGas (wad : B256) : Nat := if wad = 0 then 4440 else 13940
def withdrawRefund (bal wad : B256) : Nat := if wad ≠ 0 ∧ wad = bal then 4800 else 0
def withdrawGasUsed (bal wad : B256) : Nat :=
  withdrawIntrinsicGas wad + withdrawFrameGas wad - withdrawRefund bal wad

theorem weth9_history_tx_withdraw
    (trace : ConfiguredHistoryTrace cfg checkpoint future)            -- any valid history
    (installed : some (checkpoint.state.getCode ca).toList = weth9Sem.image)
    (sumNof : SumNof checkpoint.state.bal)                             -- total ether < 2^256
    (initial : FootInv K₀ (checkpoint.state.getStor ca) (checkpoint.state.bal ca))
    (fresh : KeysFresh K₀ (historyTouchedKeys ca trace))               -- trace-local hashes
    (hstate : benv.state = future.state) (hfork : CoveredFork benv.stat.fork)
    (htype : tx.type = .two chainId maxPriorityFee maxFee (some ca) [])
    (hvalue : tx.value = 0) (hdata : tx.data = withdrawCalldata wad)
    (hgas : withdrawIntrinsicGas wad + withdrawFrameGas wad + 811 ≤ tx.gas)
    (hcap : tx.gas ≤ 16777216)                                         -- EIP-7825
    (hroom : tx.gas ≤ benv.stat.blockGasLimit - bout.blockGasUsed)
    (hrecover : recoverSender benv.stat.chainId tx = .ok E)
    (hnonce : (benv.state.get E).nonce = tx.nonce) (hnocode : (benv.state.getCode E).size = 0)
    (hfunds : tx.gas * maxFee ≤ (benv.state.get E).bal.toNat)
    (hholder : historyKeyUniverse ca trace K₀ (.bal E))
    (hbal : wad ≤ bal) (hcbE : benv.stat.coinbase ≠ E) (hcbCa : benv.stat.coinbase ≠ ca) :
    ∃ st bout', processTransaction benv bout tx index = .ok (st, bout') ∧
      bout'.blockGasUsed = bout.blockGasUsed + withdrawGasUsed bal wad ∧
      st.getStor ca = (future.state.getStor ca).set (balSlot E) (bal - wad) ∧
      (∀ a, a ≠ ca → st.getStor a = future.state.getStor a) ∧
      (st.get ca).bal = future.state.bal ca - wad ∧
      ((future.state.bal E).toNat + wad.toNat < 2 ^ 256 →
        (st.get E).bal.toNat + withdrawGasUsed bal wad * price =
          (future.state.bal E).toNat + wad.toNat)          -- price: the effective gas price
\end{Verbatim}
\par\smallskip\normalsize
\emph{Figure
1. The
WETH9 exit
theorem,
abridged.
The full
statement
has 27
explicit
hypotheses
and nine
conclusions;
Appendix A
gives its
location.}
\end{minipage}
\end{figure}

The
hypotheses
fall into
five
groups.

\emph{The
history.}
\texttt{trace}
is any
configured
history
from
\texttt{checkpoint}
to
\texttt{future}
(Section
3.4): a
retained
run of
Jaune's
block
transition
over valid
blocks,
each under
one of the
four
covered
forks.
Apart from
the hash
premise,
nothing
restricts
which
transactions
it
contains
or how
other
contracts
call
WETH9.

\emph{The
checkpoint.}
WETH9's
bytes are
installed
at address
\texttt{ca},
the total
ether is
below
2²⁵⁶, and
the
footprint
invariant
\texttt{FootInv K₀}
holds for
a set
\texttt{K₀}
of tracked
keys:
every
nonzero
storage
word lies
at one of
the three
metadata
slots or
at a
tracked
key's
slot,
tracked
keys
occupy
distinct
slots
apart from
the
metadata
slots, and
the
tracked
balances
sum to at
most the
contract's
ether.
Executing
WETH9's
recorded
creation
input
leaves
storage
satisfying
the
invariant
with
nothing
tracked,
with no
hash
premise
(Section
3.9).

\emph{The
hashes.}
\texttt{fresh}
says that
each key
the
history
may touch,
five per
call to
the
contract
computed
from its
caller and
arguments,
is tracked
already or
lands on a
slot in no
use, and
that
distinct
such keys
occupy
distinct
slots: a
statement
about
finitely
many
concrete
hash
values,
which
anyone
holding
the
history
can
compute
(Section
3.7).

\emph{The
holder.}
\texttt{hholder}
says that
\emph{E}'s
balance
key is
tracked,
in
\texttt{K₀}
or among
the keys
the
history
touched.
\emph{E}
must have
no code,
which
excludes
contract
holders,
whose
withdrawals
carry a
premise
about the
callee
(Section
3.8), and
accounts
with an
EIP-7702
delegation
{[}EIP-7702{]}.

\emph{The
transaction.}
Everything
else
concerns
the
transaction
and its
block:
type,
value,
calldata
and gas
limits, a
signature
that
recovers
to
\emph{E},
\emph{E}'s
nonce and
funds, an
amount
\emph{w}
no larger
than
\emph{E}'s
WETH
balance,
room in
the block,
and a fee
recipient
other than
\emph{E}
and the
contract,
whose fee
would
otherwise
mix into
the
balances
the
theorem
computes.
Signature
recovery
is a
premise
because
the
theorem
ranges
over every
signed
transaction;
for a
concrete
transaction
the kernel
can
compute
it, as it
does in
the closed
instance
of Section
7.

The
theorem
executes
the
transaction
at the
state the
history
reaches,
the state
at the end
of a
block. A
companion
theorem,
\texttt{weth9\_\allowbreak{}prefix\_\allowbreak{}tx\_\allowbreak{}withdraw},
puts the
withdrawal
where real
transactions
run: after
the
block's
two system
calls to
the
EIP-4788
and
EIP-2935
contracts
{[}EIP-4788;
EIP-2935{]}
and after
any
\emph{n}
of its
transactions.
Given a
configured
history up
to the
start of a
block, the
retained
execution
of that
block, and
one
freshness
premise
that also
covers the
keys of
every
WETH9
frame
entered
before the
cut,
rolled-back
frames
included,
it derives
the
installed
code, the
bound on
total
ether and
the
footprint
invariant
at the
cut; and
if the
block's
next
transaction
is a
withdrawal
meeting
the
transaction
premises
there, it
is
admitted
with the
same exact
conclusions,
and its
result is
the one
the block
actually
produced.
The
theorem
concerns a
block
whose
whole
execution
is given:
it does
not
construct
a block
from
transactions
not yet
validated,
and it is
proved for
WETH9
only.

\subsubsection{2.2
The
conclusion}\label{the-conclusion}

The
conclusion
is the
holder's
guarantee:
the
transaction
is
admitted,
the
holder's
balance
slot is
debited by
exactly
\emph{w},
no other
account's
storage
changes,
the
contract's
ether
falls by
\emph{w},
and, net
of fees,
the holder
receives
exactly
\emph{w}:
provided
their
ether
balance
plus
\emph{w}
stays
below
2²⁵⁶,
their new
balance
plus
\texttt{gasUsed}
times the
effective
gas price
equals
their old
balance
plus
\emph{w}.

Every term
of the gas
used can
be read
off the
bytecode
and the
fork
rules. The
intrinsic
cost is
the 21,000
every
transaction
pays plus
four gas
per
calldata
token. The
frame
cost,
13,940 for
a nonzero
withdrawal,
is the
instructions'
fixed
costs
(2,040), a
cold and
then a
warm read
of the
balance
slot
(2,100 and
100), a
write that
changes a
nonzero
original
value
(2,900),
and the
ether
transfer
to the
already-warm
sender
(6,800).
When
\emph{w} =
0, the
write and
the
transfer
cost 100
each and
the frame
costs
4,440.
When a
nonzero
withdrawal
empties
the
balance,
clearing
the slot
earns the
4,800-gas
refund of
EIP-3529
{[}EIP-3529{]}.
The refund
cap of one
fifth of
the gas
used and
the
calldata
floor of
EIP-7623
{[}EIP-7623{]}
never bind
for this
call,
which the
proof
establishes
rather
than
assumes.
The margin
of 811 gas
in the
gas-limit
premise is
gas the
transaction
must hold
but does
not use.
For a
nonzero
withdrawal
it comes
from the
ether
transfer:
the
\texttt{CALL}
is charged
9,100 gas
before the
2,300-gas
stipend it
gives the
recipient
comes back
unused, so
the frame
must then
hold 2,300
gas more
than the
transfer's
net cost
of 6,800.
Of the
frame's
cost,
1,489 gas,
mostly the
\texttt{Withdrawal}
event's
\texttt{LOG2},
falls
after the
call, and
2,300 −
1,489 =
811. A
zero
withdrawal
transfers
no ether
and needs
a smaller
margin,
set by the
rule that
a storage
write
fails
unless
more than
2,300 gas
remains
{[}EIP-2200{]};
the
premise
keeps 811
for both.

For the
one-wei
withdrawal
that
empties
the
balance in
the closed
instance
of Section
7, the
terms are
these:

{\def\LTcaptype{none} % do not increment counter
\par\noindent\begin{minipage}{\linewidth}\footnotesize
\begin{tabular}{@{}
  >{\raggedright\arraybackslash}p{(\linewidth - 2\tabcolsep) * \real{0.8881}}
  >{\raggedleft\arraybackslash}p{(\linewidth - 2\tabcolsep) * \real{0.1119}}@{}}
\toprule\noalign{}
\begin{minipage}[b]{\linewidth}\raggedright
Quantity
\end{minipage}
&
\begin{minipage}[b]{\linewidth}\raggedleft
Gas
\end{minipage} \\
\midrule\noalign{}

Intrinsic:
21,000 +
4·31 zero
bytes +
16·5
nonzero
bytes &
21,204 \\
Frame &
13,940 \\
Consumed
before the
refund &
35,144 \\
Refund for
clearing
the slot &
4,800 \\
\texttt{gasUsed},
as the
receipt
records it
&
30,344 \\
Gas limit
the
theorem
requires,
at least:
35,144 +
811 &
35,955 \\
\bottomrule\noalign{}
\end{tabular}
\end{minipage}\par
}

A gas
limit
equal to
the
receipt's
30,344
does not
meet the
theorem's
premise,
and 35,955
is
sufficient,
not shown
to be
minimal.

\subsubsection{2.3
What the
answers
rest
on}\label{what-the-answers-rest-on}

Two
answers
deserve
care. The
theorem is
formally
about
whatever
account
holds the
3,124
lifted
bytes, and
reaches
WETH9
because
three
providers
return
those
bytes for
\texttt{0xC02a\pfvellipsis{}6Cc2},
which we
have not
proved
from a
block
header.
And the
machine is
the whole
machine as
Jaune
defines it
and the
fixtures
test it:
validation,
intrinsic
gas, the
warm sets
of
EIP-2929
{[}EIP-2929{]},
the gas
cap of
EIP-7825
{[}EIP-7825{]}
from Osaka
on,
refunds
and the
fee paid
to the
block's
fee
recipient
all take
part in
the proof.

Besides
the
installed
code and
the bound
on total
ether,
three
hypotheses
are facts
about a
particular
chain
rather
than about
the
transaction:
the
invariant
at the
checkpoint,
the
freshness
of the
history's
keys, and
the
tracking
of the
holder.
Freshness
and
tracking
are finite
properties
of a
concrete
history.
The
invariant
is
established
by our
deployment
theorem
for a
deployment
executed
under
current
rules; for
the live
contract
it would
have to be
checked on
its
storage,
as Section
7
describes.

\subsubsection{2.4
The
formula
against
the
chain}\label{the-formula-against-the-chain}

A gas
formula
proved
against a
model is
only as
good as
the
model's
gas
schedule,
which the
fixture
corpus
tests; we
also
compared
the WETH9
formula
with
mainnet
directly.
Through
WETH9's
\texttt{Withdrawal}
events we
found
every
successful
transaction
that
called its
\texttt{withdraw}
directly
in the
30,000
blocks
from
26,111,230
to
26,141,229
(3 October
2026,
10:31 UTC,
to 7
October,
14:53
UTC),
6,823 in
all. Those
outside
the shape
the
theorem
describes
were set
aside by
criteria
that do
not look
at gas:
1,859
carry
calldata
other than
the 36
bytes of
\texttt{withdraw(w)},
1,070 come
from
accounts
with code,
and 114
are of
type 0.
The
remaining
3,780 are
of type 2
and carry
no value
and no
access
list. Each
sender
holds no
code at
either end
of the
block, and
its nonce
rose only
by its own
transactions,
so no
EIP-7702
delegation
was set or
removed
within the
block. The
prediction
also needs
the
balance
the
withdrawal
starts
from, to
tell
whether it
empties
it.
WETH9's
source
logs an
event for
every
change to
a balance,
so when no
other
event of
the block
names the
sender,
that
balance is
the
sender's
balance at
the end of
the
previous
block.
This holds
for 3,748
of the
3,780,
from 2,255
senders;
the other
32 share
their
block with
another
WETH9
event
naming the
sender,
and we set
them
aside.
Every one
of the
3,748
receipts
equals the
formula's
prediction;
none
withdraws
zero.

{\def\LTcaptype{none} % do not increment counter
\begin{table}[!htbp]
\begin{minipage}{\linewidth}\footnotesize
\begin{tabular}{@{}
  >{\raggedleft\arraybackslash}p{(\linewidth - 8\tabcolsep) * \real{0.1464}}
  >{\raggedleft\arraybackslash}p{(\linewidth - 8\tabcolsep) * \real{0.2620}}
  >{\raggedleft\arraybackslash}p{(\linewidth - 8\tabcolsep) * \real{0.1843}}
  >{\raggedleft\arraybackslash}p{(\linewidth - 8\tabcolsep) * \real{0.2231}}
  >{\raggedleft\arraybackslash}p{(\linewidth - 8\tabcolsep) * \real{0.1843}}@{}}
\toprule\noalign{}
\begin{minipage}[b]{\linewidth}\raggedleft
Calldata
tokens
\end{minipage}
&
\begin{minipage}[b]{\linewidth}\raggedleft
Predicted
gas, not
emptying
\end{minipage}
&
\begin{minipage}[b]{\linewidth}\raggedleft
Receipts
that match
\end{minipage}
&
\begin{minipage}[b]{\linewidth}\raggedleft
Predicted
gas,
emptying
\end{minipage}
&
\begin{minipage}[b]{\linewidth}\raggedleft
Receipts
that match
\end{minipage} \\
\midrule\noalign{}

60 &
35,180 & 1
of 1 &
30,380 &
12 of
12 \\
63 &
35,192 &
128 of 128
& 30,392 &
238 of
238 \\
66 &
35,204 &
481 of 481
& 30,404 &
1,097 of
1,097 \\
69 &
35,216 &
465 of 465
& 30,416 &
900 of
900 \\
72 &
35,228 &
150 of 150
& 30,428 &
250 of
250 \\
75 &
35,240 & 2
of 2 &
30,440 &
24 of
24 \\
\bottomrule\noalign{}
\end{tabular}
\par\smallskip\normalsize
\emph{Table
1. 3,748
mainnet
withdrawals
against
the
formula,
by
calldata
tokens and
by whether
the
withdrawal
empties
the
sender's
balance.
Appendix C
gives one
transaction
for each
case.}
\end{minipage}
\end{table}
}

Two
excluded
groups
still
admit a
prediction,
and their
98
admissible
calls of
type 0 and
1,590 with
calldata
of 37 to
76 bytes
match too;
we count
them
apart.
Calls from
accounts
with code
cannot be
predicted,
since the
ether
transfer
runs the
sender's
code, and
the scan
sees
successful
withdrawals
only.

The
comparison
is not
part of
the proof.
It checks,
on these
transactions,
something
no theorem
can, that
the
formula
agrees
with what
Ethereum's
clients
charged;
it
exercises
only
nonzero
withdrawals
and one
fork,
BPO2, and
says
nothing
about the
theorem's
other
hypotheses,
which we
did not
establish
for these
transactions.

\subsection{3
From bytes
to
theorems}\label{from-bytes-to-theorems}

Blanc's
results
are built
in layers:
Jaune's
semantics,
a
kernel-checked
reading of
the bytes,
generic
tools for
proving
things
about that
reading, a
per-contract
model and
invariant,
and
theorems
at four
levels of
execution,
from a
single
call frame
up to an
arbitrary
chain
history.
Figure 2
shows the
path.

\begin{figure}[!htbp]
\begin{minipage}{\linewidth}
\begin{Verbatim}[fontsize=\footnotesize]
 runtime bytes ──► untrusted producer ──► certificate: entries, abstract stacks, trees
      │                                         │
      └──► Cert.check, Cert.jumpsOk (kernel) ◄──┘
                          │
   lift_sound   every successful run of the bytes is a run of the trees    (safety)
   cursor       every step of a run of any outcome sits at a checked node  (all outcomes)
   lift_exact   every gas-exact run of the trees is a run of the bytes     (liveness)
                          │
   per-contract model, invariant and deployment theorem
                          │
   frame ──► message ──► transaction ──► configured history
\end{Verbatim}
\par\smallskip\normalsize
\emph{Figure
2. The
lift
path.}
\end{minipage}
\end{figure}

\subsubsection{3.1
The
semantics:
Jaune}\label{the-semantics-jaune}

Jaune is
an
executable
formal
specification
of the EVM
and of
Ethereum
block
processing
in Lean 4
{[}Jaune{]},
first
written
from the
Yellow
Paper
{[}YellowPaper{]}
and then
rebuilt on
a pinned
2025
revision
of
Ethereum's
executable
specification,
\texttt{execution-specs},
with Osaka
and the
two
blob-parameter
forks
implemented
from their
EIPs. Its
library of
about
45,500
lines
defines
the
machine,
all
transaction
types,
block
validation,
the
processing
of system
calls,
withdrawals
and
requests,
every
precompile
from
\texttt{ecrecover}
to the
BLS12-381
operations
and
\texttt{P256VERIFY},
and the
selection
of fork
rules by a
chain's
activation
schedule.
The same
definitions
run
against
test
fixtures
and appear
in proofs.

\textbf{Conformance.}
Jaune's
current-mainnet
lane runs
the
blockchain
tests of
the
fixture
release
\texttt{tests@v20.0.2}
and
requires
every one
whose
forks lie
among
Prague,
Osaka,
BPO1 and
BPO2,
transitions
included,
to pass.
We re-ran
the lane
for this
paper at
the pinned
revision:
all 5,006
selected
files
passed,
holding
34,205
cases,
among them
the 47
that cross
into BPO1
or BPO2
and so
exercise
those
forks
(Appendix
E).

\textbf{Forks.}
Four forks
have
governed
mainnet
since 7
May 2025:
Prague
from that
day, Osaka
from 3
December
2025, BPO1
from 9
December
2025 and
BPO2 from
7 January
2026
{[}EIP-7600;
EIP-7607{]}.
Jaune
implements
each as a
complete
record of
rule data,
selected
per block
by the
chain's
schedule.
It also
runs a
prerelease
of
Amsterdam,
the next
upgrade,
against
the
fixtures
of the
Glamsterdam
devnet-8.
Amsterdam
activated
on the
Sepolia
test
network on
6 October
2026 and,
as of 10
October
2026, has
no mainnet
date
{[}Glamsterdam{]};
no theorem
in this
paper is
stated for
it.

\textbf{One
transition
core for
tests and
theorems.}
Blanc's
history
theorems
quantify
over runs
of
\texttt{stateTransit\penalty200{}ionU\penalty200{}sing\penalty200{}},
Jaune's
configured
block
transition,
and the
fixture
runner
imports
each block
through a
function
whose core
is
\texttt{stateTransit\penalty200{}ionE\penalty200{}},
which
\texttt{stateTransit\penalty200{}ionU\penalty200{}sing\penalty200{}}
also
reaches,
so the
fixtures
test, as
compiled
code, the
per-block
transition
the
theorems
describe;
Appendix E
says which
steps of
the
configured
transition
only the
transition
cases
exercise.

\textbf{Trust
surface.}
Jaune's
library
contains
no axiom,
\texttt{sorry},
\texttt{native\_decide},
\texttt{@[extern]},
\texttt{opaque},
\texttt{partial}
or
\texttt{implemented\_by},
as a
continuous-integration
check
enforces,
with one
allowlisted
\texttt{@[csimp]}
substitution
proved
equal to
the
definition
it
replaces
{[}Jaune{]}.
For
SHA-256, a
theorem
proves
Jaune's
optimized
implementation
equal on
every
input to a
transcription
of NIST
FIPS 180-4
{[}FIPS180-4{]};
it lies in
a Jaune
module
outside
the Blanc
library's
import
closure,
whose
axioms no
gate
audits
(Appendix
E). The
replay of
Section
3.11
includes
the
module.
For
Keccak-256,
only the
optimized
permutation
is proved
equal to a
reference
transcription;
the
transcription's
conformance
to the
standard
rests on
test
vectors
and the
fixtures,
one reason
no theorem
here
relies on
Keccak's
quality as
a hash
function
(Section
3.7).

\subsubsection{3.2
Reading
bytes as a
program}\label{reading-bytes-as-a-program}

EVM
bytecode
has no
structured
control
flow: a
\texttt{JUMP}
takes its
destination
from the
stack, and
compilers
implement
internal
functions
by jumping
through
pushed
return
addresses.
Recovering
the
structure
is a
heuristic
task, long
studied by
decompilers
{[}Vandal;
Gigahorse;
Elipmoc;
EtherSolve{]},
and Blanc
does not
trust
whatever
performs
it.

A
\emph{certificate}
is a list
of
entries.
Each entry
gives a
program
counter,
an
abstract
stack of
the
operand
words
above the
current
function's
base (each
a known
constant,
the
function's
own return
address,
or
unknown),
and a tree
over a
small
instruction
language
whose
nodes
correspond
one to one
with
instructions
of the
bytes:
straight-line
instructions,
branches,
gotos,
calls of
internal
functions
with their
continuations,
returns
and halts.
A
2,836-line
Python
program
proposes
certificates
from the
raw bytes,
and
nothing it
does can
make an
unsound
certificate
pass. The
eight
artifacts'
runtime
certificates
range from
7 entries,
for the
EIP-7002
contract,
to 211,
for the
vulnerable
Vyper
pool.

The trust
boundary
is a Lean
function,
\texttt{Cert.check},
which
walks
every tree
against
the
literal
bytes. It
checks
that each
node's
instruction
is the one
encoded at
that
position,
tracks the
abstract
stack with
a transfer
function
proved
sound
against
Jaune's
semantics,
and
requires
every jump
to land on
the
constant
destination
its tree
names and
every call
and return
to pair up
through a
constant
return
address. A
companion
check,
\texttt{Cert.jumpsOk},
confirms
that every
jump
target is
a valid
jump
destination,
which
liveness
needs. The
checks are
closed by
evaluation
in Lean's
kernel
(\texttt{decide +kernel}),
split into
parts that
each fit
in memory,
and a gate
regenerates
every
registered
certificate
from its
committed
input and
compares
it byte
for byte.

New
producers
of
bytecode
have
needed
small,
general
extensions
rather
than new
trust,
each
proved
sound once
for all
certificates:
a
conditional
jump whose
taken side
is a back
edge, for
WETH9's
loops;
constant
folding,
nodes that
push their
own
program
counter,
decided
conditional
jumps and
a variant
that
tracks a
constant
memory
map, for
the
vulnerable
Vyper
pool; and
folding of
the
\texttt{AND}
with which
solc
0.5.16
masks
internal
function
addresses,
for the
Uniswap
pair. The
hand-written
EIP-7002
contract
needed no
extension.

Generic
theorems,
proved
once for
all
certificates,
turn a
passing
check into
reasoning
power:
\texttt{lift\_sound}
says that
every
successful
execution
of the
bytes from
program
counter 0,
on a
covered
fork, is a
run of the
certificate's
program,
so that
safety
properties
of the
program
hold of
the bytes.
\texttt{lift\_exact}
is the
converse
for
liveness:
every
gas-exact
run of the
program is
a
successful
execution
of the
bytes. A
third, the
cursor of
Section
3.3,
covers
executions
of any
outcome.
The design
is
translation
validation
{[}Pnueli98{]}
run
backwards,
validating
a
decompilation
instead of
a
compilation
{[}Myreen08{]}:
the
producer
is free to
be clever,
and to
change,
without
entering
the
trusted
base.

\subsubsection{3.3
Proving
things
about
lifted
programs}\label{proving-things-about-lifted-programs}

Per-contract
proofs use
three
generic
tools, and
a fourth
for the
reentrancy
lock.
\emph{Inversion
walks}
establish
safety.
From a
successful
run of a
tree node,
a lemma
for each
kind of
instruction
recovers
the
successor
state;
chaining
these
lemmas
along a
function's
tree
replays
every path
the bytes
can take,
and a
contract's
safety
proof is
such a
walk,
tracking
what each
path does
to
storage.
Loops are
walked by
induction,
among them
the
EIP-7002
fee series
and the
Uniswap
pair's
Babylonian
square
root.

\emph{Exact
walks}
establish
liveness
and its
cost,
building a
gas-exact
run
forward,
one node
at a time.
Each step
lemma is
stated
backwards,
with its
charge
written as
a numeral
wherever
the charge
is fixed:
the run of
a node
from a
state with
gas
\emph{G +
c} follows
from the
run of its
continuation
from gas
\emph{G}.
Applied to
a goal
whose gas
is \emph{g
+ N},
Lean's
unification
of
natural-number
offsets
solves the
continuation's
gas as
\emph{g} +
(\emph{N}
−
\emph{c}),
so the gas
obligations
of a walk
disappear
into
unification,
and
\texttt{lift\_exact}
turns the
finished
run into a
Jaune
execution.

\emph{The
certificate
cursor}
extends
\texttt{lift\_sound}
to
executions
of any
outcome.
It places
every node
of a
frame's
chain of
execution
at a
checked
point of
the
certificate,
whatever
the frame
eventually
does, and
no
statement
about it
has a
success
premise.
The
reentrancy
theorems
need it,
because a
lock must
stop
reentry
even into
a frame
that later
reverts,
and the
committed-replay
theorems
use it to
account
for every
call a
frame can
make.

\emph{The
lock
certificate}
serves the
fixed
Vyper
pool: a
second
Boolean
pass over
the
certificate,
with the
lock state
at join
points
supplied
by a
444-line
untrusted
annotator,
decides
that every
guarded
body is
reached
only after
the lock
was found
free, that
every
mutating
body
starts
with the
lock held,
and that
inside it
only
release
points
write the
lock's
slot; a
soundness
theorem,
proved
along the
cursor,
connects
the pass
to every
frame that
runs the
code from
program
counter 0
on a
covered
fork,
provided
the hashes
the frame
computes
avoid the
lock's
slot.

\subsubsection{3.4
Four
levels and
the
configured
history}\label{four-levels-and-the-configured-history}

Blanc
states
results at
four
levels: a
\emph{frame}
theorem
concerns
one
execution
of the
contract's
code from
program
counter 0;
a
\emph{message}
theorem,
Jaune's
message
processing,
including
child
frames,
reverts
and
settlement;
a
\emph{transaction}
theorem,
\texttt{processTrans\penalty200{}acti\penalty200{}on},
with
validation,
intrinsic
gas,
refunds
and fees;
and a
\emph{history}
theorem, a
chain. A
\texttt{ConfiguredHi\penalty200{}stor\penalty200{}yTra\penalty200{}ce cfg checkpoint future}
is a
sequence
of block
transitions
under the
chain
configuration
\texttt{cfg},
from
\texttt{checkpoint}
to
\texttt{future},
each
retaining
the
evidence
of its own
execution:
the fork
the
block's
timestamp
selected,
a proof
that this
fork is
covered,
the
successful
run of
\texttt{stateTransit\penalty200{}ionU\penalty200{}sing\penalty200{}},
and the
retained
trace of
the block
body, with
its system
calls,
transactions,
withdrawals
and
requests,
down to
every
interpreter
frame,
including
frames
whose
effects
were later
rolled
back.
Quantifying
over
traces
loses
nothing,
since
every
reachability
derivation
of the
configured
chain
under
covered
forks has
a trace,
and
everything
on such a
chain is
in scope,
including
other
contracts,
reentrant
and
reverted
calls,
system
calls and
withdrawals.

\subsubsection{3.5
Exactly
what was
committed}\label{exactly-what-was-committed}

A history
contains
calls
whose
effects
did not
survive: a
frame that
reverted,
or one
that
succeeded
under an
ancestor
that
reverted.
Jaune's
semantics
records
settlement,
and Blanc
extracts
from a
configured
history
the list
of
\emph{settlement-committed}
calls to a
contract,
in
execution
order.
This is
what makes
the
history
theorems
exact. For
WETH9, the
ledger
read from
storage
after any
history
equals the
model's
ledger run
over
exactly
the
committed
non-static
writer
calls:
rolled-back
frames are
absent,
view calls
contribute
nothing,
and a call
re-entered
from
inside a
withdrawal's
ether
transfer
is placed
after the
withdrawal
that made
it. The
Uniswap
pair needs
a nested
form,
because it
calls
tokens and
a callback
that may
call it
back: its
replay
consists
of
\emph{steps},
one for
each
outermost
committed
non-static
call to
the pair,
each
consuming
the
committed
calls to
the pair
made
during it,
and an
equation
places
every
committed
non-static
call to
the pair
in exactly
one step,
so no step
is
invented
and none
is lost.
These
statements
extract
the
history;
they do
not
abstract
it.

The
opposite
choice is
right for
locks. The
reentrancy
theorems
of Section
4.7
quantify
over
Jaune's
\emph{raw}
execution
tree,
every
frame
whatever
its
outcome: a
callee
that later
reverts
has still
executed,
and
read-only
reentrancy
needs no
commitment
at all.

\subsubsection{3.6
Other
contracts'
answers as
inputs}\label{other-contracts-answers-as-inputs}

A contract
that calls
others
cannot be
proved
correct,
or live,
for every
callee: a
token can
report any
balance,
and a
callee can
revert or
consume
all the
gas it is
given. The
question
is where
premises
about
callees
go. The
Uniswap
pair,
which
calls its
tokens,
its
factory, a
callback
recipient
and the
\texttt{ecrecover}
precompile,
keeps two
layers
apart. In
the first,
the
answers
are
inputs.
The model
runs one
\emph{entry}
at a time,
a call to
one of the
contract's
public
functions
with its
arguments
decoded,
and
consumes a
transcript
of the
answers
that call
received.
The
refinement
theorems
say that
each
successful
frame of
the bytes
agrees
with the
model's
run of the
entry the
frame
decoded,
fed
exactly
the
answers
the
frame's
own calls
received.
And every
step of a
history
replay is
\emph{authenticated}:
its entry
is the one
its own
frame
decoded,
and its
transcript
is
consumed
\emph{positionally}.
Each
answer is
tied to
the actual
call that
returned
it, with
its reply,
whether
that call
entered a
frame or
was
answered
by a
precompile,
and its
nested
calls to
the pair,
each
admitted
in the
same way
in turn.
The
replay,
ledger,
oracle and
refinement
theorems
are stated
at this
layer and
take no
premise
about any
callee. In
the second
layer, the
economic
statements
take a
named
premise
about the
answers
that
ranges
only over
the steps
of the
history at
hand: a
checkable
property
of a
concrete
history,
not an
assumption
about
every
token that
might ever
exist.

\subsubsection{3.7
Hashes
without
added
axioms}\label{hashes-without-added-axioms}

Solidity
and Vyper
keep
mappings
at hashed
storage
slots,
such as
\texttt{keccak256(\allowbreak{}a ‖ 3)}
for
\texttt{balanceOf[a]},
and a
proof
about such
storage
must know
that the
slots it
reasons
about are
distinct,
from each
other and
from the
fixed
slots. One
shortcut
is to
assume
that
Keccak is
injective,
which is
false,
since the
hash maps
an
unbounded
domain
into 2²⁵⁶
values;
DeepSEA's
authors,
who adopt
such an
axiom,
call it
``actually
false, but
`true
enough'\,''
{[}DeepSEA{]}.
Others
assume
facts
about
particular
hashes
{[}EquiVM{]}
or
constrain
an
uninterpreted
hash by
injectivity
assumptions
{[}hevm{]}.
Blanc adds
no axiom
about
Keccak:
each
theorem
states as
a premise
exactly
the
separation
facts its
proof
uses, over
finitely
many slots
wherever
it can.

\emph{Trace-local
premises.}
A
footprint
names a
set of
tracked
keys whose
slots are
distinct
and off
the fixed
slots, and
says that
every
nonzero
word of
storage is
at a fixed
slot or a
tracked
slot. A
history
extends
the
footprint
by the
keys it
may touch,
which
needs one
premise:
each such
key is
tracked
already or
lands on a
slot in no
use, and
distinct
such keys
land on
distinct
slots. For
WETH9 the
keys are
five per
call,
computed
from its
caller and
the first
two words
of its
arguments
whatever
the
function
or
outcome,
an
over-approximation
that one
pass over
a concrete
history
can check.
For WETH9
deployed
from
scratch,
it says
that no
two of
those keys
land on
one slot
and none
on slots
0, 1 or 2:
a fact
about
specific
hash
values
that
computational
collision
and
preimage
resistance
do not
imply, and
for which
no
reduction
is proved.
It is also
necessary:
a written
allowance
slot
coinciding
with a
tracked
balance
slot can
break
backing.
3Crv and
the
Uniswap
pair take
premises
of the
same kind,
over the
mapping
rows their
histories
touch. The
fixed
Vyper
pool's one
hash
premise is
local too,
to a
single
execution:
no
Keccak-256
digest the
pool
computes
in that
execution
equals 0,
the lock's
slot.

\emph{Decidable
premises.}
When the
keys are
concrete,
separation
becomes
computation:
in the
CircuitBreaker's
finite
tier every
separation
premise is
a
decidable
check on
explicit
key lists,
with an
exact
soundness
theorem.
The kernel
can
evaluate
such a
check,
Keccak
included,
as it does
for one
example
state,
just as
the 3Crv
deployment
theorem
computes
the hashed
slot its
constructor
writes.
The
deposit
and
EIP-7002
contracts
keep
nothing at
hashed
slots and
need no
hash
premise.

\emph{Universal
premises.}
Among the
results
reported
here,
exactly
one family
still
quantifies
separation
over all
2¹⁶⁰
addresses:
the
history
tier of
the
CircuitBreaker
(Section
4.6). The
lift holds
one more
such
family, a
WETH9
solvency
invariant
(\texttt{weth9\_\allowbreak{}history\_\allowbreak{}preserves\_\allowbreak{}solvent})
and the
frame
lemmas
beneath
it, which
no result
reported
here uses.
Under a
random-function
model of
Keccak,
the chance
that such
a premise
fails is
about
\emph{q}·2⁻⁹⁴
per frame
for
\emph{q}
written
slots,
from the
registry's
four key
families,
each
ranging
over 2¹⁶⁰
addresses
or
indices,
among 2²⁵⁶
slots
(4·2¹⁶⁰/2²⁵⁶
= 2⁻⁹⁴),
but this
estimate
is a
heuristic:
no
reduction
or bound
is proved.

\subsubsection{3.8
Exact gas
and the
liveness
ladder}\label{exact-gas-and-the-liveness-ladder}

Safety is
proved by
induction
over
executions
that
exist;
liveness
requires
exhibiting
one, and
on the EVM
whether
one exists
depends on
gas. Jaune
meters gas
exactly,
memory
expansion,
warm and
cold
access,
refunds
and the
63/64 rule
included,
and
Blanc's
liveness
theorems
state
costs
exactly:
for WETH9
a closed
formula;
for the
deposit
contract
an
explicit
sum whose
terms
depend on
the
deposit's
position
in the
tree; for
the
Uniswap
pair a
closed
expression
in the
access
charges of
the actual
slots and
the gas
forwarded
to each
callee;
and for
the
EIP-7002
contract a
fixed
charge
plus 87
gas for
each
iteration
of the fee
loop plus
the
charges of
the actual
storage
accesses.
For 3Crv
the cost
is exact
but
existential:
for each
accepted
call other
than
\texttt{set\_name}
there is a
constant
\emph{c}
such that,
for every
\emph{G}
above the
2,300-gas
call
stipend,
the call
succeeds
when given
\emph{G +
c} and
leaves
exactly
\emph{G};
the proofs
construct
each
\emph{c},
but the
exported
theorem
names no
formula.

Liveness
is proved
on four
rungs: a
fresh
\emph{frame}
with
enough gas
succeeds
and costs
exactly
the stated
amount; a
call the
\emph{model}
accepts is
realized
by an
execution
that ends
in the
model's
next
state;
this holds
at the
final
state of
any
configured
history,
the
\emph{reachable-state}
rung; and
a signed
\emph{transaction}
is
admitted
by
\texttt{processTrans\penalty200{}acti\penalty200{}on}
and
succeeds,
with
\texttt{gasUsed}
in closed
form.
WETH9
reaches
all four;
the
deposit
contract,
3Crv, the
Uniswap
pair and
the
EIP-7002
contract
reach the
third.

A function
whose
success
depends on
an
external
call can
be live
only
conditionally,
since the
callee can
revert or
consume
everything
forwarded
to it.
Blanc
states
each such
premise
explicitly,
in the
theorem
that needs
it:
WETH9's
withdrawal
to a
contract
assumes
\texttt{SendOk},
that the
call
succeeds,
leaves at
least
1,489 gas
and
changes
none of
WETH9's
storage;
3Crv's
\texttt{set\_name}
assumes
\texttt{OwnerCallOk}
of the
minter's
\texttt{owner()};
and the
Uniswap
pair's
\texttt{sync},
\texttt{mint},
\texttt{burn},
\texttt{swap},
\texttt{skim}
and
\texttt{permit}
take the
replies
and
returned
gas of
their
calls.
Withdrawal
to an
account
without
code needs
no such
premise.
The
EIP-7002
contract
needs
liveness
in a
different
sense: its
end-of-block
call is
made by
the
client,
and Blanc
proves
that it
succeeds
whatever
the
contract's
storage
holds
(Section
4.5).

\subsubsection{3.9
Deployment}\label{deployment}

An
invariant
stated at
a
checkpoint
is useful
only if
some real
state
satisfies
it. For
WETH9, the
deposit
contract,
3Crv and
the
EIP-7002
contract,
Blanc
executes
the
recorded
creation
input of
the actual
deployment
transaction
as a
CREATE
message,
under
every
covered
fork, and
proves
that it
succeeds,
installs
exactly
the
certified
runtime,
and leaves
storage
satisfying
the
history
theorems'
checkpoint
invariant,
with no
hash
premise.
These
deployment
theorems,
and the
CircuitBreaker's,
also prove
in the
kernel, by
evaluating
the RLP
encoding
and
Keccak-256,
that the
recorded
deployer
and nonce
yield the
contract's
mainnet
address.
The
Uniswap
pair's
deployment
and
initialization,
and the
CircuitBreaker's
two tiers,
have
deployment
theorems
of their
own
(Sections
4.4 and
4.6). All
of these
are
modeled
deployments:
they run
the
historical
input
under
current
rules and
do not
prove that
the
historical
transaction
was
included
in the
chain;
only the
CircuitBreaker,
deployed
under
BPO2, was
deployed
under
rules
Jaune
covers.
For the
two Curve
pool
implementations,
deployment
is the
first step
of
executed
message
sequences
(Section
4.7).

\subsubsection{3.10
Premises
that
bite}\label{premises-that-bite}

A premise
never
needed, or
a
statement
that would
survive a
change to
its code,
says less
than it
seems, so
Blanc also
proves
controls:
closed
counterexamples,
or
theorems
about
every run
in which a
premise
fails. For
WETH9, a
written
allowance
slot that
coincides
with a
tracked
balance
slot
breaks
backing.
For the
EIP-7002
contract,
a
two-block
history
that meets
every
other
premise of
the
request
theorem
but holds
code at
the system
address
violates
its
conclusion,
and a
stored
excess
near 2²⁵⁶
makes the
bytecode's
bookkeeping
depart
from the
EIP's
recurrence.
For the
Uniswap
pair, the
premise on
token
answers,
the callee
premises
of
liveness,
the
trace-local
hash
premise,
the
112-bit
guard and
the 32-bit
timestamp
wrap are
each shown
to matter,
some in
the typed
model and
some at
the bytes
(Appendix
A); in the
typed
model, a
mint that
rounds up
breaks the
share-value
statement,
a burn
that
rounds
toward the
user
changes an
accepted
run, and a
different
fee
constant
accepts a
swap the
transcribed
model
rejects;
and a
deployment
address
one bit
off in its
salt or
init-code
hash is a
different
address.
For the
fixed
Vyper
pool,
three
executed
witnesses
show that
the
exclusion
theorem's
antecedent
can hold.

Two
controls
record
differences
between
code and
what was
believed
of it. An
earlier
family of
deposit-contract
statements
assumed
that no
system
contract's
code
contains a
call- or
create-type
instruction
byte at
any
offset,
and Blanc
proves
that
premise
false of
the
canonical
EIP-7002
code
(Section
4.2). And
the
guarantee
that every
committed
EIP-7002
submission
paid at
least the
fee the
EIP
specifies,
which the
contract's
auditors
had
already
shown to
fail in
principle,
is false
of the
deployed
bytes by a
closed
counterexample
(Section
4.5).

\subsubsection{3.11
What the
kernel
checks,
and what
it does
not}\label{what-the-kernel-checks-and-what-it-does-not}

\textbf{Axioms.}
A script
walks the
dependency
graph of
every
declaration
in the
library,
not only
of its
theorems,
from
scratch,
and fails
unless the
union of
the axioms
reached is
within
Lean's
three. One
walk from
all
102,116
constants
of the
library's
1,484
modules
reached
\texttt{propext},
\texttt{Classical.\allowbreak{}choice}
and
\texttt{Quot.sound}
and
nothing
else, and
so bounds
\texttt{sorryAx},
the axiom
behind
native
evaluation,
and any
bespoke
axiom, for
every
declaration
at once
(Appendix
E). We do
not rely
on Lean's
\texttt{\#print axioms},
which can
under-report
axioms
reached
through
imported
inductive
types, an
open issue
we
reported
{[}lean4\#15226{]}
that
concerns
the
completeness
of that
report,
not the
soundness
of the
kernel.

\textbf{Replay.}
The walk
bounds
what the
proofs
rest on;
it does
not check
them
again. For
that, we
replayed
the
compiled
library
with
\texttt{leanchecker},
the
environment
checker
distributed
with
Lean's
toolchain
{[}leanchecker{]}.
In a
separate
process,
it reads
each
module's
compiled
declarations
and checks
every one
again with
Lean's
kernel,
against
the
compiled
declarations
of the
module's
imports.
We ran it,
one module
at a time,
over all
1,484
modules of
the
library,
naming
explicitly
the two
that its
root
module
does not
import,
over the
21 Jaune
modules in
the
library's
import
closure,
and over
Jaune's
SHA-256
specification
module,
which
holds the
equivalence
of Section
3.1 but
lies
outside
that
closure;
every
declaration
passed.
The replay
uses the
same
kernel,
not an
independent
implementation,
so it
detects
checks
that
elaboration
skipped or
that a
tactic
bypassed,
not bugs
in the
kernel;
the other
libraries
Blanc
imports,
Lean's
core and
Mathlib
among
them, are
outside
its scope.
Together,
the replay
says that
every
declaration
it checks
is
well-typed,
and the
walk that
the
library's
declarations
rest on
Lean's
three
axioms.

\textbf{Claims.}
Two gates
hold the
public
account of
the
results to
the code.
The
claim-map
gate
checks
that the
account of
the
deployed
artifacts,
from which
Appendix A
is
condensed,
cites
names that
resolve to
declarations
at the
cited
files and
lines, and
it
recomputes
the size
and code
hash of
every
artifact
from the
bytes the
repository
holds. The
statement-pin
gate
restates
758
statements
exactly,
the 49
headline
theorems
among
them, so
that
weakening
any of
them fails
to
elaborate.
Both pass
in the
catalogue
of Section
5.2;
neither
can judge
whether a
sentence
fairly
summarizes
its
theorem,
so the
statements
remain the
authority
(Appendix
E).

\textbf{What
the kernel
does not
check.}
Two things
are
trusted:
that
Jaune's
definitions
match
Ethereum,
which the
fixtures
support
but no
theorem
can
establish,
the
fixture
runs also
trusting
Lean's
compiler
and
runtime;
and that
the lifted
bytes are
the
chain's,
which is
recorded
provenance,
re-checked
for this
paper
(Appendix
B).
Separately,
whoever
applies a
theorem to
a chain
must
discharge
its
hypotheses,
which
Section 7
gathers.

\subsection{4
The
corpus}\label{the-corpus}

Table 2
lists the
eight
artifacts.
None was
written in
a
verifier's
language
or
compiled
by a
verifying
compiler,
and none
was shaped
for our
tools.
Together
they span
two source
languages
and
hand-written
assembly,
eight
toolchains,
and
deployments
from 2017
to 2026.

{\def\LTcaptype{none} % do not increment counter
\begin{table}[!htbp]
\begin{minipage}{\linewidth}\footnotesize
\begin{tabular}{@{}
  >{\raggedright\arraybackslash}p{(\linewidth - 8\tabcolsep) * \real{0.3836}}
  >{\raggedright\arraybackslash}p{(\linewidth - 8\tabcolsep) * \real{0.1469}}
  >{\raggedright\arraybackslash}p{(\linewidth - 8\tabcolsep) * \real{0.2378}}
  >{\raggedleft\arraybackslash}p{(\linewidth - 8\tabcolsep) * \real{0.1199}}
  >{\raggedright\arraybackslash}p{(\linewidth - 8\tabcolsep) * \real{0.1119}}@{}}
\toprule\noalign{}
\begin{minipage}[b]{\linewidth}\raggedright
Artifact
\end{minipage}
&
\begin{minipage}[b]{\linewidth}\raggedright
Address
\end{minipage}
&
\begin{minipage}[b]{\linewidth}\raggedright
Produced
by
\end{minipage}
&
\begin{minipage}[b]{\linewidth}\raggedleft
Runtime
bytes
\end{minipage}
&
\begin{minipage}[b]{\linewidth}\raggedright
Deployed
\end{minipage} \\
\midrule\noalign{}

WETH9 &
\texttt{0xC02a\pfvellipsis{}6Cc2}
& solc
0.4.19,
optimizer
off &
3,124 & 12
Dec
2017 \\
Beacon
deposit
contract &
\texttt{0x0000\pfvellipsis{}05Fa}
& solc
0.6.11 &
6,358 & 14
Oct
2020 \\
Curve 3Crv
LP token &
\texttt{0x6c3F\pfvellipsis{}E490}
& Vyper
0.2.4 &
2,276 & 6
Sep
2020 \\
Uniswap V2
pair
(exhibit:
USDC/WETH)
&
\texttt{0xB4e1\pfvellipsis{}C9Dc}
& solc
0.5.16 &
11,293 & 5
May
2020 \\
EIP-7002
withdrawal-request
contract &
\texttt{0x0000\pfvellipsis{}7002}
& assembly
(geas) &
504 & 19
Feb
2025 \\
Lido
CircuitBreaker
&
\texttt{0x6019\pfvellipsis{}B2F7}
& solc
0.8.34 &
4,584 & 30
Apr
2026 \\
Curve pool
implementation,
fixed (V+)
&
\texttt{0x847e\pfvellipsis{}4ed9}
& Vyper
0.3.7 &
18,320 &
23 Apr
2023 \\
Curve pool
implementation,
vulnerable
(V−) &
\texttt{0x6326\pfvellipsis{}B43E}
& Vyper
0.2.15 &
17,535 &
26 Jul
2021 \\
\bottomrule\noalign{}
\end{tabular}
\par\smallskip\normalsize
\emph{Table
2. The
artifacts.
The
Uniswap
runtime is
the same
at every
pair the
factory
creates;
the date
is that of
the
exhibit
pair, the
first the
factory
created.
The pool
implementations
are
reached
through
45-byte
ERC-1167
forwarders
{[}ERC-1167{]}:
V+ through
the
stETH-ng
factory
pool
\texttt{0x21E2\pfvellipsis{}843a},
whose
coins are
ether and
stETH; V−
through
the
pETH/ETH
pool
\texttt{0x9848\pfvellipsis{}85C5},
and two
other
pools
drained in
2023. Full
addresses,
code
hashes and
creation
transactions
are in
Appendix
B.}
\end{minipage}
\end{table}
}

The
sections
that
follow
take the
artifacts
in the
order of
what each
one adds.
WETH9, the
subject of
Section 2,
is the
baseline:
a ledger
in hashed
storage,
and an
exit at an
exact
cost. The
deposit
contract
keeps a
Merkle
tree that
commits to
its whole
history.
3Crv shows
what a
model
proved to
match the
bytes is
worth as
documentation.
The
Uniswap
pair
cannot be
described
without
the
answers of
the
contracts
it calls.
The
EIP-7002
contract
was
written
without a
compiler,
and a
block
whose call
to it
fails is
invalid.
The
CircuitBreaker
trades a
premise
about all
addresses
for a
computation
over
finitely
many. And
the two
Curve pool
implementations
return to
the
incident
of Section
1.

Each
result is
stated
with the
premises
that set
it apart.
Most also
carry
premises
of the
kinds
introduced
in Section
3, which
the text
names only
where they
matter:
trace-local
hash
premises
(Section
3.7),
callee
premises
for
liveness
(Section
3.8), and
conditions
on the
environment,
such as
canonical
code at
the system
contracts
that every
block
calls.
Table 3
sets the
results
side by
side, and
Appendix F
gives each
in more
detail,
with what
it leaves
out. The
statements
themselves,
which
Appendix A
locates,
are the
authority.

\subsubsection{4.1
WETH9:
backing,
exact
replay and
exit}\label{weth9-backing-exact-replay-and-exit}

WETH9
turns
ether into
a token: a
deposit
credits
the sender
with as
much WETH
as the
ether
sent, WETH
can be
transferred
or
approved
for
another
account to
spend, and
a
withdrawal
debits it
and
returns
the ether.
Its
storage is
two hashed
mappings,
balances
and
allowances,
beside
three
metadata
slots.

Beneath
the exit
theorem of
Section 2
lie two
history
results,
under the
history,
checkpoint
and hash
premises
of Section
2.1. After
any
configured
history
from a
checkpoint
satisfying
the
footprint
invariant,
the code
is intact
and the
storage
still
satisfies
the
invariant,
so every
tracked
holder's
balance is
backed by
the
contract's
ether, in
total and
individually.
And the
ledger
read from
the final
storage is
the
model's
ledger,
run over
exactly
the
settlement-committed
calls to
the
contract's
writers,
in order
(Section
3.5).
After any
history,
too, a
tracked
holder
without
code can
withdraw,
anyone can
deposit,
and a
holder can
transfer
what its
stored
balance
covers.

\subsubsection{4.2
The Beacon
deposit
contract:
exactly
the
deposits
made}\label{the-beacon-deposit-contract-exactly-the-deposits-made}

Every
validator
deposit on
Ethereum
mainnet
has been a
call to
this
contract,
which has
recorded
2,565,727
deposits.\footnote{The
  count is
  the
  value
  \texttt{get\_\allowbreak{}deposit\_\allowbreak{}count(\allowbreak{})}
  returned
  at block
  26,141,229.}
It keeps
an
incremental
Merkle
tree of
depth 32
over the
deposit
data and
publishes
the tree's
root and
the count.
Since
Prague,
the beacon
chain
takes
deposits
from the
contract's
\texttt{DepositEvent}
logs
without
checking
them
against
the root
{[}EIP-6110{]};
the root
remains
the
contract's
public
commitment
to its
history.

After any
configured
history
from a
checkpoint
whose
storage is
the
accumulator
of some
list of
deposits,
empty for
a fresh
deployment,
the
storage is
the
accumulator
of that
list
followed
by exactly
the
deposits
whose
frames
were
committed,
in order.
The count
is then
the
combined
list's
length,
and the
root its
reference
root, a
recursively
defined
Merkle
root:
Blanc
proves
that the
contract's
incremental
algorithm
computes
that root
for every
hash
function,
and the
history
theorems
instantiate
the hash
with
Jaune's
SHA-256,
which a
Jaune
theorem
proves
equal to
FIPS 180-4
(Section
3.1). No
hash
premise
appears,
since the
layout is
fixed and
the tree
uses
SHA-256.
At the
final
state, the
views
return the
count and
the root
at exact
gas, and a
deposit
that the
model and
the
deployed
decoder
accept
executes
at an
exact
cost,
extends
the tree
by exactly
the new
node, and
emits
exactly
the
model's
\texttt{DepositEvent}.
The
storage
result
holds
across a
history,
but the
event
result
holds only
of a
single
call: no
theorem
states
that the
\texttt{DepositEvent}
logs a
configured
history
leaves in
its
receipts
are
exactly
its
committed
deposits,
one each,
in order,
although
that list,
not the
root, is
what the
beacon
chain now
applies.

One
premise of
our own
did not
survive.
An earlier
family of
statements
in the
repository
assumed
that the
code of
the system
contracts
called at
every
block
contains
no call-
or
create-type
instruction
byte at
any
offset.
The
canonical
EIP-7002
contract
has such
bytes
inside its
push data,
so that
premise is
false on
mainnet,
and Blanc
proves it
false; the
current
statements
assume
instead
that the
canonical
code is
installed,
which
mainnet
satisfies
(Section
5.1).
Earlier
versions
of the
contract,
its
original
Vyper code
and a solc
0.6.8
build of
its
Solidity
rewrite,
were
verified
in K
{[}KEVM-Deposit;
RV-DepositSol{]},
and its
tree
algorithm
in Dafny
{[}Cassez21{]};
the
deployed
bytes were
compiled
by solc
0.6.11
{[}Sourcify{]}.

\subsubsection{4.3
Curve
3Crv: a
token's
real
policy}\label{curve-3crv-a-tokens-real-policy}

A model
proved to
match the
bytes is
also
reliable
documentation,
and 3Crv
shows what
that is
worth. It
is the
liquidity-provider
token of
Curve's
3pool for
DAI, USDC
and USDT,
and the
pool is
its
minter: it
mints 3Crv
to
depositors
and burns
it on
withdrawal.

Every
successful
call to
3Crv, for
every
selector,
is exactly
a step of
a token
model,
with its
storage
abstraction,
events and
return
bytes.
After any
configured
history,
the final
storage
represents
the model
state
reached by
running
the model
over
exactly
the
committed
writer
calls, and
that state
conserves
supply:
\texttt{totalSupply}
equals the
sum of
balances.
At the
state any
configured
history
reaches,
every call
the model
accepts
succeeds
on the
bytes,
with the
model's
output, at
an exact
cost
(Section
3.8);
\texttt{set\_name}
needs a
premise on
its call
to the
minter's
\texttt{owner()}.

The model
records a
policy
that
differs
from what
a modern
ERC-20
user might
assume.
Only the
minter
changes
the supply
or
replaces
the
minter. A
non-minter's
\texttt{transferFrom}
spends
allowance
even when
the
allowance
is the
maximum
value, so
there is
no
infinite
approval;
the
minter's
own
\texttt{transferFrom}
leaves
every
allowance
untouched,
as the
deployed
source's
comment
``minter
is allowed
to
transfer
anything''
says; and
\texttt{approve}
can set a
nonzero
allowance
only when
the
current
one is
zero. Each
of these
is a
theorem
about the
model,
carried to
the bytes
by the
refinement
theorems.
We found
no earlier
machine-checked
verification
of this
contract.

\subsubsection{4.4
The
Uniswap V2
pair: one
runtime,
every
pair}\label{the-uniswap-v2-pair-one-runtime-every-pair}

Uniswap V2
is a
constant-product
market
maker. One
11,293-byte
runtime,
compiled
by solc
0.5.16
from the
\texttt{v1.0.0}
sources
{[}UniswapV2{]},
runs at
each of
the
524,109
pairs that
its
factory
has
created
from one
creation
code
(Appendix
B). A pair
holds
reserves
of two
tokens and
issues
liquidity-provider
(LP)
tokens
against
them;
\texttt{mint}
and
\texttt{burn}
add and
remove
liquidity,
\texttt{swap}
trades one
token for
the other
for a
0.3\% fee,
and two
price
accumulators
serve as a
time-weighted
oracle.
The
theorems
quantify
over the
pair's
address,
its tokens
and its
factory,
so they
hold of
every
account
running
these
bytes; the
USDC/WETH
pair
\texttt{0xB4e1\pfvellipsis{}C9Dc}
is our
exhibit.

The pair
is the
first
artifact
that
cannot be
described
on its
own. It
learns its
balances
only from
each
token's
\texttt{balanceOf},
and
whether a
protocol
fee is due
only from
the
factory's
\texttt{feeTo},
and a swap
sends its
output
first and
may call
the
recipient,
which may
call the
pair back.
The bytes
are proved
to refine
a
\emph{model}:
a
transcription
of those
sources
into Lean
that takes
these
answers as
inputs,
read from
a
transcript,
instead of
calling
the tokens
or the
factory
(Section
3.6). The
model runs
one
\emph{entry}
at a time:
a call to
one of the
runtime's
27 entry
points,
with its
arguments
decoded.
Nothing
was
recompiled,
and the
tie is
proved
from both
sides:
every
successful
frame of
the bytes
ends as
the
model's
run of the
entry it
decoded
ends, with
the same
storage,
logs and
return
bytes, and
every
state-changing
call the
model
accepts,
given its
callees'
answers,
has a
successful
execution
of the
bytes. So
no model
that
departed
from the
bytes on a
successful
call, or
accepted a
state-changing
call that
the bytes
reject,
could have
been
proved to
match them
(Section
3.10);
whether
the model
reads as
the
sources
do, a
reader can
settle by
comparing
the two.

After any
configured
history
from a
checkpoint
at which
the bytes
are
installed
and the
storage
represents
a model
state, the
storage
represents
the model
replayed
over
exactly
the
outermost
committed
non-static
calls to
the pair,
in order,
each
against
the
answers
its own
run
observed;
calls that
reach the
pair from
inside one
of these,
through a
token
transfer
or the
flash-swap
callback,
are
replayed
inside the
step that
contains
them
(Section
3.5). No
premise
concerns
the tokens
or the
factory.
In the
replayed
state, the
model's LP
balances
still sum
to the
total
supply if
they did
at the
checkpoint,
and each
price
accumulator
has
advanced
by exactly
the
increments
of the
committed
updates,
modulo
2²⁵⁶.

Only the
economic
statement
needs a
premise
about the
tokens. If
the
factory's
answers
left the
protocol
fee off,
and the
tokens'
answers
satisfy a
premise
called
\emph{NoShrink},
every
committed
step that
starts
with
positive
supply
\emph{T}
and
reserves
r₀, r₁
ends with
r₀·r₁·T′²
≤
r₀′·r₁′·T²:
while the
supply
stays
positive,
the
geometric
mean of
the
reserves
per LP
token,
√(r₀·r₁)/T,
never
decreases.
With the
fee on,
the bound
becomes
r₀·r₁·T′²
≤
r₀′·r₁′·(T
+ F)²,
where
\emph{F}
is the
step's
exact fee
mint; this
is the
form that
matters on
mainnet,
where the
factory
has named
a fee
recipient
since 27
December
2025. The
measure is
not market
value: a
share can
gain by it
while
losing
against
simply
holding
the tokens
(Appendix
F).
NoShrink
concerns
only the
history's
own
\texttt{sync}
and
\texttt{burn}
steps:
each
balance a
token
reports to
a
\texttt{sync},
and the
first it
reports to
a
\texttt{burn},
is at
least the
stored
reserve,
and the
balance it
reports
after the
burn's
payout
falls
short of
the first
by at most
that
payout. It
is a
premise
about the
answers
the tokens
actually
gave: a
token that
takes its
transfer
fee from
the
recipient
can
satisfy
it, one
that
rebases
downwards
may not,
and no
theorem
here
proves it
of any
real
token.

Further
results
keep the
1,000 LP
tokens of
minimum
liquidity
locked at
address 0,
given a
supply
floor at
the
checkpoint
and two
conditions
on address
0, and
make each
of the ten
state-changing
functions
live after
any
history,
whenever
the model
accepts
the call,
at a
closed
cost,
given the
replies
and
returned
gas of the
calls it
makes
(Appendix
F). The
checkpoint
itself
comes from
the
factory's
\texttt{CREATE2}
deployment
and the
pair's
initialization,
and for
the
exhibit
the kernel
computes
the
address
\texttt{0xB4e1\pfvellipsis{}C9Dc}.
No theorem
here
composes
the pair
with a
token or
with its
factory
(Section
7).
dapp.org's
audit
proved
reachability
claims in
K, with
exact gas,
for each
method of
the same
bytes, one
call at a
time
{[}dapp-UniV2{]};
to our
knowledge,
no earlier
machine-checked
result
states the
pair's
properties
across
calls of
its
bytecode.

\subsubsection{4.5
The
EIP-7002
contract:
the
protocol's
own
code}\label{the-eip-7002-contract-the-protocols-own-code}

Since
Prague, a
validator's
exit or
partial
withdrawal
can be
triggered
from the
execution
layer
through a
504-byte
contract
at
\texttt{0x0000\pfvellipsis{}7002},
written by
hand for
the geas
assembler
{[}EIP-7002;
sys-asm{]}.
A user
submits a
request, a
validator's
public key
and an
amount,
with a
fee, and
the
contract
queues it
with its
caller. At
the end of
every
block the
client
calls the
contract
from a
reserved
system
address;
the
contract
removes up
to 16
requests
from the
head of
the queue,
returns
them as
the
block's
withdrawal
requests,
committed
to in its
header
{[}EIP-7685{]},
and
updates an
\emph{excess}
counter
that sets
the next
fee. A
block
whose call
to the
contract
fails is
invalid,
and a
fault here
could
drop,
duplicate
or reorder
exits.
With no
compiler,
ABI or
selector
dispatch,
the
contract
also tests
whether
the method
depends on
a
compiler's
conventions.
The model
its bytes
refine is
transcribed
from the
EIP's text
and
pseudocode,
not read
off the
bytes.

\textbf{The
system
call
succeeds.}
Whatever
the
contract's
storage
holds, as
long as
its
canonical
code is
installed,
the
end-of-block
call
succeeds
without
error, at
an exact,
state-dependent
cost that
never
exceeds
210,000
gas,
against
the
30,000,000
the
protocol
supplies,
and no
configured
history
from a
checkpoint
holding
that code
removes
it. In
Jaune's
semantics,
then,
nothing a
history
does to
this
contract's
storage
can make
this call
fail and
the block
invalid.

\textbf{Submissions
in,
requests
out.} Take
a
configured
history
from a
checkpoint
at which
the
contract's
storage is
as its
deployment
leaves it:
an empty
queue, and
an excess
that
blocks
submissions
until the
first
system
call, as
mainnet's
storage
was when
Prague
activated
(Section
5.1). For
every
block of
such a
history,
the
requests
the block
hands to
the
consensus
layer are
exactly
the first
queued
entries,
at most
16, in
first-in-first-out
order.
Each comes
from
exactly
one
committed
submission,
which paid
at least
the fee
the
bytecode
computed,
and
carries
that
submission's
caller,
public key
and
amount. No
committed
submission
is
dropped,
duplicated,
reordered
or emitted
before it
was made.
An entry
at
position
\emph{q}
of the
queue when
a block's
requests
are
computed
is emitted
⌊\emph{q}/16⌋
blocks
later, in
every
configured
extension
of the
history
that long.
While the
excess
holds the
deployment's
inhibitor,
no user
call
succeeds,
not even a
read of
the fee.
At any
state a
configured
history
reaches
once the
inhibitor
is gone,
every
submission
that pays
at least
the EIP's
fee
succeeds
given
enough
gas, at an
exact
cost, and
so does
every read
of the
current
fee
(Section
3.8).

\textbf{What
it
assumes.}
There is
no hash
premise.
The
premises
are:
canonical
code at
the system
contracts
at the
checkpoint;
at most
2²⁵⁴
committed
submissions;
throughout
the
history,
no
transaction
sent from
the system
address,
no
EIP-7702
authorization
signed by
it, and no
contract
creation
at it,
even one
that
reverts;
and no
code at
the system
address at
the
checkpoint.
A closed
theorem
shows the
last
premise is
needed: in
a
two-block
configured
history
that meets
every
other
premise,
code at
that
address
calls the
contract
from
inside a
user
transaction,
and the
contract,
which
knows the
system
call only
by its
caller,
drains the
queue
before the
block's
own system
call.

\textbf{The
fee the
bytes
compute.}
The EIP
specifies
the fee in
Python
over
unbounded
integers;
the
bytecode
computes
the same
series in
256-bit
words,
whose
intermediate
products
can
overflow,
a
departure
that two
pre-deployment
audits had
noted
{[}Blackthorn-7002;
SigmaPrime-7002{]}.
Blanc
proves
that the
bytecode's
fee never
exceeds
the EIP's,
and equals
it
precisely
when no
intermediate
value of
the series
wraps. A
closed
theorem
then
builds a
three-block
configured
history,
meeting
the
premises
of the
request
theorem,
in which a
submission
is
committed
after
paying
less than
the EIP's
fee. The
guarantee
that every
committed
submission
paid at
least the
EIP's fee
is
therefore
not a
theorem
about the
deployed
bytes,
which is
why the
request
theorem
speaks of
the fee
the
bytecode
computes.
The
counterexample
is far
from
mainnet:
by direct
computation
outside
the
proofs,
the two
fees agree
at every
excess up
to 2,892;
the
counterexample's
submission
comes at
an excess
of 2,893
and pays
2²⁴⁵ wei,
more than
10⁴⁷ times
all the
ether in
existence,\footnote{About
  1.2·10⁸
  ether at
  block
  26,141,495:
  the
  execution-layer
  balances
  plus the
  beacon-chain
  balances,
  less the
  deposits
  in
  transit,
  as
  ultrasound.money
  reports
  them
  {[}ultrasound{]}.}
and its
chain's
block gas
limit is
about nine
times
mainnet's.
On
mainnet,
the excess
has so far
peaked at
666, on 19
December
2025.

\textbf{Deployment.}
Unlike the
older
contracts
in the
corpus,
this one
has only
ever been
used under
forks
Jaune
covers. It
was
deployed
before
Prague by
a keyless
transaction,
whose
recorded
creation
input
Blanc
executes
to the
model's
initial
state, and
first used
when
Prague
activated,
at which
moment
mainnet's
storage
still held
that state
(Section
5.1). Four
audits and
a
loop-bounded
Halmos
check
examined
the
contract
before
deployment
{[}Pectra-Audits;
Halmos-Pectra{]};
we know of
no earlier
proof
about it
that is
unbounded
in the
excess and
in the
history,
or that
follows a
request
from its
submission
to the
block that
emits it.

\subsubsection{4.6
Lido's
CircuitBreaker:
trading a
universal
premise
for a
computation}\label{lidos-circuitbreaker-trading-a-universal-premise-for-a-computation}

The
CircuitBreaker
is a
pauser
registry:
Lido's DAO
registers
a pauser
for each
protected
contract,
and the
pauser can
pause it
for a
bounded
time
{[}LIP-34{]}.
The
registry
is a
Solidity
array of
protected
contracts
with three
mappings,
from
contract
to pauser,
from
contract
to array
index, and
from
pauser to
the number
of
contracts
it guards.

Its
\emph{history
tier} is
the
exception
among the
results
reported
here: its
hash
premises,
assumed at
every
frame,
range over
all 2¹⁶⁰
addresses,
whereas
every
other
result's
cover only
the keys a
history
touches,
are
decidable
checks, or
are absent
(Section
3.7).
After any
configured
history
from a
checkpoint
holding
the code
and an
empty
registry,
the
storage is
a
consistent
registry,
whose
assignments,
indices,
counts and
array
agree, and
every
committed
removal of
a target,
including
one
re-entered
from
inside
\texttt{pause}'s
external
call,
repairs
the array
by swap
and pop.

The
\emph{finite
tier}
trades
that
premise
for a
computation,
at a price
in scope.
It
describes
the
registry
through an
explicit
list of
entries
and the
slots of a
chosen
list of
probe
addresses,
and every
separation
premise
becomes a
decidable
check on
those
concrete
keys: the
modeled
deployment
leaves
storage
that
passes the
check, and
a
successful
registration
of a
nonzero
pauser for
an
existing
target
takes
storage
that
passes it
to storage
that
passes it
for the
updated
entries.
The tier
covers one
such
update at
a time and
nothing
outside
the
probes,
but its
check can
be
computed
from the
live
contract's
storage,
which
passes it
on 29
probes
(Section
7). Cyfrin
checked
properties
of an
earlier
commit of
the source
with the
Certora
Prover,
and
compiled
the
deployed
commit and
compared
the output
with the
deployed
bytecode,
apart from
its
metadata
and
immutable
values
{[}Cyfrin-CB;
Cyfrin-Audit{]};
we found
no
machine-checked
verification
of the
deployed
bytecode.

\subsubsection{4.7
The two
Curve pool
implementations:
the bug
that
compiled,
and the
fix}\label{the-two-curve-pool-implementations-the-bug-that-compiled-and-the-fix}

Three of
the four
pools
drained in
the 2023
incident
are
45-byte
ERC-1167
forwarders
that
delegate
every call
to one
implementation.
The corpus
contains
that
implementation
and a
later one,
both of
Curve's
StableSwap
pool for
ether and
one other
coin: one
built by a
compiler
with the
bug, which
affected
Vyper
0.2.15,
0.2.16 and
0.3.0
{[}VyperPM{]},
and one
built
after the
fix
(Figure
3). Both
mark their
liquidity
functions
with the
same key,
\texttt{@nonreentrant\penalty200{}(\allowbreak{}'lock')}.

\begin{figure}[!htbp]
\begin{minipage}{\linewidth}
\begin{Verbatim}[fontsize=\footnotesize]
                     V−: Vyper 0.2.15 (0x6326…)          V+: Vyper 0.3.7 (0x847e…)
 remove_liquidity    lock at slot 2 (bytes 6900–6911)    one lock: slot 0, held value 2,
 add_liquidity       lock at slot 0 (bytes 88–99)        shared by all seven guarded functions
\end{Verbatim}
\par\smallskip\normalsize
\emph{Figure
3. The
same
source-level
key,
\texttt{'lock'},
compiled
two ways.}
\end{minipage}
\end{figure}

\textbf{The
vulnerable
implementation.}
\texttt{0x6326\pfvellipsis{}},
deployed
on 26 July
2021, less
than three
days after
Vyper
0.2.15 was
published
{[}Vyper-0.2.15{]},
is the
implementation
behind
three of
the pools
drained on
30 July
2023,
pETH/ETH,
msETH/ETH
and
alETH/ETH,
from which
attackers
extracted
about \$37
million,
the sum of
LlamaRisk's
per-pool
estimates
{[}LlamaRisk{]}.
Its
\texttt{remove\_\allowbreak{}liquidity}
reads the
total
supply,
sends
ether to
the
receiver,
and only
afterwards
writes the
reduced
supply
back.
Blanc
executes
the
attack's
mechanism
on these
bytes,
from
deployment
(Figure
4). On
every
covered
fork,
eight
messages
run from a
world
holding
only a
funded
creator
without
code: they
deploy the
implementation
from its
preserved
creation
input, set
up a
forwarder,
the pool,
a token
and an
attacker
contract,
and add
liquidity,
leaving a
sound LP
ledger
whose
whole
supply of
2,000 the
attacker
holds; the
eighth
message is
the
attack.

\begin{figure}[!htbp]
\begin{minipage}{\linewidth}
\begin{Verbatim}[fontsize=\footnotesize]
attacker ──► pool.remove_liquidity      slot 2: 0 → 1; reads totalSupply = 2,000
               └─ ether to attacker ──► attacker ──► pool.add_liquidity
                                          slot 0 free: enters the body, mints 106,
                                          balanceOf[attacker] = 2,106
             remove_liquidity resumes: burns 200, writes back totalSupply = 2,000 − 200
 result:     totalSupply = 1,800  <  balanceOf[attacker] = 1,906
\end{Verbatim}
\par\smallskip\normalsize
\emph{Figure
4. The V−
violation,
executed
by Jaune
on the
deployed
bytes
after
deployment,
initialization
and
funding.}
\end{minipage}
\end{figure}

The outer
call
writes
back a
supply it
read
before the
callback,
which
erases the
106 tokens
minted
during the
callback
from the
supply but
not from
the
attacker's
balance. A
second
result
carries
the same
corruption
into a
signed
transaction
that
\texttt{processTrans\penalty200{}acti\penalty200{}on}
admits,
with every
admission
check,
signature
recovery
included,
discharged
by kernel
evaluation
and only
room in
the block
assumed.
It starts
from a
separate
synthetic
pre-state,
not the
world the
eight
messages
reach, so
the two
results do
not
compose
into one
run from
deployment
to an
admitted
transaction.
Neither
replays
the
historical
transactions,
which ran
under
Shanghai,
a fork
Jaune does
not
implement.

\textbf{The
fixed
implementation.}
\texttt{0x847e\pfvellipsis{}},
compiled
by Vyper
0.3.7 and
deployed
in April
2023,
guards
seven
functions
with one
lock, and
Blanc
proves
that the
lock
excludes
reentry.
In any
execution,
whatever
its
outcome,
while a
frame of
the pool
is inside
one of the
five
mutating
guarded
bodies and
has not
released
the lock,
no frame
below it
enters any
of the
seven
guarded
bodies of
that pool,
whatever
code the
frames in
between
run. No
premise
mentions
callbacks,
reentry or
the value
of the
lock as a
callee
sees it:
exclusion
is
derived,
not
assumed
(Section
3.3). The
one hash
premise is
that no
Keccak-256
digest the
pool
computes
in that
execution
equals 0,
the lock's
slot. The
theorem
holds for
the
implementation
called
directly
and
through
the
stETH-ng
forwarder,
for
messages
and for
transactions,
and for
every
execution
recorded
in a
configured
history,
given
premises
for each
execution
(Appendix
F).

Three
executed
witnesses
show that
the
exclusion
is not
vacuous.
The
strongest
mirrors
the attack
from
deployment:
in its
ninth
message,
\texttt{remove\_\allowbreak{}liquidity}
holds the
shared
lock while
it sends
100 wei to
a receiver
whose code
attempts
\texttt{add\_liquidity}
through
the pool's
forwarder,
a clone
that the
messages
create,
with the
same 45
bytes as
the
stETH-ng
pool's
forwarder.
That frame
reads the
held lock
and
reverts
without
reaching a
guarded
body, the
outer call
commits
with the
ledger
still
sound, and
the
exclusion
theorem is
instantiated
on this
execution
with all
of its
premises
proved.

Both
implementations
promise
the same
thing in
their
source,
and an
argument
about the
bytes
tells
apart the
compiler
that kept
the
promise
from the
one that
did not.
They are
not a
controlled
experiment
on the
fix, since
their
sources
differ in
other ways
too; to
our
knowledge
neither
has had a
machine-checked
treatment
before.

\subsubsection{4.8
Summary}\label{summary}

Table 3
sets the
eight
results
side by
side, by
how close
each comes
to the
live
chain. A
conditional
theorem
holds of
every
history
that meets
its
premises,
which the
first two
columns
summarize,
and says
nothing
about
mainnet
until
those
premises
are shown
to hold
there. A
closed
instance
is a
theorem
with every
premise
discharged,
about one
concrete
execution
in a
modeled,
partly
synthetic
world. An
empirical
comparison
checks a
theorem's
prediction
against
data read
from
mainnet,
without
proof that
the data
is the
chain's. A
fourth
kind, a
theorem
applied to
mainnet
state
authenticated
against a
block's
state
root,
would come
closest,
but no
result
reaches it
yet, so
the table
has no
column for
it.

{\def\LTcaptype{none} % do not increment counter
\begin{longtable}[]{@{}
  >{\raggedright\arraybackslash}p{(\linewidth - 12\tabcolsep) * \real{0.1500}}
  >{\raggedright\arraybackslash}p{(\linewidth - 12\tabcolsep) * \real{0.1300}}
  >{\raggedright\arraybackslash}p{(\linewidth - 12\tabcolsep) * \real{0.1500}}
  >{\raggedright\arraybackslash}p{(\linewidth - 12\tabcolsep) * \real{0.1300}}
  >{\raggedright\arraybackslash}p{(\linewidth - 12\tabcolsep) * \real{0.1500}}
  >{\raggedright\arraybackslash}p{(\linewidth - 12\tabcolsep) * \real{0.1790}}
  >{\raggedright\arraybackslash}p{(\linewidth - 12\tabcolsep) * \real{0.1110}}@{}}
\toprule\noalign{}
\begin{minipage}[b]{\linewidth}\raggedright
\end{minipage}
&
\begin{minipage}[b]{\linewidth}\raggedright
Hash
premises
\end{minipage}
&
\begin{minipage}[b]{\linewidth}\raggedright
Checkpoint
established
by
\end{minipage}
&
\begin{minipage}[b]{\linewidth}\raggedright
Conditional
theorem:
strongest
history
result
\end{minipage}
&
\begin{minipage}[b]{\linewidth}\raggedright
Conditional
theorem:
liveness
\end{minipage}
&
\begin{minipage}[b]{\linewidth}\raggedright
Closed
instances
\end{minipage}
&
\begin{minipage}[b]{\linewidth}\raggedright
Empirical
comparison
with
mainnet
\end{minipage} \\
\midrule\noalign{}
\endhead
\bottomrule\noalign{}
\endlastfoot
WETH9 &
trace-local
& modeled
deployment
& backing;
exact
replay of
committed
calls & a
\texttt{withdraw}
transaction
after a
history or
a block
prefix,
exact
\texttt{gasUsed}
& the exit
theorem in
a
synthetic
world,
from
deployment
& 3,748
withdrawal
receipts
equal the
gas
formula \\
Deposit
contract &
none &
modeled
deployment
& exact
committed
deposits;
count and
root &
deposit
after a
history,
exact gas
& --- &
canonical
code at
the system
contracts
and none
at
\texttt{0x02},
at block
26,141,229 \\
3Crv &
trace-local
& modeled
deployment
& exact
replay;
conservation
& every
accepted
call after
a history,
exact
existential
cost;
\texttt{set\_name}
given the
minter's
answer &
--- &
--- \\
Uniswap V2
pair &
trace-local
& modeled
\texttt{CREATE2}
deployment
and
initialization
& nested
exact
replay;
oracle
law; LP
ledger and
minimum
liquidity;
share
value
given
NoShrink &
all ten
state-changing
functions
after a
history,
closed
cost, six
given
their
callees'
answers &
--- & 542
\texttt{approve}
receipts
equal the
frame cost
plus the
protocol's
charges \\
EIP-7002
contract &
none &
modeled
deployment
& exact
replay;
requests
are the
oldest
pending
submissions,
in order;
delivery
bound &
the system
call
whatever
the
storage
holds;
submissions
and fee
reads
after a
history
that
leaves the
contract
active,
exact gas
& two
histories:
a fee
below the
EIP's; a
drained
queue &
storage at
Prague's
activation
is the
checkpoint
state \\
CircuitBreaker
&
decidable
(finite
tier);
universal
(history
tier) &
modeled
deployment,
under two
universal
premises
for the
history
tier &
registry
witness;
swap-and-pop
removal &
none & a
constructed
state
meeting
the finite
tier's
premises,
not a call
& the live
registry
passes the
finite
check on
29 probes,
evaluated
by a
script \\
V+ &
trace-local,
per
execution
& not
applicable
&
exclusion
in every
execution
of a
history,
with
premises
per
execution
& none &
three
executions,
on every
covered
fork; one
from
deployment
& --- \\
V− & none
& not
applicable
& not
applicable
& not
applicable
& the
ledger-corrupting
reentrancy,
from
deployment
as
messages
and as a
synthetic
transaction
& --- \\
\end{longtable}
}

\nopagebreak\noindent\begin{minipage}{\linewidth}
\emph{Table
3. Every
result is
about the
lifted
bytes,
which
match the
deployed
code
(Appendix
B),
depends on
no axiom
beyond
Lean's
three, is
stated
under
Jaune's
full
semantics,
and covers
all four
forks that
have
governed
mainnet
since May
2025,
except two
kinds of
closed
instance
that run
on one
fork: the
EIP-7002
counterexample
histories,
on a
synthetic
Prague
chain, and
the WETH9
closed
instance,
in one
BPO2
block. The
empirical
comparisons
are read
over RPC
and are
not
authenticated
against a
state
root.}
\end{minipage}

\subsection{5
Evaluation}\label{evaluation}

\subsubsection{5.1
Fidelity
to the
chain}\label{fidelity-to-the-chain}

Beyond the
conformance
results of
Section
3.1, three
kinds of
evidence
connect
the
theorems
to
Ethereum
as its
clients
run it.

\emph{The
bytes.} At
block
26,141,229,
three
separately
operated
providers
returned
identical
code for
every
address in
Table 2,
for the
forwarders
of the
stETH-ng
pool and
of the
three
drained
pools that
delegate
to V−, and
for the
four
system
contracts.
Each
implementation's
code is
byte-identical
to its
lifted
input,
each
forwarder
is the
ERC-1167
template
naming its
implementation,
and each
system
contract's
code is
the
canonical
code that
Blanc's
premises
name. The
deployment
theorems
compute
five of
the
mainnet
addresses
from the
recorded
deployers
and
nonces,
and the
Uniswap
exhibit's
from the
factory,
the salt
and the
creation
code
(Section
3.9). The
factory's
single
\texttt{CREATE2}
site
installs
exactly
the lifted
runtime,
so every
pair it
creates
holds that
runtime;
eleven
sampled
pairs
confirm it
(Appendix
B).

\emph{The
environment.}
At block
26,141,229,
mainnet's
four
system
contracts
held their
canonical
code, the
SHA-256
precompile's
address
and the
system
address
held none,
and the
system
address
had nonce
0, so no
transaction
has ever
been sent
from it.
For the
EIP-7002
contract
we also
looked
back to
the moment
its use
began. At
block
22,431,083,
the last
before
Prague
activated,
its
storage
held
exactly
the
model's
initial
state, the
inhibitor
in the
excess
slot and
zeros in
the count,
head and
tail, with
the
storage
root it
had at
deployment,
549,694
blocks
earlier,
and it had
logged no
request;
the four
system
contracts
held their
canonical
code, and
the system
address
none.
Mainnet's
state when
Prague
activated,
as the
providers
report it,
therefore
met the
checkpoint
premises
of the
request
theorems.
What
separates
those
theorems
from the
live
contract
is the
premise
that
mainnet's
history is
a
configured
one, which
rests on
Jaune's
fidelity,
and the
exclusions
at the
system
address
(Section
4.5).

\emph{The
behavior.}
Beyond the
WETH9
receipts
of Section
2.4, the
Uniswap
pair
allows a
comparison
for a
different
compiler's
output,
across
hundreds
of
contracts.
Its frame
theorem
for
\texttt{approve},
2,342 gas
plus the
charge of
its one
store,
with the
protocol's
intrinsic
gas and
refund
added,
predicts
the
receipt of
a
top-level
\texttt{approve}
transaction
(Appendix
D).
Through
the
\texttt{Approval}
events of
the 30,000
blocks
ending at
block
26,141,229
we found
542 such
transactions
whose
target
held the
lifted
runtime,
sent to
476
different
pairs, 57
of them
from
accounts
with
EIP-7702
delegation
code,
which
\texttt{approve}
never
executes;
every
receipt
equals the
prediction,
in each of
the
store's
four cases
(Appendix
D). With
the 3,748
WETH9
receipts
of Section
2.4, that
makes
4,290
comparisons,
all equal.
Since the
pair has
no
transaction-level
theorem,
this tests
the frame
theorem's
cost
together
with the
protocol's
transaction
rules.

\subsubsection{5.2
The cost
of
checking}\label{the-cost-of-checking}

Kernel
checking
is not
free, but
it is
routine.
Table 4
gives the
cost of
checking
the
library at
\texttt{5048a51f}
on the
development
machine, a
10-core
Apple M5
with 24
GiB of
memory.
Four rows
were
measured
afresh at
that
commit.
The
elaboration
row mixes
fresh and
cached
timings,
the
certificate
row is
cached,
and the
gate
catalogue's
time is
mixed too,
because
its
elaboration
gate
measured
only part
of the
library
afresh; a
cached
timing is
one taken
earlier on
the same
machine of
a file
whose
content
and
imports
are those
at
\texttt{5048a51f}.
Each row
answers a
different
question,
and the
rows do
not add.
The clean
build is
what a
reader
waits for
to rebuild
and check
the
library
from
source,
with
modules
built in
parallel.
The
elaboration
row is the
elaboration
gate's own
measurement:
each file
elaborated
alone, one
after
another,
with its
imports
already
built, so
that a
module
whose cost
grows is
caught and
the cost
can be
located;
it is a
sum over
files, not
a wall
time, and
it repeats
the
build's
work
rather
than
adding to
it.

{\def\LTcaptype{none} % do not increment counter
\begin{table}[!htbp]
\begin{minipage}{\linewidth}\footnotesize
\begin{tabular}{@{}
  >{\raggedright\arraybackslash}p{(\linewidth - 6\tabcolsep) * \real{0.1808}}
  >{\raggedright\arraybackslash}p{(\linewidth - 6\tabcolsep) * \real{0.3327}}
  >{\raggedleft\arraybackslash}p{(\linewidth - 6\tabcolsep) * \real{0.1538}}
  >{\raggedright\arraybackslash}p{(\linewidth - 6\tabcolsep) * \real{0.3327}}@{}}
\toprule\noalign{}
\begin{minipage}[b]{\linewidth}\raggedright
Check
\end{minipage}
&
\begin{minipage}[b]{\linewidth}\raggedright
Scope
\end{minipage}
&
\begin{minipage}[b]{\linewidth}\raggedleft
Time
\end{minipage}
&
\begin{minipage}[b]{\linewidth}\raggedright
Notes
\end{minipage} \\
\midrule\noalign{}

Clean
build &
the 1,482
modules
the
library's
root
imports,
rebuilt
from
source
with
Blanc's
artifact
cache
disabled;
toolchain
and pinned
dependencies
prebuilt &
4,380 s &
batches of
2 to 5
threads
under the
host's
memory
admission,
waits and
two
automatic
retries
included;
peak 11.2
GiB,
sampled
over the
process
tree \\
Per-module
elaboration
& all
1,484
library
files and
the
executable's
entry
point,
each with
its
imports
built &
5,733 s,
summed;
mixed &
239 files
timed at
\texttt{5048a51f}
(568 s),
the other
1,246
earlier on
this
machine
with
identical
content
and
imports;
the lift's
912
modules
take 3,492
s \\
Certificate
checks &
the 34
registered
certificates,
in 95
modules &
735 s,
summed;
cached &
all 95
modules
timed
earlier on
this
machine,
unchanged
since;
regenerating
every
certificate
from its
input and
comparing,
at
\texttt{5048a51f},
takes 3.5
s \\
Axiom walk
and leaf
census &
1,484
modules,
128,938
constants
& 397 s &
one
command,
its
controls
included \\
Kernel
replay &
the 1,484
library
modules,
the 21
Jaune
modules
they
import and
Jaune's
SHA-256
specification
(Section
3.11) &
3,900 s &
\texttt{leanchecker},
one module
at a time
on one
thread;
largest
process
11.9
GiB \\
Conformance
& 5,006
fixture
files,
Jaune
\texttt{2737c8eb}
& 1,150 s
& one
worker;
largest
process
2.3 GiB \\
Gate
catalogue
& 86 rows,
every
verdict
computed
afresh &
2,338 s;
mixed &
the
elaboration,
axiom,
certificate
and claim
gates
among
them; the
elaboration
gate timed
239 files
afresh and
reused the
1,246
cached
timings
above \\
\bottomrule\noalign{}
\end{tabular}
\par\smallskip\normalsize
\emph{Table
4. The
cost of
checking
the
library at
\texttt{5048a51f},
measured
at that
commit
except
where the
Time
column
says mixed
or
cached.}
\end{minipage}
\end{table}
}

The
largest
certificate,
the
vulnerable
Vyper
pool's 211
entries,
accounts
for 279
seconds of
the
certificate
row, in
nine
modules,
but the
slowest
module of
the
elaboration
row was a
proof
about the
two token
calls of
the
Uniswap
pair's
\texttt{sync},
at 139
seconds.
The proofs
are long
compared
with the
code they
describe:
WETH9's
15,221
lines of
Lean cover
3,124
bytes,
about five
lines per
byte, and
the
Uniswap
pair's
95,581
lines
cover
11,293
bytes,
about
eight and
a half.

\subsubsection{5.3
Fifteen
days}\label{fifteen-days}

The lift
was
written by
AI agents.
We report
it as a
case study
of one
program,
not as a
forecast.
The first
commit to
the lift
is dated
25
September
2026 and
the last,
the commit
this paper
describes,
9 October;
the 1,053
commits
that touch
\texttt{Blanc/Lift}
or
\texttt{scripts/lift}
fall on
each of
those
fifteen
calendar
days.
\texttt{Blanc/Lift}
holds
236,372
lines of
Lean in
912 files,
from 9,745
for 3Crv
to 95,581
for the
Uniswap
pair, with
37,035 of
shared
machinery;
modules
elsewhere
in the
library
that grew
during
those days
are not
counted.
Not all of
it was
written as
proof: 84
files with
23,132
lines,
about a
tenth,
declare
themselves
generated,
75 of them
by the
certificate
producer
and the
rest by
per-artifact
lifting
and data
scripts.
Six
artifacts,
all but
the
Uniswap
pair and
the
EIP-7002
contract,
were
finished
in the
first six
days;
those two
and the
reachable
Vyper
witnesses
took the
following
seven; and
the last
two days
went
mostly to
work
prompted
by review.

\textbf{Who
did what.}
Commits
are made
under the
author's
Git
identity,
which the
agents
also use,
or under a
Codex
identity
(\texttt{codex@openai.\allowbreak{}com}
or
\texttt{codex@local}),
so Git
alone does
not record
who wrote
what. Of
the 888
non-merge
commits,
450 carry
a
co-author
line
naming a
model:
Claude
Opus 5.5
on 362 of
them,
Claude
Sonnet 5.5
on 72,
Meta's
Muse Spark
1.3 on 11,
Claude
Fable 5.1
on 7,
Google's
Gemini 3.8
Flash on
3, Claude
Sonnet 5
on 3 and
OpenAI's
GPT-5.6
Luna on 1,
some
jointly.
The other
438 carry
none, 52
of them
under a
Codex
identity.
Agents
made
almost all
of these
as well,
on the
author's
instructions,
but no
record
attributes
each
commit, so
the split
cannot be
stated
more
precisely.
The goals'
dispatch
records,
which are
partial
and not
public,
name
OpenAI
models run
through
Codex for
parts of
the work:
GPT-5.6
Luna for
the first
proofs of
the two
generic
lifting
theorems,
and Sol,
another
OpenAI
model, for
the
Uniswap
pair's
model, the
statements
of the
reachable
Vyper
witnesses
and the
CircuitBreaker's
finite
tier, one
module of
which Muse
Spark 1.3
wrote. The
tallies
describe
trailers,
not
measured
shares of
the work.
The work
was
organized
by the
coordinator--worker
workflow
of Creme
{[}Creme{]}:
a
coordinating
session
dispatches
bounded
units of
work to
worker
agents,
each in
its own
Git
worktree;
it accepts
a result
only on
the gate
catalogue's
verdict on
the exact
candidate
commit,
the diff,
and a
report
that ties
each
condition
to its
evidence;
and it
sends
review,
where
possible,
to a
different
model from
the one
that wrote
the work.

\textbf{The
author and
the cost.}
The author
set the
program
and its
goals and
made the
decisions
the
workflow
reserves
for a
human,
among them
the
product
semantics
each goal
fixes, the
wording of
published
claims and
counts,
and any
change to
a pinned
reference
such as
the Jaune
revision,
and
estimated
spending
four to
five hours
a day at
the
screen,
about 60
to 75
hours in
all, a
self-estimate
rather
than a
measurement.
The
author's
flat-rate
agent
subscriptions,
\$200 a
month each
for
Anthropic
and
OpenAI,
\$50 for
Meta and
\$20 for
Google,
prorate to
about
\$235 for
the
fifteen
days. The
work ran
on two
machines,
and the
figure
covers
both;
since the
same
subscriptions
served
other work
too, it is
an upper
bound as
an
allocation
of them,
and it is
not a
measure of
compute,
which we
did not
meter. It
is also
conditional
on an
existing
stack:
Jaune, the
history
and
settlement
machinery,
and
Blanc-native
implementations
of WETH9,
the
deposit
contract
and the
CircuitBreaker
were built
over the
preceding
months,
while
3Crv, the
Uniswap
pair, the
EIP-7002
contract
and the
two Vyper
pools had
no native
counterpart.

\textbf{Knowing
what was
proved.}
When
agents do
this much
of the
verification,
the main
bottleneck,
in our
experience,
is not
proof but
meaning:
making
sure that
each
proved
statement
means what
its
informal
description
says and,
more
important,
what it
\emph{should}
mean if
the
theorem is
to be of
value to
its users.
Defects of
this kind
occurred
in this
work: an
earlier
form of
the
Uniswap
replay
matched
answers to
calls by
target and
input
rather
than by
position,
so that a
\texttt{burn}
could be
replayed
against
the wrong
balances
(Appendix
F.4), and
an earlier
family of
deposit-contract
statements
assumed a
property
of the
system
contracts'
code that
the
canonical
EIP-7002
contract
lacks
(Section
4.2). Both
were
repaired,
but we
found no
satisfactory
way to
make sure
that no
others
remain.
The
controls
of Section
3.10 show
that key
premises
are needed
and that
statements
would not
survive a
change to
the code,
and the
statement
pins keep
a
published
statement
from
changing
unnoticed,
but
neither
says
whether a
statement
is the one
that
should
have been
proved;
that was
left to
reading.
Every
proof in
the corpus
was
produced
by agents
working
from
informal
instructions
about what
to prove.
The author
examined
the
resulting
theorems
by reading
their
statements
and
inspecting
their key
components,
expanding
definitions
and asking
the agents
for
clarification.
This
probed
only a few
levels
beneath
the
headline
theorems
and did
not reach
their
primitives,
so the
author
lacks the
complete,
ground-up
grasp of
each
statement
that
proving it
by hand
would have
given.

\subsection{6
Related
work}\label{related-work}

Formal
verification
of
Ethereum
contracts
has
explored
these
trade-offs
from many
directions,
and
several
projects
reach
further
than this
work along
their own
axes. We
describe
each in
its own
terms,
from its
documentation,
paper or
code as
read
between 1
and 7
October
2026, at
the
repository
commits
named in
the
references;
Table 5
summarizes
the
interactive-theorem-proving
and
K-based
systems.

\textbf{Executable
EVM
semantics.}
KEVM gave
the EVM a
complete
formal
semantics
in K,
tested
against
the
official
tests, and
its
schedules
now extend
to
Amsterdam
{[}KEVM{]};
earlier
work gave
mechanized
semantics
{[}Hirai17;
Grishchenko18{]},
a program
logic
{[}Amani18{]}
and a
sound
analyzer
{[}eThor{]}.
EVMYulLean,
in Lean 4,
reports
passing
22,330 of
22,332
Cancun
tests and
executes
Keccak,
SHA-256
and BLAKE2
through C
foreign
functions
and some
precompiles
through
Python
subprocesses
{[}EVMYulLean{]};
Verifereum,
in HOL4,
reports
17,129 of
17,423
Osaka
execution-spec
tests
{[}Verifereum{]};
Isabelle/EVM,
without
precompiles,
1,966 of
2,669
files for
Shanghai
and Cancun
{[}IsabelleEVM{]};
and
powdr's
semantics
no
failures
on its
Osaka
suites
{[}powdr{]}.
These
figures
come from
different
suites,
releases
and forks
and are
not
directly
comparable;
Jaune's
lane over
the
current-mainnet
release
allows no
expected
failures,
and its
library
uses no
foreign
functions
or native
evaluation.

\textbf{Verified
compilation.}
DeepSEA
{[}DeepSEA{]},
Elle
{[}Elle{]},
powdr's
Yul
compiler
{[}powdr{]},
Solidus
{[}Solidus{]}
and Verity
{[}Verity{]}
verify or
build on a
compiler,
and their
theorems
cover the
stages
they
verify:
Verity's
from its
language
to Yul,
with solc
trusted
below, and
powdr's
and
Solidus's
from Yul
to
bytecode.
Blanc's
own
language
and
compiler
belong to
this
family.
evm-smith
proves
properties
of
hand-written
bytecode,
including
an 86-byte
WETH, over
a fork of
EVMYulLean;
its ERC-20
report
relates
its
hand-written
programs
to
compiled
bytecode
by
inspection,
not by
proof
{[}evm-smith{]}.

\textbf{Existing
bytecode.}
EquiVM is
the
closest
peer along
the axis
of
exogenous
code
{[}EquiVM{]}.
It proves
that solc
and Vyper
output
refines
specifications
in a small
imperative
language,
for
twenty-three
real-world
contracts
in its
preprint,
including
most of
MakerDAO's
Dai
system,
with
proofs by
language
models,
and its
repository
has since
completed
the proof
for a
local
build of
the
Uniswap V2
pair,
which the
preprint
reported
as
incomplete.
Its
benchmark
bytecode
is
compiled
locally
from
pinned
sources:
its WETH9
is 1,763
bytes of
solc
0.5.16
output,
against
the
deployed
3,124
bytes of
solc
0.4.19
output.
The
approaches
are
complementary:
EquiVM's
specifications
cover
whole
interfaces
at scale,
ours are
bespoke
properties
with exact
gas and
liveness,
about the
bytes on
chain.
Verifereum
pairs its
semantics
with a
WETH9
example
whose
bytes are
those at
the
mainnet
address,
among its
experimental
developments
{[}Verifereum{]}.
With the K
proofs of
the
deposit
contract
and the
Uniswap V2
pair,
Cyfrin's
check of
the
CircuitBreaker's
source and
the
bounded
Halmos
check of
the
EIP-7002
contract
(Section
4), these
are the
earlier
machine-checked
results we
know of on
artifacts
in our
corpus;
for the
deposit
contract
they
concern
its
original
Vyper
version
and a solc
0.6.8
build of a
pre-deployment
commit of
its
source.
Kontrol,
Runtime
Verification's
prover for
Foundry
property
tests over
KEVM,
computes
gas on
request
{[}Kontrol{]}.

\textbf{Other
strands.}
Constant-product
markets
have a
mathematical
theory
{[}Bartoletti22{]}
and a Lean
4
formalization
as an
economic
model
{[}Pusceddu24{]},
and four
functions
of the
Vyper-based
Uniswap V1
were
checked by
symbolic
execution
of their
bytecode
in KEVM,
without
gas, in
what its
authors
call a
lightweight
verification
{[}RV-UniV1{]}.
Our
certificates
play the
role of
decompilation
into logic
{[}Myreen08{]},
with a
checker in
place of a
proof-producing
decompiler,
and EVM
decompilers
solve the
structural
problem at
scale
without
proof
{[}Vandal;
Gigahorse;
Elipmoc;
EtherSolve{]}.
Gas has
been
analyzed
statically
{[}GASTAP;
GASOL;
MadMax;
Marescotti18{]},
and
reentrancy
has a
theory of
its own
{[}Grossman18;
Albert20{]}
and
automated
verification
{[}VerX{]};
our costs
are exact
per path
and
proved,
and the V+
theorem is
one safety
property,
for one
lock, on
the
deployed
bytes. Of
the
symbolic
tools,
hevm
treats
Keccak as
an
uninterpreted
function
constrained
by
injectivity
assumptions
{[}hevm{]},
halmos
replaces
Keccak-256
of
symbolic
inputs
with an
uninterpreted
function
{[}halmos{]},
and the
Certora
Prover
models
hashing as
an
injective
function
{[}Certora{]};
Act has
backends
for
equivalence
checking
with hevm
and for
Rocq
{[}Act{]}.

{\def\LTcaptype{none} % do not increment counter
\begin{longtable}[]{@{}
  >{\raggedright\arraybackslash}p{(\linewidth - 14\tabcolsep) * \real{0.0971}}
  >{\raggedright\arraybackslash}p{(\linewidth - 14\tabcolsep) * \real{0.1221}}
  >{\raggedright\arraybackslash}p{(\linewidth - 14\tabcolsep) * \real{0.1391}}
  >{\raggedright\arraybackslash}p{(\linewidth - 14\tabcolsep) * \real{0.1552}}
  >{\raggedright\arraybackslash}p{(\linewidth - 14\tabcolsep) * \real{0.1301}}
  >{\raggedright\arraybackslash}p{(\linewidth - 14\tabcolsep) * \real{0.1131}}
  >{\raggedright\arraybackslash}p{(\linewidth - 14\tabcolsep) * \real{0.1131}}
  >{\raggedright\arraybackslash}p{(\linewidth - 14\tabcolsep) * \real{0.1301}}@{}}
\toprule\noalign{}
\begin{minipage}[b]{\linewidth}\raggedright
System
(prover)
\end{minipage}
&
\begin{minipage}[b]{\linewidth}\raggedright
Semantics
and fork
\end{minipage}
&
\begin{minipage}[b]{\linewidth}\raggedright
Stated
conformance
\end{minipage}
&
\begin{minipage}[b]{\linewidth}\raggedright
Trust
beyond the
prover's
standard
base, as
stated
\end{minipage}
&
\begin{minipage}[b]{\linewidth}\raggedright
Exogenous
bytecode
\end{minipage}
&
\begin{minipage}[b]{\linewidth}\raggedright
Gas
\end{minipage}
&
\begin{minipage}[b]{\linewidth}\raggedright
Liveness
\end{minipage}
&
\begin{minipage}[b]{\linewidth}\raggedright
Deployed
artifacts
\end{minipage} \\
\midrule\noalign{}
\endhead
\bottomrule\noalign{}
\endlastfoot
\textbf{This
work}
(Lean 4) &
Jaune;
theorems
for
Prague,
Osaka,
BPO1, BPO2
&
5,006/5,006
files,
34,205/34,205
blockchain-test
cases &
none
beyond the
three
standard
axioms;
recovery a
premise of
signature-generic
theorems;
hash
separation
trace-local,
decidable
or absent,
universal
on one
tier &
yes: the
mainnet
bytes,
seven
compiler
versions
and an
assembler
& exact;
closed for
WETH9, the
Uniswap
pair and
EIP-7002,
explicit
sums for
the
deposit
contract,
existential
for 3Crv &
frame,
model,
reachable
state,
transaction
& eight
(Section
4) \\
EquiVM
(Lean 4) &
EVMLean, a
port of
EVMYulLean;
Cancun &
inherits
EVMYulLean's
99.99\% &
\texttt{native\_decide};
two WETH9
return-size
axioms
(repository);
preprint:
an opaque
Keccak
with
per-contract
selector-hash
axioms &
yes: solc
and Vyper
output,
compiled
locally
from
pinned
sources &
not
compared;
out-of-gas
runs
excluded &
termination
``a
separate
proof
obligation''
& local
builds of
WETH9,
most of
MakerDAO's
Dai
system,
the Nouns
auction
and the
Uniswap V2
pair \\
Verifereum
(HOL4) &
own; Osaka
&
17,129/17,423
EEST tests
& README
silent;
one
\texttt{cheat},
in a lemma
used only
by
commented-out
WETH
specifications
& yes:
WETH9's
mainnet
bytes &
gas
monotonicity
theorems
(README) &
not stated
& WETH9: a
deployment
theorem;
deposit
and
fallback
specifications
commented
out;
experimental \\
KEVM
deposit
contract
(K) & KEVM
& KEVM's
own test
runs
{[}KEVM{]}
& KEVM,
the K
prover,
the
bytecode
specification
& yes:
Vyper
output;
solc 0.6.8
build of a
pre-deployment
commit &
per-function
bounds
(Solidity
version) &
deposits
accepted
given
enough gas
(Vyper
version);
success
and revert
paths
(Solidity
version) &
the
contract's
source,
not the
deployed
solc
0.6.11
bytes \\
dapp.org,
Uniswap V2
(K) & KEVM
(2020);
fork not
stated &
KEVM's own
test runs
{[}KEVM{]}
& klab's
act
implementation,
KEVM, the
K prover,
Z3 and
documented
lemmas; a
token
matching
the pair's
own
assumed &
yes:
bytecode
built from
the
\texttt{v1.0.0}
sources,
whose
published
build
matches
the
deployed
runtime &
exact
symbolic
gas for
each
method's
success &
success
given at
least that
gas & the
Uniswap V2
pair and
factory,
method by
method \\
DeepSEA
(Coq) &
Hirai's
Lem EVM;
fork not
named &
inherits
Hirai's
tests & a
Keccak
injectivity
axiom & no
& target
uses at
most the
source's
gas &
target
succeeds
given
enough
gas;
forward
simulation
& none;
own
contracts \\
powdr
(Lean 4) &
own
relational
semantics;
Osaka & no
failures
on its
EEST Osaka
suites &
standard
axioms
only;
Keccak
\texttt{opaque}
with
\texttt{implemented\_by};
\texttt{ExternalsRea\penalty200{}lize\penalty200{}d}
hypothesis
& no &
existential
bound &
forward
simulation
& none;
compiler
correctness \\
Solidus
(Lean 4) &
EVMYulLean
fork;
London to
Osaka &
none
stated for
its fork;
Foundry
differential
tests &
standard
axioms for
headline
theorems;
solc run
trusted &
no &
gasful
refinement;
out-of-gas
halts
allowed;
no bound &
none
claimed &
none;
compiler
correctness \\
Verity
(Lean 4) &
EVMYulLean
fork,
native Yul
dispatcher;
Cancun &
inherits;
520
Foundry
tests (per
its paper)
& one
axiom on
mapping-slot
injectivity,
per its
AXIOMS.md
(none in
the
compiler
core, per
its
README);
\texttt{native\_decide}
in
per-contract
lowering
witnesses,
which its
SimpleStorage
end-to-end
proof
uses, per
its CI
axiom
report;
solc
trusted
from Yul
to
bytecode &
no & not
modeled &
none (its
paper:
correctness
does not
imply
gas-bounded
liveness)
& none;
own
contracts \\
evm-smith
(Lean 4) &
EVMYulLean
fork, no
longer
public;
SELFDESTRUCT
before
EIP-6780,
per its
documentation
&
inherits;
Foundry &
two
axioms;
\texttt{native\_decide}
& no &
tracked;
out-of-gas
runs
vacuous &
none
(partial
correctness)
& none;
hand-written
bytecode \\
\end{longtable}
}

\nopagebreak\noindent\begin{minipage}{\linewidth}
\emph{Table
5.
Interactive-theorem-proving
and
K-based
systems
for EVM
contracts,
each
described
in its own
terms,
with
repository-derived
cells as
of the
commits in
the
references.
The
columns
follow
this
paper's
questions
and so
favor it;
the prose
above
describes
what the
other
systems do
that we do
not,
notably
scale and
automation
(the
symbolic
tools),
whole-interface
specifications
(EquiVM)
and
compiler
theorems
(the
verified
compilers).
``Not
stated''
means that
a
project's
public
materials
say
nothing
either
way.}
\end{minipage}

\subsection{7
Limitations}\label{limitations}

We gather
here the
limits a
reader
should
carry
away; most
are also
stated
where they
arise.

\begin{itemize}
\tightlist
\item
  \textbf{The
  live
  contracts
  are not
  yet
  reached.}
  The
  deployment
  theorems
  execute
  the
  recorded
  creation
  inputs
  under
  current
  rules;
  they do
  not
  prove
  historical
  transaction
  inclusion.
  For
  WETH9,
  the
  deposit
  contract,
  3Crv and
  the
  Uniswap
  pair,
  whose
  histories
  began
  before
  Prague,
  reaching
  today's
  state
  means
  checking
  the
  checkpoint
  invariant
  on
  mainnet's
  storage
  at a
  block
  boundary.
  Each
  check is
  a finite
  computation
  over
  public
  data, in
  two
  parts.
  The rows
  recovered
  from the
  contract's
  history
  are read
  and
  checked
  against
  the
  invariant;
  and,
  because
  every
  invariant
  here
  requires
  complete
  support,
  every
  nonzero
  word at
  a fixed
  slot or
  a
  tracked
  row, the
  storage
  must be
  shown to
  hold
  nothing
  else.
  Reading
  listed
  slots,
  even
  with an
  authenticated
  proof
  for
  each,
  never
  establishes
  the
  absence
  of
  unlisted
  rows, so
  the
  second
  part
  needs an
  enumeration
  of the
  contract's
  storage,
  from an
  archive
  node's
  storage-range
  query or
  a state
  snapshot.
  Read
  over
  RPC, the
  result
  is
  evidence,
  not a
  proof.
  We
  describe
  the
  first
  four
  checks
  but have
  not run
  them;
  the
  fifth we
  ran:

  \begin{itemize}
  \tightlist
  \item
    \emph{WETH9.}
    The
    contract's
    logs
    and
    calldata
    yield
    every
    address
    that
    has
    held a
    balance
    or an
    allowance;
    their
    slots,
    with
    slots
    0 to
    2,
    must
    be
    distinct
    and
    the
    only
    nonzero
    ones,
    and
    the
    tracked
    balances
    must
    sum to
    at
    most
    the
    contract's
    ether.
    Freshness
    for a
    later
    history
    is one
    more
    pass:
    five
    keys,
    seven
    Keccak
    evaluations,
    per
    call.
  \item
    \emph{The
    deposit
    contract.}
    Its
    \texttt{DepositEvent}
    logs
    rebuild
    the
    list
    of all
    2,565,727
    deposits,
    whose
    recomputed
    accumulator
    must
    match
    the
    branch
    slots
    and
    count
    in
    storage.
  \item
    \emph{3Crv.}
    Its
    \texttt{Transfer}
    and
    \texttt{Approval}
    logs
    recover
    the
    holders
    and
    allowance
    pairs,
    whose
    slots,
    with
    the
    minter
    and
    metadata
    slots,
    must
    represent
    a
    model
    state,
    with
    no
    other
    slot
    nonzero.
  \item
    \emph{The
    Uniswap
    pair.}
    Per
    pair,
    the
    fixed
    slots
    and
    the
    mapping
    rows
    recovered
    from
    its
    \texttt{Transfer}
    and
    \texttt{Approval}
    logs
    and
    \texttt{permit}
    calls
    must
    represent
    a
    model
    state
    over
    distinct,
    unused
    rows,
    with
    no
    other
    slot
    nonzero;
    the
    ledger
    result
    also
    needs
    the
    model's
    LP
    balances
    to sum
    to the
    supply,
    and
    the
    minimum-liquidity
    result
    the
    supply
    floor
    of
    Appendix
    F.
  \item
    \emph{The
    CircuitBreaker.}
    It has
    only
    ever
    been
    used
    under
    forks
    Jaune
    covers.
    At
    block
    26,141,229,
    29
    explicit
    probes
    over
    107
    storage
    slots,
    on
    which
    three
    providers
    agree,
    satisfy
    every
    field
    of the
    finite
    tier's
    decidable
    \texttt{RegistryOn}
    check
    for
    the 19
    live
    entries,
    which
    replaying
    all 23
    \texttt{PauserSet}
    events
    since
    deployment
    reproduces.
    The
    probes
    cover
    finitely
    many
    addresses
    and no
    state
    root
    authenticates
    the
    reads;
    a
    kernel-checked
    version
    of
    this
    check
    has
    not
    been
    built.
  \end{itemize}

  The
  Vyper
  pools
  have no
  checkpoint
  invariant,
  since
  their
  exclusion
  theorem
  takes
  its
  premises
  per
  execution.
  For the
  EIP-7002
  contract
  the
  check is
  done:
  mainnet's
  state at
  Prague's
  activation
  meets
  the
  checkpoint
  premises
  (Section
  5.1),
  and what
  separates
  the
  theorems
  from the
  live
  contract
  is the
  premise
  that
  mainnet's
  history
  is a
  configured
  one and
  the
  trace
  premises.
  In no
  case is
  semantics
  missing.
\item
  \textbf{Mid-block
  positions
  are
  covered
  for
  WETH9
  only.}
  The
  prefix
  machinery
  of
  Section
  2.1 is
  contract-neutral,
  but only
  WETH9's
  exit
  uses it;
  the
  other
  contracts'
  liveness
  theorems
  apply at
  the end
  of a
  block.
\item
  \textbf{Few
  closed
  instances
  of the
  history
  theorems.}
  \texttt{weth9\_\allowbreak{}closed\_\allowbreak{}exit\_\allowbreak{}instance}
  discharges
  every
  premise
  of the
  WETH9
  exit
  theorem
  on a
  connected
  synthetic
  execution:
  deployment
  from the
  recorded
  creation
  input, a
  signed
  deposit
  in a
  valid
  configured
  BPO2
  block,
  and a
  signed
  withdrawal
  of the
  holder's
  whole
  one-wei
  balance,
  using
  30,344
  gas. It
  shows
  that the
  premises
  can hold
  together
  in the
  model,
  but its
  funding
  and the
  programs
  at the
  system-contract
  addresses
  are
  synthetic,
  and it
  neither
  authenticates
  mainnet
  inclusion
  nor
  places
  the
  withdrawal
  inside a
  valid
  successor
  block.
  The
  deposit-contract,
  3Crv and
  Uniswap
  history
  theorems
  still
  lack
  complete
  applications;
  the
  mainnet
  receipts
  of
  Sections
  2.4 and
  5.1 are
  empirical
  examples
  of
  transactions
  of the
  theorems'
  shape.
\item
  \textbf{The
  statements
  are the
  specification.}
  The
  kernel
  checks
  proofs,
  not
  whether
  a
  statement
  says
  what it
  should.
  These
  statements
  were
  written
  by
  agents
  from
  informal
  instructions,
  and the
  author's
  reading
  of them,
  partly
  through
  agents'
  explanations,
  stopped
  short of
  the
  bottom
  of their
  definitions
  (Section
  5.3).
  The
  statements'
  adequacy
  rests on
  what a
  reader
  can
  check:
  the
  claim-map
  gate,
  which
  checks
  that the
  public
  account
  of the
  results
  cites
  declarations
  that
  exist
  where it
  says;
  the
  statement
  pins,
  which
  fix the
  statements
  its
  claims
  rest on;
  and the
  statements
  themselves,
  which
  Appendix
  A
  locates
  and
  which
  have to
  be read.
  Section
  5.3
  gives
  two
  defects
  of the
  kind
  such
  reading
  must
  catch.
\item
  \textbf{Amsterdam
  is not
  covered.}
  Jaune
  runs a
  prerelease
  of
  Amsterdam
  (Section
  3.1),
  but no
  theorem
  here is
  stated
  for it.
  Every
  history
  theorem
  requires
  each
  block of
  the
  history
  to run
  under
  one of
  the four
  covered
  forks
  (Section
  3.4), so
  once
  Amsterdam
  activates
  on
  mainnet
  the
  theorems
  apply to
  histories
  that end
  before
  its
  activation
  and say
  nothing
  about
  histories
  that
  cross
  it, the
  safety
  results
  included.
  Amsterdam
  also
  reprices
  gas
  {[}Jaune{]},
  so the
  exact
  costs
  proved
  here
  need not
  carry
  over in
  any
  case.
\item
  \textbf{Premises
  of
  individual
  results.}
  Some
  results
  carry
  premises
  of their
  own:
  signature
  recovery
  (Section
  2.1),
  the
  CircuitBreaker
  history
  tier's
  hash
  premises
  over all
  addresses
  (Section
  3.7),
  the
  fixed
  pool's
  premises
  for each
  execution
  (Section
  4.7) and
  the
  EIP-7002
  results'
  bound of
  2²⁵⁴
  committed
  submissions
  (Section
  4.5).
\item
  \textbf{Callee
  premises.}
  WETH9's
  withdrawal
  to a
  contract,
  3Crv's
  \texttt{set\_name}
  and six
  of the
  Uniswap
  pair's
  functions
  assume
  the
  callee's
  behavior
  for
  liveness,
  as any
  such
  statement
  must
  (Section
  3.8).
  The
  Uniswap
  share-value
  theorems
  assume a
  premise
  about
  the
  tokens'
  actual
  answers
  over the
  history's
  own
  steps,
  and the
  fee-off
  form one
  about
  the
  factory's;
  no
  theorem
  says
  that a
  real
  token
  satisfies
  them.
\item
  \textbf{Contracts
  are
  verified
  one at a
  time.}
  At the
  exhibit
  pair,
  the
  Uniswap
  theorems
  take
  WETH's
  answers
  as
  inputs
  constrained
  by
  NoShrink,
  while
  the
  WETH9
  theorems
  describe
  WETH9's
  own
  storage;
  a
  statement
  joining
  them can
  be posed
  over the
  same
  configured
  histories,
  but
  proving
  it needs
  an
  invariant
  relating
  the
  pair's
  reserves
  to
  WETH9's
  ledger,
  which we
  have not
  proved.
\item
  \textbf{Reachability
  is at
  message
  level.}
  The
  Vyper
  witnesses
  from
  deployment
  are
  sequences
  of
  messages,
  not
  admitted
  transactions:
  implementations
  are
  placed
  at their
  addresses
  as
  supplied
  creation
  targets,
  messages
  start
  with
  empty
  warm
  sets, no
  fees or
  block-level
  system
  calls
  intervene,
  and the
  token,
  attacker,
  receiver
  and
  clone-deployer
  code is
  synthetic.
  That
  admitted
  transactions
  can
  produce
  equivalent
  messages
  is
  argued,
  not
  proved.
\item
  \textbf{Static
  and
  rolled-back
  frames.}
  For
  WETH9
  and the
  CircuitBreaker,
  committed-call
  lists
  exclude
  static
  frames
  by the
  form of
  the
  statement
  or by a
  premise,
  and
  statements
  about
  committed
  frames
  say
  nothing
  about
  rolled-back
  ones,
  although
  the
  history
  invariants
  hold of
  the
  final
  storage
  regardless.
\item
  \textbf{Byte
  identity
  is
  recorded,
  not
  proved.}
  Agreement
  among
  providers
  is
  evidence
  about
  which
  bytes
  are
  deployed,
  not a
  theorem;
  a proof
  anchored
  in a
  block
  header's
  state
  root
  would
  make it
  one.
\item
  \textbf{Certificates
  need
  constant
  jump
  targets.}
  \texttt{Cert.check}
  accepts
  a jump
  only to
  a known
  constant
  destination
  or the
  current
  function's
  own
  return
  address,
  extended
  to words
  at fixed
  memory
  offsets
  by the
  memory-map
  variant;
  code
  that
  jumps to
  computed
  targets,
  such as
  function
  pointers
  loaded
  from
  storage,
  cannot
  be
  certified
  in this
  form.
\item
  \textbf{The
  proof-checking
  trusted
  base is
  small,
  not
  minimal.}
  It is
  Lean's
  kernel,
  with its
  arbitrary-precision
  arithmetic,
  and
  Lean's
  three
  standard
  axioms,
  all of
  which
  the
  library
  reaches.
  The
  replay
  of
  Section
  3.11
  checks
  the
  compiled
  library
  again
  with
  that
  same
  kernel,
  so it
  shows
  that no
  check
  was
  skipped,
  not that
  the
  kernel
  is
  correct.
\item
  \textbf{Coverage
  is
  demonstrated,
  not
  promised.}
  Eight
  artifacts
  show
  that the
  method
  reaches
  ordinary
  compiler
  output
  and
  hand-written
  assembly
  across
  languages
  and
  eras,
  not that
  every
  contract
  can be
  done in
  fifteen
  days.
\end{itemize}

\subsection{8
Conclusion}\label{conclusion}

The six
questions
of Section
1 are
usually
answered a
few at a
time,
because
each
answer has
a cost;
for
contracts
like
these,
they can
be
answered
together.
A theorem
can be
about the
bytes at a
mainnet
address,
have a
proof that
rests on
nothing
but a
kernel and
its three
axioms,
concern
compiler
output
from 2017
or
assembly
that the
protocol
itself
runs, be
stated
against
the whole
machine,
guarantee
something
a holder
cares
about, and
say that,
under its
premises,
a
transaction
will be
admitted
and
exactly
what it
will cost;
and the
receipts
we
compared
for the
output of
two
compilers,
4,290 in
our main
comparisons
and 1,688
more that
we counted
apart
(Section
2.4), all
agree with
the stated
costs to
the unit.

The corpus
now
includes a
contract
whose
failure
would
invalidate
blocks,
and a
market
whose
correctness
depends on
what other
contracts
answer.
For the
first, the
proofs
show that,
in Jaune's
semantics
and with
its
canonical
code
installed,
no
configured
history
can make
the
end-of-block
call fail,
that the
requests
it returns
are
exactly
the oldest
committed
submissions,
and
exactly
where the
deployed
arithmetic
agrees
with the
specification
it
implements;
by
computation,
the first
disagreement
comes
where the
specified
fee
already
exceeds
all the
ether in
existence.
For the
second,
the
answers
became
inputs to
the
statements,
and the
premises
about them
range over
one
history's
answers
only.

What
remains is
work of
three
kinds. The
theorems'
starting
points
must be
tied to
the
contracts'
actual
state, by
checking
each
checkpoint
invariant
against
the
contract's
complete,
live
storage;
the
EIP-7002
contract's
unchanged
storage
root at
Prague's
activation
already
does so,
and the
CircuitBreaker's
live
registry
passes its
finite
check,
evaluated
by a
script.
Contracts
must be
composed,
beginning
with the
tokens
whose
answers
the
Uniswap
theorems
take as
premises.
And the
3Crv,
deposit-contract
and
Uniswap
history
theorems
need
closed
instances,
as WETH9's
exit
theorem
has.

\subsection{Appendix
A. Where
the
theorems
are}\label{appendix-a.-where-the-theorems-are}

All paths
are in the
Blanc
repository
at
\texttt{5048a51f};
every
declaration
below
resolves
at its
file
there. The
repository's
\texttt{docs/\allowbreak{}DEPLOYED\_\allowbreak{}BYTECODE\_\allowbreak{}CLAIM\_\allowbreak{}MAP.\allowbreak{}md}
gives the
premises
of the
deployed-artifact
theorems
below,
with line
numbers
checked by
its gate,
and the
full set
of
disclosures;
the
EIP-7002
excess
bound is
cited from
the source
only.
Within
each run
of rows
whose
Result
names the
same
artifact,
\texttt{\pfvellipsis{}}
stands for
the
namespace
spelled
out in the
run's
first row
(for the
V± rows,
\texttt{Blanc.\allowbreak{}Lift.\allowbreak{}VyperNonreen\penalty200{}tran\penalty200{}tDep\penalty200{}loye\penalty200{}d});
a file
name
beginning
with
\texttt{\pfvellipsis{}}
lies in
the
directory
of the
row's
first
file, and
\texttt{\pfvellipsis{}/}
stands for
\texttt{Blanc/\allowbreak{}Lift/\allowbreak{}LidoCircuitB\penalty200{}reak\penalty200{}erDe\penalty200{}ploy\penalty200{}ed/\allowbreak{}}
in the
CircuitBreaker
rows and
for
\texttt{Blanc/\allowbreak{}Lift/\allowbreak{}VyperNonreen\penalty200{}tran\penalty200{}tDep\penalty200{}loye\penalty200{}d/\allowbreak{}}
in the V±
rows.

{\def\LTcaptype{none} % do not increment counter
\begin{longtable}[]{@{}
  >{\raggedright\arraybackslash}p{(\linewidth - 4\tabcolsep) * \real{0.2238}}
  >{\raggedright\arraybackslash}p{(\linewidth - 4\tabcolsep) * \real{0.4056}}
  >{\raggedright\arraybackslash}p{(\linewidth - 4\tabcolsep) * \real{0.3706}}@{}}
\toprule\noalign{}
\begin{minipage}[b]{\linewidth}\raggedright
Result
\end{minipage}
&
\begin{minipage}[b]{\linewidth}\raggedright
Declaration
\end{minipage}
&
\begin{minipage}[b]{\linewidth}\raggedright
File
\end{minipage} \\
\midrule\noalign{}
\endhead
\bottomrule\noalign{}
\endlastfoot
Lifting,
safety &
\texttt{Blanc.\allowbreak{}Lift.\allowbreak{}lift\_\allowbreak{}sound}
&
\texttt{Blanc/\allowbreak{}Lift/\allowbreak{}Sound.\allowbreak{}lean} \\
Lifting,
gas-exact
converse &
\texttt{Blanc.\allowbreak{}Lift.\allowbreak{}lift\_\allowbreak{}exact}
&
\texttt{Blanc/\allowbreak{}Lift/\allowbreak{}Exact.\allowbreak{}lean} \\
Certificate
cursor &
\texttt{Blanc.\allowbreak{}Lift.\allowbreak{}cursor\_\allowbreak{}of\_\allowbreak{}parentPrefix\penalty200{}}
&
\texttt{Blanc/\allowbreak{}Lift/\allowbreak{}Cursor.\allowbreak{}lean} \\
Lock
certificate
and its
soundness
&
\texttt{Blanc.\allowbreak{}Lift.\allowbreak{}LockCheck.\allowbreak{}lockCert},
\texttt{Blanc.\allowbreak{}Lift.\allowbreak{}LockCheck.\allowbreak{}dominance}
&
\texttt{Blanc/\allowbreak{}Lift/\allowbreak{}LockCheck.\allowbreak{}lean},
\texttt{Blanc/\allowbreak{}Lift/\allowbreak{}LockCheckFlo\penalty200{}w.\allowbreak{}lean} \\
Configured
history &
\texttt{Blanc.\allowbreak{}ExecutionTra\penalty200{}ce.\allowbreak{}ConfiguredHi\penalty200{}stor\penalty200{}yTra\penalty200{}ce}
&
\texttt{Blanc/\allowbreak{}ExecutionHis\penalty200{}tory\penalty200{}.\allowbreak{}lean} \\
WETH9 exit
transaction
after any
history &
\texttt{Blanc.\allowbreak{}Lift.\allowbreak{}Weth9.\allowbreak{}weth9\_\allowbreak{}history\_\allowbreak{}tx\_\allowbreak{}withdraw}
&
\texttt{Blanc/\allowbreak{}Lift/\allowbreak{}Weth9/\allowbreak{}LiveTx.\allowbreak{}lean} \\
WETH9 exit
transaction
at an
invariant
state &
\texttt{\pfvellipsis{}weth9\_\allowbreak{}tx\_\allowbreak{}withdraw}
&
\texttt{Blanc/\allowbreak{}Lift/\allowbreak{}Weth9/\allowbreak{}LiveTx.\allowbreak{}lean} \\
WETH9
invariant
after any
prefix of
a block &
\texttt{\pfvellipsis{}weth9\_\allowbreak{}prefix\_\allowbreak{}footprint\_\allowbreak{}universe}
&
\texttt{Blanc/\allowbreak{}Lift/\allowbreak{}Weth9/\allowbreak{}FootPrefix.\allowbreak{}lean} \\
WETH9 exit
transaction
after any
prefix of
a block &
\texttt{\pfvellipsis{}weth9\_\allowbreak{}prefix\_\allowbreak{}tx\_\allowbreak{}withdraw}
&
\texttt{Blanc/\allowbreak{}Lift/\allowbreak{}Weth9/\allowbreak{}PrefixTx.\allowbreak{}lean} \\
WETH9 exit
theorem, a
closed
instance &
\texttt{\pfvellipsis{}ClosedInsta\penalty200{}nce.\allowbreak{}weth9\_\allowbreak{}closed\_\allowbreak{}exit\_\allowbreak{}instance}
&
\texttt{Blanc/\allowbreak{}Lift/\allowbreak{}Weth9/\allowbreak{}ClosedExit.\allowbreak{}lean} \\
WETH9
footprint
and
backing &
\texttt{\pfvellipsis{}weth9\_\allowbreak{}history\_\allowbreak{}footprint},
\texttt{\pfvellipsis{}weth9\_\allowbreak{}history\_\allowbreak{}backed}
&
\texttt{Blanc/\allowbreak{}Lift/\allowbreak{}Weth9/\allowbreak{}FootHistory.\allowbreak{}lean} \\
WETH9
exact
committed
replay &
\texttt{\pfvellipsis{}weth9\_\allowbreak{}history\_\allowbreak{}committed}
&
\texttt{Blanc/\allowbreak{}Lift/\allowbreak{}Weth9/\allowbreak{}CommittedHis\penalty200{}tory\penalty200{}.\allowbreak{}lean} \\
WETH9
liveness
after any
history &
\texttt{\pfvellipsis{}weth9\_\allowbreak{}history\_\allowbreak{}\{withdraw,\allowbreak{}deposit,\allowbreak{}transfer\}\_\allowbreak{}live}
&
\texttt{Blanc/\allowbreak{}Lift/\allowbreak{}Weth9/\allowbreak{}LiveHistory.\allowbreak{}lean} \\
WETH9 hash
premise is
needed &
\texttt{\pfvellipsis{}trackedSum\_\allowbreak{}collision\_\allowbreak{}breaks\_\allowbreak{}backing}
&
\texttt{Blanc/\allowbreak{}Lift/\allowbreak{}Weth9/\allowbreak{}FootFrame.\allowbreak{}lean} \\
WETH9
deployment
establishes
the
invariant
&
\texttt{\pfvellipsis{}Creation.\allowbreak{}weth9\_\allowbreak{}deploy\_\allowbreak{}init\_\allowbreak{}covered}
&
\texttt{Blanc/\allowbreak{}Lift/\allowbreak{}Weth9/\allowbreak{}Creation/\allowbreak{}DeployInit.\allowbreak{}lean} \\
Deposit
contract:
committed
history,
count,
root &
\texttt{Blanc.\allowbreak{}Lift.\allowbreak{}BeaconDeposi\penalty200{}t.\allowbreak{}configuredHi\penalty200{}stor\penalty200{}y\_\allowbreak{}\{solInv,\allowbreak{}count,\allowbreak{}root\}\_\allowbreak{}sys}
&
\texttt{Blanc/\allowbreak{}Lift/\allowbreak{}BeaconDeposi\penalty200{}t/\allowbreak{}BeaconEnv.\allowbreak{}lean} \\
Deposit
contract:
frame
refinement
&
\texttt{\pfvellipsis{}deposit\_\allowbreak{}frame\_\allowbreak{}refines}
&
\texttt{Blanc/\allowbreak{}Lift/\allowbreak{}BeaconDeposi\penalty200{}t/\allowbreak{}Safe.\allowbreak{}lean} \\
Deposit
contract:
liveness
after any
history &
\texttt{\pfvellipsis{}configuredH\penalty200{}isto\penalty200{}ry\_\allowbreak{}deposit\_\allowbreak{}live}
&
\texttt{Blanc/\allowbreak{}Lift/\allowbreak{}BeaconDeposi\penalty200{}t/\allowbreak{}Liveness.\allowbreak{}lean} \\
Deposit
contract:
deployment
&
\texttt{\pfvellipsis{}Creation.\allowbreak{}beacon\_\allowbreak{}deploy\_\allowbreak{}covered}
&
\texttt{Blanc/\allowbreak{}Lift/\allowbreak{}BeaconDeposi\penalty200{}t/\allowbreak{}Creation/\allowbreak{}Deploy.\allowbreak{}lean} \\
Incremental
root
equals
reference
root &
\texttt{Blanc.\allowbreak{}BeaconDeposi\penalty200{}t.\allowbreak{}root\_\allowbreak{}correct}
&
\texttt{Blanc/\allowbreak{}BeaconDeposi\penalty200{}tCor\penalty200{}rect\penalty200{}ness\penalty200{}.\allowbreak{}lean} \\
The
earlier
premise is
false of
the
EIP-7002
code &
\texttt{Blanc.\allowbreak{}withdrawalRe\penalty200{}ques\penalty200{}tCod\penalty200{}e\_\allowbreak{}not\_\allowbreak{}spawnFree}
&
\texttt{Blanc/\allowbreak{}SystemContra\penalty200{}cts.\allowbreak{}lean} \\
3Crv:
exact
replay and
conservation
&
\texttt{Blanc.\allowbreak{}Lift.\allowbreak{}Curve3Crv.\allowbreak{}c3crv\_\allowbreak{}history\_\allowbreak{}committed\_\allowbreak{}derived}
&
\texttt{Blanc/\allowbreak{}Lift/\allowbreak{}Curve3Crv/\allowbreak{}CommittedHis\penalty200{}tory\penalty200{}.\allowbreak{}lean} \\
3Crv:
frame
refinement,
every
selector &
\texttt{\pfvellipsis{}c3crv\_\allowbreak{}frame\_\allowbreak{}refines}
&
\texttt{Blanc/\allowbreak{}Lift/\allowbreak{}Curve3Crv/\allowbreak{}Safe.\allowbreak{}lean} \\
3Crv:
liveness
after any
history &
\texttt{\pfvellipsis{}c3crv\_\allowbreak{}history\_\allowbreak{}live}
&
\texttt{Blanc/\allowbreak{}Lift/\allowbreak{}Curve3Crv/\allowbreak{}Liveness.\allowbreak{}lean} \\
3Crv:
\texttt{set\_name}
live given
the owner
call &
\texttt{\pfvellipsis{}c3crv\_\allowbreak{}history\_\allowbreak{}setName\_\allowbreak{}live},
\texttt{\pfvellipsis{}OwnerCallOk}
&
\texttt{Blanc/\allowbreak{}Lift/\allowbreak{}Curve3Crv/\allowbreak{}Liveness.\allowbreak{}lean},
\texttt{\pfvellipsis{}LiveBodies.\allowbreak{}lean} \\
3Crv:
token-model
properties
&
\texttt{Blanc.\allowbreak{}Curve3Crv.\allowbreak{}step\_\allowbreak{}conserved}
and
neighbors
&
\texttt{Blanc/\allowbreak{}Curve3Crv/\allowbreak{}Properties.\allowbreak{}lean} \\
Uniswap:
nested
committed
replay &
\texttt{Blanc.\allowbreak{}Lift.\allowbreak{}UniswapV2Pai\penalty200{}r.\allowbreak{}pair\_\allowbreak{}history\_\allowbreak{}committed},
\texttt{\pfvellipsis{}pair\_\allowbreak{}history\_\allowbreak{}initialized}
&
\texttt{Blanc/\allowbreak{}Lift/\allowbreak{}UniswapV2Pai\penalty200{}r/\allowbreak{}PairHistory.\allowbreak{}lean} \\
Uniswap:
LP ledger
from any
represented
checkpoint,
its
initialized
form, and
the oracle
&
\texttt{\pfvellipsis{}pair\_\allowbreak{}history\_\allowbreak{}ledger\_\allowbreak{}from\_\allowbreak{}checkpoint},
\texttt{\pfvellipsis{}pair\_\allowbreak{}history\_\allowbreak{}ledger},
\texttt{\pfvellipsis{}pair\_\allowbreak{}history\_\allowbreak{}oracle}
&
\texttt{Blanc/\allowbreak{}Lift/\allowbreak{}UniswapV2Pai\penalty200{}r/\allowbreak{}PairHistory.\allowbreak{}lean} \\
Uniswap:
share
value, fee
off and
fee on &
\texttt{\pfvellipsis{}pair\_\allowbreak{}history\_\allowbreak{}feeOff\_\allowbreak{}product},
\texttt{\pfvellipsis{}pair\_\allowbreak{}history\_\allowbreak{}feeOn\_\allowbreak{}product}
&
\texttt{Blanc/\allowbreak{}Lift/\allowbreak{}UniswapV2Pai\penalty200{}r/\allowbreak{}PairHistory.\allowbreak{}lean} \\
Uniswap:
minimum
liquidity
and the
ratio form
from any
represented
checkpoint,
and their
initialized
forms &
\texttt{\pfvellipsis{}pair\_\allowbreak{}history\_\allowbreak{}minimum\_\allowbreak{}liquidity\_\allowbreak{}from\_\allowbreak{}checkpoint},
\texttt{\pfvellipsis{}pair\_\allowbreak{}history\_\allowbreak{}feeOff\_\allowbreak{}ratio\_\allowbreak{}from\_\allowbreak{}checkpoint},
\texttt{\pfvellipsis{}pair\_\allowbreak{}history\_\allowbreak{}minimum\_\allowbreak{}liquidity},
\texttt{\pfvellipsis{}pair\_\allowbreak{}history\_\allowbreak{}feeOff\_\allowbreak{}ratio}
&
\texttt{Blanc/\allowbreak{}Lift/\allowbreak{}UniswapV2Pai\penalty200{}r/\allowbreak{}PairHistoryM\penalty200{}inLi\penalty200{}quid\penalty200{}ity.\allowbreak{}lean} \\
Uniswap:
frame
refinement,
for
example
\texttt{swap}
and the
views &
\texttt{\pfvellipsis{}swap\_\allowbreak{}bytecode\_\allowbreak{}exact\_\allowbreak{}consumes},
\texttt{\pfvellipsis{}staticView\_\allowbreak{}source\_\allowbreak{}handler\_\allowbreak{}inv}
&
\texttt{Blanc/\allowbreak{}Lift/\allowbreak{}UniswapV2Pai\penalty200{}r/\allowbreak{}SwapPosition\penalty200{}alRe\penalty200{}fine\penalty200{}ment\penalty200{}.\allowbreak{}lean},
\texttt{Blanc/\allowbreak{}Lift/\allowbreak{}UniswapV2Pai\penalty200{}r/\allowbreak{}StaticViewSo\penalty200{}urce\penalty200{}.\allowbreak{}lean} \\
Uniswap:
positional
frame
refinement
of the
mutable
writers &
\texttt{\pfvellipsis{}burnRaw\_\allowbreak{}source\_\allowbreak{}authentic},
\texttt{\pfvellipsis{}skim\_\allowbreak{}bytecode\_\allowbreak{}exact\_\allowbreak{}consumes},
\texttt{\pfvellipsis{}mint\_\allowbreak{}bytecode\_\allowbreak{}exact\_\allowbreak{}consumes}
&
\texttt{Blanc/\allowbreak{}Lift/\allowbreak{}UniswapV2Pai\penalty200{}r/\allowbreak{}BurnPosition\penalty200{}alRe\penalty200{}fine\penalty200{}ment\penalty200{}.\allowbreak{}lean},
\texttt{\pfvellipsis{}SkimPositio\penalty200{}nalR\penalty200{}efin\penalty200{}emen\penalty200{}t.\allowbreak{}lean},
\texttt{\pfvellipsis{}MintPositio\penalty200{}nalR\penalty200{}efin\penalty200{}emen\penalty200{}t.\allowbreak{}lean} \\
Uniswap:
the
admitted
dispatcher
over all
eleven
families
(the ten
state-changing
entry
points and
the views
as one),
and the
admission
it carries
&
\texttt{\pfvellipsis{}pairAdmitte\penalty200{}dSup\penalty200{}ply},
\texttt{\pfvellipsis{}PairAdmitte\penalty200{}dCon\penalty200{}sume\penalty200{}s},
\texttt{\pfvellipsis{}PairEntryAuth}
&
\texttt{Blanc/\allowbreak{}Lift/\allowbreak{}UniswapV2Pai\penalty200{}r/\allowbreak{}PairPosition\penalty200{}alSu\penalty200{}pply\penalty200{}.\allowbreak{}lean},
\texttt{\pfvellipsis{}PairPositio\penalty200{}nalA\penalty200{}dmis\penalty200{}sion\penalty200{}.\allowbreak{}lean},
\texttt{\pfvellipsis{}PairPositio\penalty200{}nalE\penalty200{}ntry\penalty200{}.\allowbreak{}lean} \\
Uniswap:
the
square-root
loop &
\texttt{\pfvellipsis{}sqrt\_of\_run}
&
\texttt{Blanc/\allowbreak{}Lift/\allowbreak{}UniswapV2Pai\penalty200{}r/\allowbreak{}SqrtWalk.\allowbreak{}lean} \\
Uniswap:
laws of
mint, burn
and swap &
\texttt{\pfvellipsis{}runTyped\_\allowbreak{}mint\_\allowbreak{}initial},
\texttt{\pfvellipsis{}runTyped\_\allowbreak{}mint\_\allowbreak{}later},
\texttt{\pfvellipsis{}runTyped\_\allowbreak{}burn\_\allowbreak{}payout},
\texttt{\pfvellipsis{}runTyped\_\allowbreak{}swap\_\allowbreak{}canonical\_\allowbreak{}success}
&
\texttt{Blanc/\allowbreak{}Lift/\allowbreak{}UniswapV2Pai\penalty200{}r/\allowbreak{}Properties.\allowbreak{}lean},
\texttt{\pfvellipsis{}PropertiesM\penalty200{}intB\penalty200{}urn.\allowbreak{}lean},
\texttt{\pfvellipsis{}PropertiesS\penalty200{}wap.\allowbreak{}lean} \\
Uniswap:
liveness
after any
history &
\texttt{\pfvellipsis{}pair\_\allowbreak{}history\_\allowbreak{}\{writer,\allowbreak{}sync,\allowbreak{}mint,\allowbreak{}swap,\allowbreak{}burn,\allowbreak{}skim\}\_\allowbreak{}live},
\texttt{\pfvellipsis{}pair\_\allowbreak{}history\_\allowbreak{}\{permit,\allowbreak{}initialize\}\_\allowbreak{}live}
&
\texttt{Blanc/\allowbreak{}Lift/\allowbreak{}UniswapV2Pai\penalty200{}r/\allowbreak{}PairHistoryL\penalty200{}ive.\allowbreak{}lean},
\texttt{\pfvellipsis{}PairHistory\penalty200{}Live\penalty200{}Admi\penalty200{}n.\allowbreak{}lean} \\
Uniswap:
deployment
and the
exhibit's
address &
\texttt{\pfvellipsis{}Creation.\allowbreak{}pair\_\allowbreak{}create2\_\allowbreak{}initialize\_\allowbreak{}live},
\texttt{\pfvellipsis{}Creation.\allowbreak{}exhibit\_\allowbreak{}create2},
\texttt{\pfvellipsis{}Creation.\allowbreak{}pairAddress\_\allowbreak{}eq}
&
\texttt{Blanc/\allowbreak{}Lift/\allowbreak{}UniswapV2Pai\penalty200{}r/\allowbreak{}Creation/\allowbreak{}DeployInit.\allowbreak{}lean},
\texttt{\pfvellipsis{}Facts.lean} \\
Uniswap:
controls &
\texttt{\pfvellipsis{}ModelContro\penalty200{}ls.\allowbreak{}noShrink\_\allowbreak{}required},
\texttt{\pfvellipsis{}ModelContro\penalty200{}ls.\allowbreak{}mintRoundUp\_\allowbreak{}breaks\_\allowbreak{}feeOff\_\allowbreak{}product},
\texttt{\pfvellipsis{}sync\_\allowbreak{}liveness\_\allowbreak{}refuted}
&
\texttt{Blanc/\allowbreak{}Lift/\allowbreak{}UniswapV2Pai\penalty200{}r/\allowbreak{}ModelControl\penalty200{}s.\allowbreak{}lean},
\texttt{\pfvellipsis{}CalleeContr\penalty200{}olsR\penalty200{}each\penalty200{}.\allowbreak{}lean} \\
EIP-7002:
requests
are the
oldest
pending
submissions
&
\texttt{Blanc.\allowbreak{}Lift.\allowbreak{}WithdrawalRe\penalty200{}ques\penalty200{}t.\allowbreak{}block\_\allowbreak{}word\_\allowbreak{}fifo}
&
\texttt{Blanc/\allowbreak{}Lift/\allowbreak{}WithdrawalRe\penalty200{}ques\penalty200{}t/\allowbreak{}WordFifo.\allowbreak{}lean} \\
EIP-7002:
delivery
bound &
\texttt{\pfvellipsis{}block\_\allowbreak{}word\_\allowbreak{}delivery}
&
\texttt{Blanc/\allowbreak{}Lift/\allowbreak{}WithdrawalRe\penalty200{}ques\penalty200{}t/\allowbreak{}WordDelivery\penalty200{}.\allowbreak{}lean} \\
EIP-7002:
exact
replay &
\texttt{\pfvellipsis{}history\_\allowbreak{}word\_\allowbreak{}storage\_\allowbreak{}replay},
\texttt{\pfvellipsis{}history\_\allowbreak{}word\_\allowbreak{}model}
&
\texttt{Blanc/\allowbreak{}Lift/\allowbreak{}WithdrawalRe\penalty200{}ques\penalty200{}t/\allowbreak{}WordReplay.\allowbreak{}lean},
\texttt{\pfvellipsis{}WordFifo.lean} \\
EIP-7002:
frame
classification
&
\texttt{\pfvellipsis{}exec\_\allowbreak{}frame\_\allowbreak{}effect}
&
\texttt{Blanc/\allowbreak{}Lift/\allowbreak{}WithdrawalRe\penalty200{}ques\penalty200{}t/\allowbreak{}FrameEffects\penalty200{}.\allowbreak{}lean} \\
EIP-7002:
the system
call
succeeds &
\texttt{\pfvellipsis{}checked\_\allowbreak{}system\_\allowbreak{}totality},
\texttt{\pfvellipsis{}history\_\allowbreak{}checked\_\allowbreak{}system\_\allowbreak{}totality}
&
\texttt{Blanc/\allowbreak{}Lift/\allowbreak{}WithdrawalRe\penalty200{}ques\penalty200{}t/\allowbreak{}SystemProtoc\penalty200{}ol.\allowbreak{}lean},
\texttt{\pfvellipsis{}SystemHisto\penalty200{}ry.\allowbreak{}lean} \\
EIP-7002:
balance
and
payments &
\texttt{\pfvellipsis{}history\_\allowbreak{}balance\_\allowbreak{}nondecreasin\penalty200{}g},
\texttt{\pfvellipsis{}history\_\allowbreak{}submission\_\allowbreak{}payments}
&
\texttt{Blanc/\allowbreak{}Lift/\allowbreak{}WithdrawalRe\penalty200{}ques\penalty200{}t/\allowbreak{}BalanceHisto\penalty200{}ry.\allowbreak{}lean} \\
EIP-7002:
liveness
after any
history &
\texttt{\pfvellipsis{}history\_\allowbreak{}submission\_\allowbreak{}nat\_\allowbreak{}live},
\texttt{\pfvellipsis{}history\_\allowbreak{}fee\_\allowbreak{}getter\_\allowbreak{}word\_\allowbreak{}live}
&
\texttt{Blanc/\allowbreak{}Lift/\allowbreak{}WithdrawalRe\penalty200{}ques\penalty200{}t/\allowbreak{}NatLiveness.\allowbreak{}lean} \\
EIP-7002:
where the
fees agree
&
\texttt{\pfvellipsis{}word\_\allowbreak{}fee\_\allowbreak{}le\_\allowbreak{}nat},
\texttt{\pfvellipsis{}word\_\allowbreak{}fee\_\allowbreak{}eq\_\allowbreak{}iff\_\allowbreak{}natFeeDomain\penalty200{}}
&
\texttt{Blanc/\allowbreak{}Lift/\allowbreak{}WithdrawalRe\penalty200{}ques\penalty200{}t/\allowbreak{}ExactFeeDoma\penalty200{}in.\allowbreak{}lean} \\
EIP-7002:
bookkeeping
wrap
control &
\texttt{\pfvellipsis{}wordSystem\_\allowbreak{}excess\_\allowbreak{}wraps}
&
\texttt{Blanc/\allowbreak{}Lift/\allowbreak{}WithdrawalRe\penalty200{}ques\penalty200{}t/\allowbreak{}WordReplay.\allowbreak{}lean} \\
EIP-7002:
excess and
count
bounded by
submissions
&
\texttt{\pfvellipsis{}WordHistory\penalty200{}.\allowbreak{}margin}
&
\texttt{Blanc/\allowbreak{}Lift/\allowbreak{}WithdrawalRe\penalty200{}ques\penalty200{}t/\allowbreak{}WordHistory.\allowbreak{}lean} \\
EIP-7002:
deployment
&
\texttt{\pfvellipsis{}Creation.\allowbreak{}deploy\_\allowbreak{}initial}
&
\texttt{Blanc/\allowbreak{}Lift/\allowbreak{}WithdrawalRe\penalty200{}ques\penalty200{}t/\allowbreak{}Creation/\allowbreak{}Deploy.\allowbreak{}lean} \\
EIP-7002:
the fee
guarantee
refuted &
\texttt{\pfvellipsis{}FeeCountere\penalty200{}xamp\penalty200{}le.\allowbreak{}nat\_\allowbreak{}fee\_\allowbreak{}guarantee\_\allowbreak{}refuted}
&
\texttt{Blanc/\allowbreak{}Composition/\allowbreak{}WithdrawalRe\penalty200{}ques\penalty200{}tFee\penalty200{}Refu\penalty200{}tati\penalty200{}on.\allowbreak{}lean} \\
EIP-7002:
the
system-address
control &
\texttt{\pfvellipsis{}DrainContro\penalty200{}l.\allowbreak{}systemEmpty\_\allowbreak{}loadBearing\_\allowbreak{}witness}
&
\texttt{Blanc/\allowbreak{}Composition/\allowbreak{}WithdrawalRe\penalty200{}ques\penalty200{}tDra\penalty200{}inCo\penalty200{}ntro\penalty200{}l.\allowbreak{}lean} \\
CircuitBreaker,
finite
tier &
\texttt{Blanc.\allowbreak{}Lift.\allowbreak{}LidoCircuitB\penalty200{}reak\penalty200{}erDe\penalty200{}ploy\penalty200{}ed.\allowbreak{}registerPaus\penalty200{}er\_\allowbreak{}nonzero\_\allowbreak{}finite}
&
\texttt{\pfvellipsis{}/\allowbreak{}FiniteFrame.\allowbreak{}lean} \\
CircuitBreaker,
finite
deployment
&
\texttt{\pfvellipsis{}lido\_\allowbreak{}create\_\allowbreak{}finite\_\allowbreak{}init}
&
\texttt{\pfvellipsis{}/\allowbreak{}FiniteInit.\allowbreak{}lean} \\
CircuitBreaker,
finite
premises
satisfiable
together &
\texttt{\pfvellipsis{}exampleAppl\penalty200{}icab\penalty200{}le}
&
\texttt{\pfvellipsis{}/\allowbreak{}FiniteExampl\penalty200{}e.\allowbreak{}lean} \\
CircuitBreaker,
history
tier &
\texttt{\pfvellipsis{}lido\_\allowbreak{}history\_\allowbreak{}l1\_\allowbreak{}l3},
\texttt{\pfvellipsis{}lido\_\allowbreak{}history\_\allowbreak{}l2\_\allowbreak{}committed}
&
\texttt{\pfvellipsis{}/History.lean},
\texttt{\pfvellipsis{}/\allowbreak{}L2History.\allowbreak{}lean} \\
V+
exclusion
&
\texttt{Blanc.\allowbreak{}Lift.\allowbreak{}VyperNonreen\penalty200{}tran\penalty200{}tDep\penalty200{}loye\penalty200{}d.\allowbreak{}Fixed.\allowbreak{}vplus\_\allowbreak{}exclusion}
&
\texttt{\pfvellipsis{}/\allowbreak{}Fixed/\allowbreak{}Exclusion.\allowbreak{}lean} \\
V+
exclusion
over a
history &
\texttt{Blanc.\allowbreak{}ExecutionTra\penalty200{}ce.\allowbreak{}ConfiguredHi\penalty200{}stor\penalty200{}yTra\penalty200{}ce.\allowbreak{}vplus}
&
\texttt{\pfvellipsis{}/\allowbreak{}Fixed/\allowbreak{}ExclusionTra\penalty200{}ce.\allowbreak{}lean} \\
V+ witness
from
deployment
&
\texttt{\pfvellipsis{}Fixed.\allowbreak{}Exit.\allowbreak{}vplus\_\allowbreak{}reachable\_\allowbreak{}capstone}
&
\texttt{\pfvellipsis{}/\allowbreak{}Fixed/\allowbreak{}Exit/\allowbreak{}Capstone.\allowbreak{}lean} \\
V+
witnesses
from
synthetic
pre-states
&
\texttt{\pfvellipsis{}Fixed.\allowbreak{}Witness.\allowbreak{}vplus\_\allowbreak{}witness\_\allowbreak{}covered},
\texttt{\pfvellipsis{}Fixed.\allowbreak{}Witness2.\allowbreak{}vplus\_\allowbreak{}witness2\_\allowbreak{}covered}
&
\texttt{\pfvellipsis{}/\allowbreak{}Fixed/\allowbreak{}Witness/\allowbreak{}Top.\allowbreak{}lean},
\texttt{\pfvellipsis{}/\allowbreak{}Fixed/\allowbreak{}Witness2/\allowbreak{}Top.\allowbreak{}lean} \\
V−
violation
from
deployment
&
\texttt{\pfvellipsis{}Vulnerable.\allowbreak{}Reach.\allowbreak{}Viol.\allowbreak{}vminus\_\allowbreak{}reachable\_\allowbreak{}capstone},
\texttt{\pfvellipsis{}Vulnerable.\allowbreak{}Reach.\allowbreak{}Viol.\allowbreak{}vminus\_\allowbreak{}reach\_\allowbreak{}violation}
&
\texttt{\pfvellipsis{}/\allowbreak{}Vulnerable/\allowbreak{}Reach/\allowbreak{}ViolFinal.\allowbreak{}lean} \\
V− message
witness &
\texttt{\pfvellipsis{}Vulnerable.\allowbreak{}Top.\allowbreak{}vminus\_\allowbreak{}witness\_\allowbreak{}covered}
&
\texttt{\pfvellipsis{}/\allowbreak{}Vulnerable/\allowbreak{}ForkTop.\allowbreak{}lean} \\
V−
admitted
transaction
and its
signature
&
\texttt{\pfvellipsis{}Vulnerable.\allowbreak{}TxC.\allowbreak{}vminus\_\allowbreak{}txC\_\allowbreak{}process},
\texttt{\pfvellipsis{}Vulnerable.\allowbreak{}TxC.\allowbreak{}txC\_\allowbreak{}recoveredSen\penalty200{}der}
&
\texttt{\pfvellipsis{}/\allowbreak{}Vulnerable/\allowbreak{}TxC/\allowbreak{}Envelope.\allowbreak{}lean},
\texttt{\pfvellipsis{}/\allowbreak{}Vulnerable/\allowbreak{}TxCRecover.\allowbreak{}lean} \\
Union
axiom
audit &
\texttt{\#union\_\allowbreak{}axioms\_\allowbreak{}of\_\allowbreak{}modules Blanc}
&
\texttt{scripts/\allowbreak{}AxiomCheck.\allowbreak{}lean} \\
Statement
pins & 758
\texttt{example}
pins &
\texttt{scripts/\allowbreak{}ClaimCheck.\allowbreak{}lean} \\
\end{longtable}
}

\subsection{Appendix
B.
Artifact
identity}\label{appendix-b.-artifact-identity}

The
runtime
bytes are
the lifted
inputs
under
\texttt{scripts/\allowbreak{}lift/\allowbreak{}inputs/\allowbreak{}};
for the
EIP-7002
contract
the lifted
input
equals
Blanc's
definition
\texttt{withdrawalRe\penalty200{}ques\penalty200{}tCod\penalty200{}e},
which the
system-contract
premises
also use.
The
certificates'
provenance
records
each
artifact's
origin
together
with the
evidence
that the
bytes are
the
chain's.
At block
26,141,229
(7 October
2026,
14:53
UTC), we
fetched
the code
at every
address
below from
three
separately
operated
public
providers,
dRPC, MEV
Blocker
and Blast.
They
agreed in
every
case, the
sizes and
Keccak-256
code
hashes are
those
listed,
and each
implementation's
code is
byte-identical
to its
lifted
input.

{\def\LTcaptype{none} % do not increment counter
\par\noindent\begin{minipage}{\linewidth}\footnotesize
\begin{tabular}{@{}
  >{\raggedright\arraybackslash}p{(\linewidth - 6\tabcolsep) * \real{0.2645}}
  >{\raggedright\arraybackslash}p{(\linewidth - 6\tabcolsep) * \real{0.5110}}
  >{\raggedleft\arraybackslash}p{(\linewidth - 6\tabcolsep) * \real{0.0888}}
  >{\raggedleft\arraybackslash}p{(\linewidth - 6\tabcolsep) * \real{0.1357}}@{}}
\toprule\noalign{}
\begin{minipage}[b]{\linewidth}\raggedright
Artifact
\end{minipage}
&
\begin{minipage}[b]{\linewidth}\raggedright
Address
\end{minipage}
&
\begin{minipage}[b]{\linewidth}\raggedleft
Bytes
\end{minipage}
&
\begin{minipage}[b]{\linewidth}\raggedleft
Creation
block
\end{minipage} \\
\midrule\noalign{}

WETH9 &
\texttt{0xC02aaA39b2\penalty200{}23FE\penalty200{}8D0A\penalty200{}0e5C\penalty200{}4F27\penalty200{}eAD9\penalty200{}083C\penalty200{}756C\penalty200{}c2}
& 3,124 &
4,719,568 \\
Beacon
deposit
contract &
\texttt{0x0000000021\penalty200{}9ab5\penalty200{}4035\penalty200{}6cBB\penalty200{}839C\penalty200{}be05\penalty200{}303d\penalty200{}7705\penalty200{}Fa}
& 6,358 &
11,052,984 \\
Curve 3Crv
&
\texttt{0x6c3F90f043\penalty200{}a72F\penalty200{}A612\penalty200{}cbac\penalty200{}8115\penalty200{}EE7e\penalty200{}52BD\penalty200{}e6E4\penalty200{}90}
& 2,276 &
10,809,467 \\
Uniswap V2
USDC/WETH
pair &
\texttt{0xB4e16d0168\penalty200{}e52d\penalty200{}35Ca\penalty200{}CD2c\penalty200{}6185\penalty200{}b442\penalty200{}81Ec\penalty200{}28C9\penalty200{}Dc}
& 11,293 &
10,008,355 \\
EIP-7002
withdrawal-request
contract &
\texttt{0x00000961Ef\penalty200{}480E\penalty200{}b55e\penalty200{}80D1\penalty200{}9ad8\penalty200{}3579\penalty200{}A64c\penalty200{}0070\penalty200{}02}
& 504 &
21,881,389 \\
Lido
CircuitBreaker
&
\texttt{0x6019CB5579\penalty200{}7829\penalty200{}6BA3\penalty200{}C08a\penalty200{}7B73\penalty200{}225C\penalty200{}0975\penalty200{}DFB2\penalty200{}F7}
& 4,584 &
24,993,190 \\
V+
implementation
&
\texttt{0x847ee1227A\penalty200{}9900\penalty200{}B73a\penalty200{}Eeb3\penalty200{}a47f\penalty200{}Ac92\penalty200{}c52F\penalty200{}D54e\penalty200{}d9}
& 18,320 &
17,110,099 \\
stETH-ng
pool,
forwarder
to V+ &
\texttt{0x21E27a5E55\penalty200{}13D6\penalty200{}e65C\penalty200{}4f83\penalty200{}0167\penalty200{}3909\penalty200{}97aA\penalty200{}8484\penalty200{}3a}
& 45 & \\
V−
implementation
&
\texttt{0x6326DEbBAa\penalty200{}15bC\penalty200{}FE60\penalty200{}3d83\penalty200{}1e7D\penalty200{}75f4\penalty200{}fc10\penalty200{}d9B4\penalty200{}3E}
& 17,535 &
12,904,329 \\
pETH/ETH
pool,
forwarder
to V− &
\texttt{0x9848482da3\penalty200{}Ee30\penalty200{}7616\penalty200{}5ce6\penalty200{}497e\penalty200{}DA90\penalty200{}6E66\penalty200{}bB85\penalty200{}C5}
& 45 & \\
\bottomrule\noalign{}
\end{tabular}
\end{minipage}\par
}

Keccak-256
code
hashes, in
the same
order:

\par\noindent\begin{minipage}{\linewidth}
\begin{Verbatim}[fontsize=\footnotesize]
WETH9                    0xd0a06b12ac47863b5c7be4185c2deaad1c61557033f56c7d4ea74429cbb25e23
Beacon deposit contract  0x6c029a231254fadb724d63be769f75eedd66362df034a3e663252b49d062a666
Curve 3Crv               0xb731b7a9a74c6c60d715f0c791d663f1530743a48337e7fab9cf3481afdd9feb
Uniswap V2 pair          0x5b83bdbcc56b2e630f2807bbadd2b0c21619108066b92a58de081261089e9ce5
EIP-7002 contract        0x0345a365d2f4c5975b9f1599abe0a2ee76b7a3a731bc68781bd04c84e4858f50
Lido CircuitBreaker      0x63d4da6a25804fabdc61ace21c19d1f2e1425ab9ade8d8a348a2591e801836ec
V+ implementation        0x59469db6b75b045025abb5386ae6b7ef7e1698d13ee93f3d4ed32d5f72506e0e
stETH-ng forwarder       0x9e28a09452d2354fc4e15e3244dde27cbc4d52f12a10b91f2ca755b672bfa9be
V− implementation        0xba0284a6a8a86734c1c777e3c6b5b56f2c8ba95dc0f7eba481c6d09da7885ebc
pETH/ETH forwarder       0xbed04db3507e08e5220f6eadf98d5df05bdbc74df129c73fcbab7c441e86b124
\end{Verbatim}
\end{minipage}\par

Creation
transactions.
For the
seven
contracts
created
directly,
each
receipt
records
success
and names
the
address
above as
the
created
contract;
the
deployer
nonces are
those the
deployment
theorems
use where
one exists
(446 for
WETH9, 0
for the
deposit
contract,
42 for
3Crv, 0
for the
CircuitBreaker
and 0 for
the
EIP-7002
contract's
keyless
sender
\texttt{0x8646\pfvellipsis{}9674}).
The
Uniswap
pair was
created by
a call to
the
factory
\texttt{0x5C69bEe701\penalty200{}ef81\penalty200{}4a2B\penalty200{}6a3E\penalty200{}DD4B\penalty200{}1652\penalty200{}CB9c\penalty200{}c5aA\penalty200{}6f},
whose
\texttt{PairCreated}
event in
that
transaction
names the
pair as
the first
the
factory
created.

\par\noindent\begin{minipage}{\linewidth}
\begin{Verbatim}[fontsize=\footnotesize]
WETH9                    0xb95343413e459a0f97461812111254163ae53467855c0d73e0f1e7c5b8442fa3
Beacon deposit contract  0xe75fb554e433e03763a1560646ee22dcb74e5274b34c5ad644e7c0f619a7e1d0
Curve 3Crv               0xa7d90e460bed56181e41b74d143ec98fe097c8391700e4e0788f5282053694f8
Uniswap V2 pair          0xd07cbde817318492092cc7a27b3064a69bd893c01cb593d6029683ffd290ab3a
EIP-7002 contract        0x8ded54be89448d78d4bc97782c0187b099e45380ab681742f9d3754e405c2572
Lido CircuitBreaker      0x9a1328c1f63fcdfd53d611cfe9c9a5f4c284c4c3e61af9066d90acaac2d5279f
V+ implementation        0xfaad66f8fa2b4883b9e5c91b69e31ac8c2cdbc6d383b8776d0eee94d3ee0a251
V− implementation        0xd27491757b3a4bc9287ed44ce5c43de9f32131264548ce139f0a702c3d5f389e
\end{Verbatim}
\end{minipage}\par

The
Uniswap
pair's
creation
code,
11,636
bytes,
hashes to
the pair
init-code
hash that
Uniswap
publishes,
\texttt{0x96e8ac4277\penalty200{}198f\penalty200{}f8b6\penalty200{}f785\penalty200{}478a\penalty200{}a9a3\penalty200{}9f40\penalty200{}3cb7\penalty200{}68dd\penalty200{}02cb\penalty200{}ee32\penalty200{}6c3e\penalty200{}7da3\penalty200{}4884\penalty200{}5f},
and its
bytes 261
to 11,553,
counting
from zero,
are the
lifted
runtime.
The
factory's
own code
at block
26,141,229,
13,859
bytes with
Keccak-256
hash
\texttt{0xbab145d02e\penalty200{}7005\penalty200{}f0d8\penalty200{}4c6c\penalty200{}1639\penalty200{}d39b\penalty200{}799b\penalty200{}0ea1\penalty200{}6df9\penalty200{}9ebb\penalty200{}daf5\penalty200{}a14d\penalty200{}9da8\penalty200{}20b4\penalty200{}e0},
is
identical
from the
three
providers
and equal
to the
published
build of
the
\texttt{v1.0.0}
sources;
it embeds
that
creation
code at
offset
2,171 and
has a
single
\texttt{CREATE2},
at program
counter
1,506. The
creation
code's
constructor
returns
its bytes
261 to
11,553 as
a literal,
whatever
its
inputs,
and no
executable
code of
the
factory,
the
creation
code or
the
runtime
contains
\texttt{SELFDESTRUCT},
\texttt{CALLCODE}
or
\texttt{DELEGATECALL};
push data
and the
runtime's
trailing
metadata
are not
read as
instructions.
The
identity
check
compares
the
factory's
code from
dRPC and
MEV
Blocker
with that
build, and
a one-byte
change to
it in both
responses
makes the
check fail
at exactly
that step;
Blast's
response,
read
separately,
is the
same. This
binds
every pair
the
factory
creates to
the lifted
runtime by
inspection
of the
factory's
source and
bytes at
one block,
not by a
theorem
about its
executions.
As spot
checks,
eleven
pairs from
across the
factory's
list of
524,109,
the first,
second,
third and
last among
them, hold
code with
the
runtime's
hash. The
EIP-7002
contract's
549-byte
creation
input
stores the
inhibitor
and
returns
the
runtime,
and equals
the input
listed in
the EIP
{[}EIP-7002{]}.

The
canonical
system
contracts,
at the
same
block,
also
agreed
across the
three
providers:

\par\noindent\begin{minipage}{\linewidth}
\begin{Verbatim}[fontsize=\scriptsize,breaklines,breakanywhere]
EIP-4788 beacon roots      0x000F3df6D732807Ef1319fB7B8bB8522d0Beac02    97 bytes  0xf57acd40259872606d76197ef052f3d35588dadf919ee1f0e3cb9b62d3f4b02c
EIP-2935 history storage   0x0000F90827F1C53a10cb7A02335B175320002935    83 bytes  0x6e49e66782037c0555897870e29fa5e552daf4719552131a0abce779daec0a5d
EIP-7251 consolidations    0x0000BBdDc7CE488642fb579F8B00f3a590007251   414 bytes  0x78c6cb5202685228bbcbfb992b1c4e116c7ec5ef11e25b8e92716cfc628ddd60
\end{Verbatim}
\end{minipage}\par

The two
other
pools
drained in
2023 that
forward to
V−,
alETH/ETH
\texttt{0xC4C319E2D4\penalty200{}d66C\penalty200{}cA44\penalty200{}64C0\penalty200{}c2B3\penalty200{}2c9B\penalty200{}d23e\penalty200{}be78\penalty200{}4e}
and
msETH/ETH
\texttt{0xc897b98272\penalty200{}AA23\penalty200{}7144\penalty200{}64Ea\penalty200{}2A0B\penalty200{}d518\penalty200{}0f1B\penalty200{}8C00\penalty200{}25},
hold the
same
45-byte
forwarder
code.
Compiler
versions
are those
of the
Sourcify
records of
the seven
compiled
implementations
{[}Sourcify{]};
the
forwarders
have no
Sourcify
record,
and the
EIP-7002
runtime is
the code
that the
EIP and
the
\texttt{sys-asm}
repository
publish
{[}EIP-7002;
sys-asm{]}.

\subsection{Appendix
C. The
WETH9
formula
against
mainnet
receipts}\label{appendix-c.-the-weth9-formula-against-mainnet-receipts}

For each
of the
twelve
cases of
Table 1,
the
earliest
compared
transaction
in blocks
26,111,230
to
26,141,229,
the window
of Section
2.4.
``Empties''
means that
the
withdrawn
amount
equaled
the
sender's
WETH
balance at
the end of
the
previous
block and
that the
balance at
the end of
the block
was zero;
since no
other
WETH9
event of
the block
names the
sender,
the
balance at
the end of
the
previous
block is
the
balance at
the
transaction's
entry
(Section
2.4). The
predicted
gas is
21,000 +
4·tokens +
13,940 −
(4,800 if
empties);
every
receipt
equals it.
All 3,748
compared
receipts
are listed
with their
predictions
in the
archived
verification
package
{[}Artifact{]},
as are the
98 type-0
and 1,590
longer-calldata
comparisons
of the
excluded
groups and
the 32
withdrawals
set aside,
with each
exclusion
and its
reason.

{\def\LTcaptype{none} % do not increment counter
\par\noindent\begin{minipage}{\linewidth}\footnotesize
\begin{tabular}{@{}
  >{\raggedright\arraybackslash}p{(\linewidth - 8\tabcolsep) * \real{0.6557}}
  >{\raggedleft\arraybackslash}p{(\linewidth - 8\tabcolsep) * \real{0.1124}}
  >{\centering\arraybackslash}p{(\linewidth - 8\tabcolsep) * \real{0.0846}}
  >{\raggedleft\arraybackslash}p{(\linewidth - 8\tabcolsep) * \real{0.0736}}
  >{\raggedleft\arraybackslash}p{(\linewidth - 8\tabcolsep) * \real{0.0736}}@{}}
\toprule\noalign{}
\begin{minipage}[b]{\linewidth}\raggedright
Transaction
hash
\end{minipage}
&
\begin{minipage}[b]{\linewidth}\raggedleft
Block
\end{minipage}
&
\begin{minipage}[b]{\linewidth}\centering
Empties
\end{minipage}
&
\begin{minipage}[b]{\linewidth}\raggedleft
Tokens
\end{minipage}
&
\begin{minipage}[b]{\linewidth}\raggedleft
Gas
\end{minipage} \\
\midrule\noalign{}

\texttt{0x6cde6f17d3\penalty200{}6649\penalty200{}9ac4\penalty200{}9779\penalty200{}5c06\penalty200{}1e76\penalty200{}2b73\penalty200{}1037\penalty200{}75f6\penalty200{}c956\penalty200{}1a6b\penalty200{}650d\penalty200{}731b\penalty200{}3cc6\penalty200{}bd}
&
26,115,308
& no & 60
&
35,180 \\
\texttt{0x28e261928b\penalty200{}2f69\penalty200{}0d2c\penalty200{}68ac\penalty200{}7bf8\penalty200{}4f59\penalty200{}ed66\penalty200{}f4b3\penalty200{}fca8\penalty200{}97de\penalty200{}12ba\penalty200{}16c8\penalty200{}1a15\penalty200{}33f8\penalty200{}71}
&
26,111,682
& no & 63
&
35,192 \\
\texttt{0x375c117f1c\penalty200{}730e\penalty200{}1385\penalty200{}7f42\penalty200{}2c97\penalty200{}5233\penalty200{}5d8d\penalty200{}cee4\penalty200{}13e8\penalty200{}ac11\penalty200{}3f10\penalty200{}f679\penalty200{}ef24\penalty200{}2ce1\penalty200{}0f}
&
26,111,477
& no & 66
&
35,204 \\
\texttt{0x945d74737d\penalty200{}b53f\penalty200{}e25b\penalty200{}7ef9\penalty200{}bc2d\penalty200{}ccc9\penalty200{}47fa\penalty200{}b6b3\penalty200{}d4d3\penalty200{}bc30\penalty200{}51be\penalty200{}1c9f\penalty200{}eca2\penalty200{}d05d\penalty200{}7f}
&
26,111,254
& no & 69
&
35,216 \\
\texttt{0x4e665a7260\penalty200{}cd69\penalty200{}21a0\penalty200{}a16c\penalty200{}38d7\penalty200{}6174\penalty200{}adb6\penalty200{}e6f7\penalty200{}544e\penalty200{}066d\penalty200{}eaf5\penalty200{}37c1\penalty200{}ba94\penalty200{}10fe\penalty200{}be}
&
26,111,720
& no & 72
&
35,228 \\
\texttt{0x4ce9bccb45\penalty200{}56a7\penalty200{}af37\penalty200{}0609\penalty200{}6388\penalty200{}0bad\penalty200{}da24\penalty200{}4aaf\penalty200{}60b5\penalty200{}7a36\penalty200{}a9e5\penalty200{}30c4\penalty200{}2f5e\penalty200{}e102\penalty200{}8a}
&
26,132,298
& no & 75
&
35,240 \\
\texttt{0x9d9d9ce97c\penalty200{}4f58\penalty200{}1e89\penalty200{}55ef\penalty200{}af19\penalty200{}8fbf\penalty200{}17b0\penalty200{}1d16\penalty200{}3f84\penalty200{}8e01\penalty200{}4030\penalty200{}a30b\penalty200{}fe8b\penalty200{}1708\penalty200{}e5}
&
26,111,988
& yes & 60
&
30,380 \\
\texttt{0x18536b2235\penalty200{}1124\penalty200{}87b0\penalty200{}7cf2\penalty200{}3292\penalty200{}b8e1\penalty200{}df5d\penalty200{}87aa\penalty200{}52eb\penalty200{}c7b0\penalty200{}f7b0\penalty200{}863c\penalty200{}52bb\penalty200{}836d\penalty200{}3f}
&
26,111,341
& yes & 63
&
30,392 \\
\texttt{0xb0dcb0abf1\penalty200{}2f1d\penalty200{}3241\penalty200{}1a54\penalty200{}2a06\penalty200{}f439\penalty200{}e0a8\penalty200{}e631\penalty200{}78c3\penalty200{}735b\penalty200{}1efd\penalty200{}2ee6\penalty200{}f17f\penalty200{}7411\penalty200{}da}
&
26,111,340
& yes & 66
&
30,404 \\
\texttt{0xa4a9063378\penalty200{}4e86\penalty200{}37f3\penalty200{}3400\penalty200{}9d92\penalty200{}dc78\penalty200{}5544\penalty200{}0e0d\penalty200{}c255\penalty200{}1c05\penalty200{}8dcd\penalty200{}20fc\penalty200{}ae59\penalty200{}d53c\penalty200{}53}
&
26,111,239
& yes & 69
&
30,416 \\
\texttt{0xfaa0a47a44\penalty200{}885a\penalty200{}768a\penalty200{}49bb\penalty200{}7a1c\penalty200{}768e\penalty200{}900f\penalty200{}f129\penalty200{}f8b7\penalty200{}9aa8\penalty200{}bc98\penalty200{}9ca2\penalty200{}1a8e\penalty200{}152c\penalty200{}74}
&
26,111,245
& yes & 72
&
30,428 \\
\texttt{0x594bbb9d48\penalty200{}5ba7\penalty200{}1755\penalty200{}d8d1\penalty200{}a87a\penalty200{}ce26\penalty200{}2e24\penalty200{}de56\penalty200{}5446\penalty200{}f7a8\penalty200{}cb0e\penalty200{}3e52\penalty200{}c54d\penalty200{}e611\penalty200{}98}
&
26,112,357
& yes & 75
&
30,440 \\
\bottomrule\noalign{}
\end{tabular}
\end{minipage}\par
}

\subsection{\texorpdfstring{Appendix
D. The
Uniswap
pair's
\texttt{approve}
against
mainnet
receipts}{Appendix D. The Uniswap pair's  against mainnet receipts}}\label{appendix-d.-the-uniswap-pairs-approve-against-mainnet-receipts}

\emph{Prediction.}
For a
successful
top-level
call
\texttt{approve(\allowbreak{}spender,\allowbreak{} amount)}
with 68
bytes of
input, the
receipt
should
record

\begin{quote}
gasUsed =
21,000 +
4·\emph{z}
+
16·\emph{n}
+ 2,342 +
\emph{S} −
\emph{R},
\end{quote}

where
\emph{z}
and
\emph{n}
count the
zero and
nonzero
bytes of
the input;
\emph{S},
the
store's
charge, is
2,100 for
the cold
slot plus
100 if the
amount
equals the
slot's
value
before the
transaction,
20,000 if
that value
is zero
and the
amount is
not, and
2,900
otherwise;
and
\emph{R}
is 4,800
if that
value is
nonzero
and the
amount is
zero, and
0
otherwise.
The value
before the
transaction
is the
allowance
at the end
of the
previous
block,
read from
the pair's
storage at
\texttt{keccak(\allowbreak{}spender ‖ keccak(\allowbreak{}owner ‖ 2))};
for every
candidate,
\texttt{allowance(\allowbreak{}owner,\allowbreak{} spender)}
returned
the slot's
value at
the end of
the
previous
block and
at the end
of the
transaction's
own block.
Neither
the cap on
refunds
nor the
calldata
floor of
EIP-7623
binds for
any
compared
transaction.

\emph{Selection.}
A
successful
\texttt{approve}
executed
by the
pair emits
one
\texttt{Approval}
event from
it. We
read all
1,444,195
\texttt{Approval}
events of
blocks
26,111,230
to
26,141,229
(3 October
2026,
10:31 UTC,
to 7
October,
14:53
UTC),
classified
each
emitting
account by
its code
hash, and
kept the
716
transactions
with such
an event
from an
account
holding
the pair
runtime.
In 174 of
them the
event came
from
inside
another
call, such
as a
router's
\texttt{removeLiquid\penalty200{}ityW\penalty200{}ithP\penalty200{}ermi\penalty200{}t};
the other
542 called
\texttt{approve}
on the
pair
directly.
For each
of these
we re-read
the pair's
code at
the end of
the
previous
block and
compared
it byte
for byte
with the
lifted
runtime. A
transaction
is
compared
if it is
of type 0,
1 or 2
with an
empty
access
list,
carries no
value and
exactly 68
bytes of
input,
succeeded
with
exactly
the
expected
event,
comes from
an account
without
code or
whose code
is only an
EIP-7702
delegation
designator,
which
\texttt{approve}
never
executes,
and
nothing
earlier in
its block
could have
written
the same
allowance:
no earlier
transaction
from the
same owner
to the
same pair,
and no
earlier
\texttt{Approval}
or
\texttt{Transfer}
event of
that pair
naming the
owner. All
542
qualify,
the 57
from
accounts
with
delegation
code among
them. As a
check on
the event
search, we
fetched
all 89,836
transactions
of blocks
26,140,930
to
26,141,229:
five were
top-level
\texttt{approve}
calls to a
pair, all
successful
and all
found by
the
search.

{\def\LTcaptype{none} % do not increment counter
\par\noindent\begin{minipage}{\linewidth}\footnotesize
\begin{tabular}{@{}
  >{\raggedright\arraybackslash}p{(\linewidth - 8\tabcolsep) * \real{0.3426}}
  >{\raggedleft\arraybackslash}p{(\linewidth - 8\tabcolsep) * \real{0.0886}}
  >{\raggedleft\arraybackslash}p{(\linewidth - 8\tabcolsep) * \real{0.0767}}
  >{\raggedleft\arraybackslash}p{(\linewidth - 8\tabcolsep) * \real{0.1125}}
  >{\raggedleft\arraybackslash}p{(\linewidth - 8\tabcolsep) * \real{0.3795}}@{}}
\toprule\noalign{}
\begin{minipage}[b]{\linewidth}\raggedright
Value
before →
amount
\end{minipage}
&
\begin{minipage}[b]{\linewidth}\raggedleft
\emph{S}
\end{minipage}
&
\begin{minipage}[b]{\linewidth}\raggedleft
\emph{R}
\end{minipage}
&
\begin{minipage}[b]{\linewidth}\raggedleft
Compared
\end{minipage}
&
\begin{minipage}[b]{\linewidth}\raggedleft
Receipts
equal to
the
prediction
\end{minipage} \\
\midrule\noalign{}

zero →
nonzero &
22,100 & 0
& 453 &
453 \\
nonzero →
zero &
5,000 &
4,800 & 80
& 80 \\
unchanged
& 2,200 &
0 & 8 &
8 \\
nonzero →
a
different
nonzero &
5,000 & 0
& 1 & 1 \\
all & & &
542 &
542 \\
\bottomrule\noalign{}
\end{tabular}
\end{minipage}\par
}

The 542
transactions
went to
476
different
pairs, the
USDC/WETH
exhibit
among them
five
times,
from 451
different
senders;
537 are of
type 2 and
five of
type 0.
The 57
from
delegated
accounts
are of
type 2,
carry no
access or
authorization
list and
no value,
and their
senders'
code is in
each case
a 23-byte
delegation
designator.
One of
each case:

{\def\LTcaptype{none} % do not increment counter
\par\noindent\begin{minipage}{\linewidth}\footnotesize
\begin{tabular}{@{}
  >{\raggedright\arraybackslash}p{(\linewidth - 10\tabcolsep) * \real{0.2171}}
  >{\raggedleft\arraybackslash}p{(\linewidth - 10\tabcolsep) * \real{0.1355}}
  >{\raggedleft\arraybackslash}p{(\linewidth - 10\tabcolsep) * \real{0.0448}}
  >{\raggedright\arraybackslash}p{(\linewidth - 10\tabcolsep) * \real{0.2171}}
  >{\raggedleft\arraybackslash}p{(\linewidth - 10\tabcolsep) * \real{0.2849}}
  >{\raggedleft\arraybackslash}p{(\linewidth - 10\tabcolsep) * \real{0.1006}}@{}}
\toprule\noalign{}
\begin{minipage}[b]{\linewidth}\raggedright
Transaction
\end{minipage}
&
\begin{minipage}[b]{\linewidth}\raggedleft
Block
\end{minipage}
&
\begin{minipage}[b]{\linewidth}\raggedleft
\emph{z} /
\emph{n}
\end{minipage}
&
\begin{minipage}[b]{\linewidth}\raggedright
Value
before →
amount
\end{minipage}
&
\begin{minipage}[b]{\linewidth}\raggedleft
Prediction
\end{minipage}
&
\begin{minipage}[b]{\linewidth}\raggedleft
Receipt
\end{minipage} \\
\midrule\noalign{}

\texttt{0xf5e8e34d\pfvellipsis{}f\penalty200{}3007\penalty200{}6}
(the
exhibit) &
26,137,535
& 38 / 30
& 0 →
2,297,699,900,000
& 21,000 +
632 +
2,342 +
22,100 =
46,074 &
46,074 \\
\texttt{0x0ae3736a\pfvellipsis{}3\penalty200{}2436\penalty200{}f}
&
26,139,922
& 44 / 24
& about
10³⁰ → 0 &
21,000 +
560 +
2,342 +
5,000 −
4,800 =
24,102 &
24,102 \\
\texttt{0x0d3a87bd\pfvellipsis{}3\penalty200{}8489\penalty200{}b}
&
26,123,153
& 44 / 24
& 0 → 0 &
21,000 +
560 +
2,342 +
2,200 =
26,102 &
26,102 \\
\texttt{0xe4d423a0\pfvellipsis{}7\penalty200{}6dde\penalty200{}b}
&
26,120,653
& 12 / 56
& 7.5·10¹⁷
→ 2²⁵⁶ − 1
& 21,000 +
944 +
2,342 +
5,000 =
29,286 &
29,286 \\
\bottomrule\noalign{}
\end{tabular}
\end{minipage}\par
}

The list
of all 542
candidates,
with the
scripts
that
selected
and
checked
them, is
in the
archived
package
{[}Artifact{]}.
A
separately
written
script
recomputed
twenty of
the
predictions
from the
transaction,
its
receipt
and the
chain's
state, and
found no
difference;
they
include
eight of
the nine
transactions
of the two
rarer
cases, and
the ninth,
from a
delegated
account,
joined the
comparison
later. A
further
script
recomputed
all 542
predictions
from the
recorded
inputs and
reread the
57
delegated
transactions,
their
receipts
and their
senders'
code.

\subsection{Appendix
E. Checks
and
reproduction}\label{appendix-e.-checks-and-reproduction}

Lean
toolchain
\texttt{v4.34.0};
Blanc
\texttt{5048a51f}
on its
main
branch;
Jaune
\texttt{2737c8eb};
Mathlib
\texttt{v4.34.0},
at commit
\texttt{5ed29652}.
Blanc's
\texttt{lake-\allowbreak{}manifest.\allowbreak{}json}
is the
record of
these and
of its
other
pinned
dependencies.
In Blanc,
\texttt{scripts/\allowbreak{}check.\allowbreak{}sh -\allowbreak{}-\allowbreak{}no-\allowbreak{}build}
runs the
union
axiom walk
and the
leaf
census,
\texttt{scripts/\allowbreak{}check-\allowbreak{}deployed-\allowbreak{}claim-\allowbreak{}map.\allowbreak{}sh}
checks the
claim map
against
the
repository,
\texttt{scripts/\allowbreak{}check-\allowbreak{}claims.\allowbreak{}sh}
elaborates
the
statement
pins,
\texttt{scripts/\allowbreak{}check-\allowbreak{}lift-\allowbreak{}certificates\penalty200{}.\allowbreak{}sh}
regenerates
and
compares
every
certificate,
and
\texttt{scripts/\allowbreak{}check-\allowbreak{}gates.\allowbreak{}sh -\allowbreak{}-\allowbreak{}fresh}
executes
the whole
gate
catalogue
(without
\texttt{--fresh},
it reuses
earlier
verdicts
whose
inputs are
unchanged).
The replay
of Section
3.11 is
\texttt{LEAN\_\allowbreak{}NUM\_\allowbreak{}THREADS=1 lake env leanchecker -\allowbreak{}v Blanc Blanc.\allowbreak{}ProofRecipeT\penalty200{}acti\penalty200{}c Blanc.\allowbreak{}ProofRecipes\penalty200{}Gene\penalty200{}rate\penalty200{}d}
followed
by the
names of
the 21
Jaune
modules in
the
library's
import
closure
and of
\texttt{Jaune.\allowbreak{}SHA256Spec},
run in a
built
worktree;
the two
named
Blanc
modules
are the
ones the
library's
root does
not
import. In
Jaune,
\texttt{scripts/\allowbreak{}bootstrap\_\allowbreak{}mainnet.\allowbreak{}py}
installs
\texttt{tests@v20.0.2},
and
\texttt{scripts/\allowbreak{}check-\allowbreak{}mainnet.\allowbreak{}sh -\allowbreak{}-\allowbreak{}suite full}
runs the
current-mainnet
blockchain
tests.

\textbf{Conformance.}
The lane
takes its
selection
from the
fixtures'
own
network
labels,
and the
run fails
if any
selected
case fails
or if
nothing is
selected.
Of the
34,205
cases,
16,673 are
for
Prague,
17,423 for
Osaka and
109 across
transitions;
BPO1 and
BPO2 are
exercised
through
the 47
transition
cases that
cross into
them, and
the
release's
other
formats,
such as
state and
engine
tests, are
not run.
What the
configured
transition
adds to
the
per-block
one, a
check of
the
chain's
identity
and the
selection
of a fork
by the
block's
timestamp,
the runner
performs
with the
same
functions
only for
the 109
transition
cases; a
fixture
with a
single
fork takes
it from
its label.
\texttt{tests@v20.0.2}
was the
latest
release of
\texttt{execution-specs}
when we
ran the
lane;
\texttt{tests@v21.0.1}
has since
superseded
it. On the
legacy
\texttt{ethereum/tests}
corpus,
Jaune's
documentation
records
2,978
passing
files of
2,983 and
diagnoses
each
exception.

\textbf{The
axiom
walk.} The
walk from
the
library's
102,116
constants,
auxiliary
ones
included,
visited
128,938
constants.
It covers
meta code
too,
including
the
library's
five
\texttt{partial def}s.
Neither
\texttt{native\_decide}
nor
\texttt{bv\_decide}
occurs in
the lift,
and the
walk shows
that
\texttt{bv\_decide},
which
closes
three
goals
elsewhere
in the
library,
in
\texttt{Blanc/\allowbreak{}ProxyPairOss\penalty200{}ifia\penalty200{}bleF\penalty200{}orwa\penalty200{}rdin\penalty200{}g.\allowbreak{}lean},
adds no
axiom
there. The
SHA-256
equivalence
of Section
3.1 lies
in a Jaune
module
that the
Blanc
library
does not
import, so
the walk
does not
reach it,
and no
gate of
Jaune's
audits its
axioms:
Jaune's
documentation
reports
them as
Lean's
three from
\texttt{\#print axioms},
which can
under-report
(Section
3.11)
{[}Jaune{]}.

For this
paper we
built the
library
from
source at
\texttt{5048a51f},
with
Blanc's
artifact
cache
disabled,
and then
ran the
gate
catalogue
with
\texttt{--fresh}:
all 86
rows
executed
and
passed,
among them
the axiom
walk,
which
reaches
exactly
the three
standard
axioms,
the leaf
census,
the
claim-map
gate and
the
statement-pin
gate, with
758 pins.
The replay
passed
over all
1,506
modules,
and the
Jaune
suite
passed at
its pin,
on the
machine of
Section
5.2.

\textbf{The
claim
gates.}
The
claim-map
gate
resolves
395
citations
(358
distinct),
with 49
required
headlines
and 34
required
disclosures,
and
requires
each of 25
planted
errors to
be
rejected.
Each
statement
pin is an
\texttt{example}
whose type
restates a
declaration
exactly;
the 758
pins cover
the 49
headline
theorems
and the
results
and
definitions
their
claims are
stated
through.

The
verification
package
records
how the
pin count
grew to
758
{[}Artifact{]}.

The
read-only
scripts
that
produced
the
on-chain
figures of
Sections
2.4, 4 and
5.1 and of
Appendices
B to D are
archived
with their
outputs
{[}Artifact{]},
together
with an
offline
entry
point that
recomputes
the
paper's
tables and
commit-bound
figures
from the
shipped
inputs and
the two
repositories.

\subsection{Appendix
F. The
corpus in
detail}\label{appendix-f.-the-corpus-in-detail}

Section 4
states
each
artifact's
main
results.
This
appendix
adds more
of what
the
statements
say, the
premises
that
Section 4
leaves to
its
convention,
what the
results
leave out,
and the
earlier
work in
more
detail.
The
statements
themselves,
which
Appendix A
locates,
remain the
authority.

\subsubsection{F.1
WETH9}\label{f.1-weth9}

The
footprint
that the
history
results
preserve,
under the
history,
checkpoint
and hash
premises
of Section
2.1, has
three
parts:
every
nonzero
word lies
at a
metadata
slot or at
a tracked
key's
slot,
tracked
slots are
distinct
and off
the
metadata
slots, and
the
tracked
balances
sum to no
more than
the
contract's
ether. An
address
whose
balance
slot is
neither a
metadata
slot nor a
tracked
slot is
therefore
owed
nothing.
Every
writer is
live from
a fresh
frame at a
closed
cost
(\texttt{approve},
for
instance,
2,320 gas
plus its
store),
with
\texttt{withdraw}
to a
contract
live given
the callee
premise of
Section
3.8. The
results
say
nothing
about
balances
outside
the
tracked
keys,
\texttt{approve}
and
\texttt{transferFrom}
are not
proved
live after
a history,
and static
calls are
left out
of the
committed
list by
the form
of the
statement,
not shown
to be
harmless.

\subsubsection{F.2
The Beacon
deposit
contract}\label{f.2-the-beacon-deposit-contract}

The
reference
root is a
recursively
defined
depth-32
Merkle
root with
the count
mixed in,
and a
deposit
that the
model and
the
deployed
decoder
accept
executes
at exactly
656 gas
plus the
cost of
its body.
The view
and
liveness
theorems
assume an
entry
condition
at every
frame the
history
entered at
the
contract:
the
SHA-256
precompile's
address is
warm and
holds no
delegation.
They ask
the same
of the new
call,
which must
also be
below the
call-depth
limit so
that it
can call
the
precompile,
and
non-static
for a
deposit.
Every
precompile
is warm
from the
start of a
transaction,
so inside
a
transaction
the warmth
condition
always
holds.

The
environmental
premises
are
trace-local:
the four
system
contracts
called at
every
block,
those of
EIP-4788,
EIP-2935,
EIP-7002
and
EIP-7251
{[}EIP-4788;
EIP-2935;
EIP-7002;
EIP-7251{]},
hold their
canonical
code at
the
checkpoint,
the
SHA-256
precompile's
address
holds no
code, no
EIP-7702
authorization
in the
history is
signed by
any of
these five
addresses,
and no
CREATE
frame
targets
one.

The
results
say
nothing
about
revert
routes or
the BLS
validity
of deposit
data, and
there is
no
transaction-level
liveness
theorem.
Park,
Zhang and
Roșu
verified
the
contract's
original
Vyper
version in
K
{[}KEVM-Deposit{]}
and then a
solc 0.6.8
build of
its
Solidity
rewrite,
with
per-function
gas bounds
{[}RV-DepositSol{]}.

\subsubsection{F.3
Curve
3Crv}\label{f.3-curve-3crv}

No filter
on static
frames is
needed,
because no
writer can
succeed in
a static
frame.
Every call
the model
accepts
other than
\texttt{set\_name}
executes
at an
exact but
existentially
quantified
cost
(Section
3.8);
\texttt{set\_name}
asks the
minter's
\texttt{owner()}
and
carries
the callee
premise
\texttt{OwnerCallOk}.
No
exported
theorem
gives a
closed
cost
formula,
and there
is no
history-level
statement
about logs
or return
bytes.

\subsubsection{F.4
The
Uniswap V2
pair}\label{f.4-the-uniswap-v2-pair}

\textbf{Every
frame.}
Start from
any
storage
that
represents
a model
state, and
assume
trace-local
freshness
of the
rows the
frame
touches.
Then every
successful
frame, at
any of the
27 entry
points,
decodes an
entry from
its
calldata,
and the
model, run
on that
entry with
the
answers
the
frame's
own calls
received,
succeeds
with the
same
storage,
logs and
return
bytes. No
state-changing
entry
point
succeeds
in a
static
frame. The
model's
laws, such
as the
1,000
tokens
locked at
address 0
by the
first mint
and an
exact
integer
square
root,
thereby
become
laws of
the bytes.

\textbf{Every
history.}
The replay
consumes
each
step's
answers at
the
positions
where its
own run
observed
them.
Representation
includes
complete
support:
every
nonzero
word of
the pair's
storage
lies at a
fixed slot
or at the
row of a
tracked
key. The
only hash
premise is
trace-local.
The
checkpoint
may be any
represented
state; the
deployment
theorem
supplies
one, and
proves of
it the two
checkpoint
conditions
that the
ledger and
minimum-liquidity
statements
assume. In
the
replayed
state:

\begin{itemize}
\tightlist
\item
  \emph{The
  LP
  ledger
  balances.}
  If the
  model's
  LP
  balances
  sum to
  the
  total
  supply
  at the
  checkpoint,
  they
  still
  do,
  every
  protocol-fee
  mint
  included.
  This is
  a
  statement
  about
  the
  model's
  ledger,
  whose
  rows are
  the
  tracked
  keys:
  the
  model
  reads an
  untracked
  holder's
  balance
  as zero,
  and the
  raw slot
  of an
  untracked
  key
  whose
  hash
  coincides
  with a
  tracked
  row lies
  outside
  this
  finite
  guarantee.
\item
  \emph{The
  oracle
  obeys
  its
  law.}
  Each
  accumulator
  is its
  checkpoint
  value
  plus the
  sum,
  modulo
  2²⁵⁶, of
  every
  committed
  update's
  increment,
  ⌊r₁·2¹¹²/r₀⌋·Δt
  or
  symmetrically
  when Δt
  and both
  reserves
  are
  nonzero,
  with Δt
  the
  elapsed
  time
  modulo
  2³² by
  the
  steps'
  own
  block
  timestamps.
\item
  \emph{The
  minimum
  liquidity
  stays
  locked.}
  If the
  checkpoint's
  model
  state
  also
  satisfies
  the
  supply
  floor, a
  supply
  of zero
  or at
  least
  1,000 LP
  tokens
  at
  address
  0, which
  has
  granted
  no
  allowance,
  then for
  a pair
  not at
  address
  0, and
  provided
  no step
  or
  nested
  call to
  the pair
  has
  address
  0 as its
  caller,
  the
  floor
  holds
  after
  every
  committed
  step,
  and
  every
  step
  that
  starts
  with
  positive
  supply
  ends
  with at
  least
  1,000;
  the
  fee-off
  bound
  then
  holds as
  a ratio
  over the
  rationals.
  The
  ledger
  alone
  does not
  give the
  floor: a
  model
  state
  whose
  single
  LP token
  is held
  by a
  nonzero
  address
  balances
  but
  violates
  it.
  Deployment
  establishes
  both.
\end{itemize}

The share
value of
Section
4.4,
√(r₀·r₁)/T,
is not
market
value:
moving the
reserves
from (100,
100) to
(50, 201)
at a
supply of
100 raises
√(r₀·r₁)/T
from 1 to
about
1.0025,
yet at
prices of
4 and 1
the new
reserves
are worth
401,
against
500 for
the old
ones held
outside
the pool.

Each
share-value
theorem
has the
shape ∃
\emph{steps},
(\emph{steps}
are
exactly
the
committed
calls to
the pair,
authenticated
and
admitted,
whose
replay the
final
storage
represents)
∧
(NoShrink(\emph{steps})
→
bound(\emph{steps})),
so the
premise
concerns
the
answers
the tokens
actually
gave, at
the calls
that gave
them. A
token
whose
balances
only grow
can
satisfy
it, and
one that
debits the
pair
beyond its
payouts
may not.
Mint and
swap need
no such
premise,
because
the
bytecode's
own checks
give the
inequality.
In the
fee-on
bound,
under
NoShrink
alone,
dilution
is bounded
by the fee
mint
alone.

\textbf{Every
state-changing
function
is live.}
After any
configured
history, a
call to
any of the
ten
state-changing
entry
points has
a
successful
execution
from
program
counter 0
whenever
the model
accepts
that call
at the
replayed
state, at
a closed
cost of
fixed
charges,
actual
access
charges
and
forwarded
gas
(\texttt{approve}:
exactly
2,342 gas
plus its
store),
given
trace-local
hash
premises
and, for
\texttt{sync},
\texttt{mint},
\texttt{swap},
\texttt{burn},
\texttt{skim}
and
\texttt{permit},
the
replies
and
returned
gas of the
calls they
make.
Those
callee
premises
cannot be
dropped:
if the
first
token, or
the
callback
recipient,
holds the
reverting
code
\texttt{PUSH0 PUSH0 REVERT},
no run of
\texttt{sync},
\texttt{skim},
\texttt{mint}
or a
\texttt{swap}
that pays
out the
first
token or
calls back
succeeds,
whatever
the gas.
The frame
cost of
\texttt{approve}
predicts,
with the
protocol's
intrinsic
gas and
refund
added, all
542
mainnet
receipts
of Section
5.1.

\textbf{Deployment.}
Under
every
covered
fork, the
factory's
\texttt{CREATE2}
step with
the
recorded
creation
code,
given an
empty
target and
at least
2,400,000
gas
forwarded,
installs
the
certified
runtime,
with the
factory as
\texttt{factory}
and the
EIP-712
domain
separator
over the
chain and
the new
address;
every
successful
\texttt{initialize}
from the
factory
then
leaves
exactly
the
initialized
checkpoint,
and such a
run
exists, at
a closed
gas cost.
No hash
premise is
needed.
For the
exhibit,
the kernel
evaluates
the
\texttt{CREATE2}
address
from the
factory,
the salt
\texttt{keccak(\allowbreak{}USDC ‖ WETH)}
and the
creation
code, and
obtains
\texttt{0xB4e1\pfvellipsis{}C9Dc}.

\textbf{Positional
answers.}
An earlier
version of
these
theorems
matched
answers to
calls by
target and
input, not
position,
so a
\texttt{burn}
replay
could read
the final
balances
as the
first
ones: on
the source
definitions,
a fee-off
burn of
3,000 of
4,000
shares
with
balances
of 4,001
pays out
3,000,
while a
replay
that reads
the final
balances,
1,001,
computes
750 and
leaves the
same
storage,
an
arithmetic
fact about
the
definitions,
not a
closed EVM
counterexample.
The
theorems
above
consume
the actual
call
positions,
replies
and nested
queues of
the same
run, and
the weaker
statements
have been
deleted.

\textbf{Limits.}
No theorem
says that
a real
token
satisfies
NoShrink,
that a
factory
answers
\texttt{feeTo = 0},
that the
reserves
are
matched by
real token
holdings,
or that a
price or
the pool's
economics
are fair,
and none
composes
the pair
with a
token or
the
factory.
The
factory's
own
bytecode
is not
lifted, so
that it
issues
\texttt{initialize}
is a
premise.
There is
no
transaction-level
liveness
theorem,
\texttt{permit}
takes the
precompile's
answer as
an input,
and the
refinement
covers
successful
runs only.

\textbf{Prior
work.}
dapp.org's
audit
proved K
reachability
claims for
each
method of
the pair
and
factory
against
bytecode
built from
the
\texttt{v1.0.0}
sources,
whose
published
build is
byte-identical
to the
deployed
runtime,
with an
exact
symbolic
gas cost
for
success,
making no
claim
about
invariants
across
calls and
assuming
tokens
matching
the pair's
own
{[}dapp-UniV2{]};
EquiVM's
repository
has since
proved a
local
build of
the same
sources,
8,833
bytes that
are not
the
deployed
runtime
{[}EquiVM{]}.
To our
knowledge,
no earlier
machine-checked
result
states the
pair's
properties
across
calls of
its
bytecode.

\subsubsection{F.5
The
EIP-7002
contract}\label{f.5-the-eip-7002-contract}

An exit or
partial
withdrawal
is
triggered
in this
way by the
address in
the
validator's
withdrawal
credentials,
and each
submission
carries 56
bytes, a
48-byte
public key
and an
8-byte
amount;
the
request
the system
call
returns
for it
adds the
20-byte
source
address,
for 76
bytes. The
contract
queues
each
request
with its
caller at
raw
storage
offsets,
and the
excess
sets the
next fee
through
\texttt{fake\_\allowbreak{}exponential(\allowbreak{}1,\allowbreak{} excess,\allowbreak{} 17)}.
The
model's
initial
state is
what the
deployment
leaves:
the excess
set to an
\emph{inhibitor},
2²⁵⁶ − 1,
that
blocks
submissions
until the
first
system
call, and
an empty
queue.
Under
every
covered
fork,
executing
the
recorded
creation
input from
the
recorded
sender
installs
the
certified
runtime at
the
contract's
address
and leaves
storage
representing
that
state,
with no
hash
premise.

In the
request
theorem,
the
entries
not yet
emitted
are
exactly
the queue
that
storage
holds
afterwards,
and the
queue
never
reaches
the four
bookkeeping
slots; the
delivery
bound
counts the
block
whose
requests
are
computed
as zero;
and the
exact
replay
says that
the
storage at
the end of
any such
history
represents
the EIP's
model
replayed
over
exactly
the
committed
submissions
and system
calls, in
order. The
system
address is
\texttt{0xffff\pfvellipsis{}fffe};
the
premise on
creations
addressed
to it
covers
failed
ones too,
and there
is no
entry
condition
on the
frames
that run
the
contract.
As far as
we know,
placing
code at
the system
address
would take
a
signature
by the
address's
key or a
160-bit
Keccak
preimage,
but Blanc
does not
prove
this, and
the
premise
stays.

\textbf{Every
successful
frame is
classified.}
A
successful
frame at
this code
is exactly
the system
call's
dequeue, a
submission
(not
static,
from a
caller
other than
the system
address,
with 56
bytes of
input and
a value
covering
the fee
the
bytecode
computes)
or a fee
read, with
exact
storage,
logs and
return
bytes;
while the
excess
holds the
inhibitor
no user
call
succeeds,
and which
failure a
rejected
call meets
is not
stated.
The
balance
never
decreases,
and the
committed
submissions
add
exactly
their
values to
it. After
any
history
that
leaves the
contract
active, a
submission
paying at
least the
EIP's fee
and
started
with more
than 2,300
gas beyond
its cost
executes
at exactly
1,258 +
87·\emph{k}
gas plus
its
storage
accesses,
\emph{k}
being the
iterations
of the fee
loop, and
a fee read
at 180 +
87·\emph{k}
plus one
read.

The two
audits
noted that
the fee
departs
from the
EIP's once
the excess
passes
2,892, one
judging
such an
excess
unreachable
under
current
settings
and the
other,
noting
that a
higher
block gas
limit
might
bring it
within
reach,
rating the
finding
informational
{[}Blackthorn-7002;
SigmaPrime-7002{]}.
In the
closed
counterexample,
the
submission
pays 2²⁴⁵
wei, at
least the
fee the
bytecode
computes
at an
excess of
2,893,
about
3.2·10⁷³
wei, but
less than
the EIP's,
about
8.1·10⁷³
wei; the
history
reaches
that
excess
with 2,895
submissions
in one
block, on
a
synthetic
Prague
chain
whose
block gas
limit is
2²⁹. At an
excess of
2,892, the
last at
which the
fees
agree, the
fee is
about
7.6·10⁷³
wei. The
wrapped
arithmetic
does yield
a fee of 1
wei near
an excess
of 2²⁵²,
but every
model
state a
history
reaches
from
deployment
satisfies
\texttt{effectiveExc\penalty200{}ess state + state.\allowbreak{}count ≤ submissions.\allowbreak{}length},
so that
excess
would take
about 2²⁵²
submissions.

The
theorems
say
nothing
about how
the
consensus
layer
processes
the
requests,
including
whether
the caller
is
entitled
to act on
the
validator,
or about
whether
the fee
parameters
are
economically
sound;
dequeued
slots are
left in
storage,
nothing is
proved
about
storage
growth,
and there
is no
transaction-level
liveness
theorem.
The
Ethereum
Foundation's
Protocol
Security
Research
Team
published
the four
audits'
results,
and the
Halmos
check
compared
the
bytecode
with
references
written
from the
EIP's
pseudocode
{[}Pectra-Audits;
Halmos-Pectra{]}.

\subsubsection{F.6
Lido's
CircuitBreaker}\label{f.6-lidos-circuitbreaker}

The
history
tier's
registry
witness
says that
assignment
and index
agree with
membership,
that each
count
equals the
number of
assignments,
and that
the live
counts sum
to the
array
length;
its
removal
statement
covers
every
committed,
non-static
\texttt{registerPaus\penalty200{}er(\allowbreak{}t,\allowbreak{} 0)}
frame. The
finite
tier's
observation,
\texttt{RegistryOn storage entries probes},
says that
an
explicit
list of
entries is
valid and
that the
array and,
for each
address in
a
caller-supplied
list of
probes,
the
assignment,
index and
count
words
agree with
it. A
successful
execution
of
\texttt{registerPaus\penalty200{}er(\allowbreak{}target,\allowbreak{} p)},
for an
existing
target and
a nonzero
new
pauser,
takes
storage on
which the
decidable
check for
\texttt{RegistryOn}
holds to
storage on
which it
holds for
the
updated
entries,
and the
modeled
deployment
leaves
storage
satisfying
\texttt{RegistryOn [] probes}
for
canonical
probe
lists that
pass one
decidable
check. A
kernel-checked
example
shows that
the
premises
can all
hold at
once, on a
constructed
state with
one entry
and four
probes.
The finite
tier says
nothing
about
addresses
outside
the
probes,
inserting
or
removing
targets,
or
histories.
Neither
tier
claims
that any
target is
actually
paused,
and there
is no
liveness
result.

Cyfrin
checked 41
properties
of an
earlier
commit of
the source
with the
Certora
Prover
8.8.1,
verifying
38 and
reporting
three
cases of
\texttt{registerPauser}
as known
limitations
correct by
manual
analysis
{[}Cyfrin-CB{]},
and
compiled
the
deployed
commit and
compared
the output
with the
deployed
bytecode,
apart from
its
metadata
and
immutable
values
{[}Cyfrin-Audit{]}.

\subsubsection{F.7
The two
Curve pool
implementations}\label{f.7-the-two-curve-pool-implementations}

Each of
the eight
messages
of the V−
witness
settles
without
error. The
V− theorem
also
locates
both
guards in
the
deployed
bytes
(Figure
3), and a
second
form
states the
same
violation
for every
world that
agrees
with the
reached
one on the
14 storage
slots and
6 accounts
the attack
reads. The
admitted
transaction
is a fixed
signed
type-2
transaction
with
16,043,200
gas,
accepted
on every
covered
fork.

For an
execution
recorded
in a
configured
history,
the V+
theorem's
premises
are that
the pool's
and the
implementation's
code is
installed,
that a
first
frame
addressed
to the
pool runs
the code
stored
there, and
that the
execution's
hashes
avoid the
lock's
slot. Each
of the
three V+
witnesses
runs on
every
covered
fork. The
one from
deployment
runs nine
messages
from the
same
creator-only
world as
the V−
witness,
which
deploy,
initialize
and fund
the pool
before the
ninth
attempts
the
reentry.
Besides
it, two
older
witnesses,
from
synthetic
pre-states,
refuse a
read-only
reentry
into
\texttt{get\_\allowbreak{}virtual\_\allowbreak{}price(\allowbreak{})}
and a
reentry
through
the
stETH-ng
forwarder.
The
exclusion
says
nothing
about
unguarded
functions,
reentry
after the
lock is
released,
pricing,
or
transactions
that carry
EIP-7702
authorizations,
and the V−
result is
a concrete
violation,
not the
formal
negation
of the V+
theorem.

\subsection{References}\label{references}

\begin{itemize}
\tightlist
\item
  {[}Act{]}
  Argot
  Collective.
  act: a
  formal
  specification
  language
  for EVM
  contracts.
  github.\allowbreak{}com/\allowbreak{}argotorg/\allowbreak{}act,
  commit
  b1424bd3f6d6
  (6
  October
  2026).
\item
  {[}Albert20{]}
  E.
  Albert,
  S.
  Grossman,
  N.
  Rinetzky,
  C.
  Rodríguez-Núñez,
  A.
  Rubio,
  M.
  Sagiv.
  Taming
  callbacks
  for
  smart
  contract
  modularity.
  Proc.
  ACM
  Program.
  Lang.
  4(OOPSLA),
  Article
  209,
  2020.
  DOI
  10.1145/3428277.
\item
  {[}Amani18{]}
  S.
  Amani,
  M.
  Bégel,
  M.
  Bortin,
  M.
  Staples.
  Towards
  verifying
  Ethereum
  smart
  contract
  bytecode
  in
  Isabelle/HOL.
  CPP
  2018,
  pp.~66--77.
  DOI
  10.1145/3167084.
\item
  {[}Artifact{]}
  S. Baek.
  Verification
  package
  for
  ``Verifying
  deployed
  bytecode
  against
  executable
  EVM
  semantics,
  with
  exact
  gas and
  no added
  axioms''.
  Zenodo,
  2026.
  DOI
  10.\allowbreak{}5281/\allowbreak{}zenodo.\allowbreak{}23351349.\allowbreak{}
\item
  {[}Bartoletti22{]}
  M.
  Bartoletti,
  J. H.
  Chiang,
  A.
  Lluch-Lafuente.
  A theory
  of
  Automated
  Market
  Makers
  in DeFi.
  Logical
  Methods
  in
  Computer
  Science
  18(4):12,
  2022.
  DOI
  10.\allowbreak{}46298/\allowbreak{}lmcs-\allowbreak{}18(4:\allowbreak{}12)2022.\allowbreak{}
\item
  {[}Blackthorn-7002{]}
  Blackthorn
  (Sherlock
  collaborative
  audit).
  Security
  review
  for
  Ethereum
  Foundation:
  collaborative
  audit of
  the
  Pectra
  system
  contracts,
  28
  October
  to 2
  November
  2024;
  final
  report,
  file
  dated 4
  February
  2025.
  Issue
  L-8,
  ``Overflow
  in
  fake\_expo
  leads to
  spec
  mismatch''.
  github.\allowbreak{}com/\allowbreak{}ethereum/\allowbreak{}audits,
  Pectra/\allowbreak{}2025\_\allowbreak{}02\_\allowbreak{}04\_\allowbreak{}Final\_\allowbreak{}Ethereum\_\allowbreak{}Foundation\_\allowbreak{}Collaborative\_\allowbreak{}Audit\_\allowbreak{}Report.\allowbreak{}pdf.\allowbreak{}
\item
  {[}Blanc{]}
  Blanc.
  github.\allowbreak{}com/\allowbreak{}skbaek/\allowbreak{}blanc,
  commit
  5048a51f
  on the
  main
  branch;
  archived
  by
  Software
  Heritage
  as
  swh:\allowbreak{}1:\allowbreak{}rev:\allowbreak{}5048a51fb108d8d7\allowbreak{}1dd58c89\allowbreak{}7c4c62c8\allowbreak{}bff3b437\allowbreak{}.\allowbreak{}
\item
  {[}Cassez21{]}
  F.
  Cassez.
  Verification
  of the
  incremental
  Merkle
  tree
  algorithm
  with
  Dafny.
  FM 2021,
  LNCS
  13047,
  pp.~445--462.
  DOI
  10.\allowbreak{}1007/\allowbreak{}978-\allowbreak{}3-\allowbreak{}030-\allowbreak{}90870-\allowbreak{}6\_\allowbreak{}24.\allowbreak{}
\item
  {[}Certora{]}
  Certora
  Prover
  documentation:
  Prover
  approximations,
  for
  loops
  and
  hashing.
  docs.\allowbreak{}certora.\allowbreak{}com/\allowbreak{}en/\allowbreak{}latest/\allowbreak{}docs/\allowbreak{}prover/\allowbreak{}approx/\allowbreak{}.\allowbreak{}
\item
  {[}Creme{]}
  Creme:
  the
  agent-development
  launch
  root for
  Jaune
  and
  Blanc,
  with an
  accountable
  goal-execution
  method.
  github.\allowbreak{}com/\allowbreak{}skbaek/\allowbreak{}creme.\allowbreak{}
\item
  {[}Cyfrin-Audit{]}
  Cyfrin.
  Lido
  Circuit
  Breaker
  audit
  report,
  v2.1, 1
  May
  2026.
  github.\allowbreak{}com/\allowbreak{}Cyfrin/\allowbreak{}cyfrin-\allowbreak{}audit-\allowbreak{}reports,
  reports/\allowbreak{}2026-\allowbreak{}05-\allowbreak{}01-\allowbreak{}cyfrin-\allowbreak{}lido-\allowbreak{}circuit-\allowbreak{}breaker-\allowbreak{}v2.\allowbreak{}1.\allowbreak{}pdf.\allowbreak{}
\item
  {[}Cyfrin-CB{]}
  Cyfrin.
  Formal
  verification
  report:
  CircuitBreaker,
  Certora
  Prover
  8.8.1, 6
  April
  2026.
  github.\allowbreak{}com/\allowbreak{}Cyfrin/\allowbreak{}cyfrin-\allowbreak{}audit-\allowbreak{}reports,
  reports/\allowbreak{}2026-\allowbreak{}04-\allowbreak{}08-\allowbreak{}cyfrin-\allowbreak{}lido-\allowbreak{}circuit-\allowbreak{}breaker-\allowbreak{}fv-\allowbreak{}v2.\allowbreak{}0.\allowbreak{}pdf.\allowbreak{}
\item
  {[}dapp-UniV2{]}
  dapp.org.
  Uniswap
  V2 audit
  report,
  last
  updated
  6 May
  2020.
  dapp.\allowbreak{}org.\allowbreak{}uk/\allowbreak{}reports/\allowbreak{}uniswapv2.\allowbreak{}html;
  specifications
  and
  proofs
  at
  github.\allowbreak{}com/\allowbreak{}dapp-\allowbreak{}org/\allowbreak{}k-\allowbreak{}uniswap-\allowbreak{}v2.\allowbreak{}
\item
  {[}DeepSEA{]}
  V.
  Sjöberg,
  K. Dave,
  D.
  Britten,
  M. A.
  Schett,
  X. Sun,
  Q. Wang,
  S. N.
  Anderson,
  S.
  Reeves,
  Z. Shao.
  Foundational
  verification
  of smart
  contracts
  through
  verified
  compilation.
  arXiv:2405.08348,
  2024.
\item
  {[}EIP-2200{]}
  W. Tang.
  EIP-2200:
  Structured
  definitions
  for net
  gas
  metering.
  eips.\allowbreak{}ethereum.\allowbreak{}org/\allowbreak{}EIPS/\allowbreak{}eip-\allowbreak{}2200.\allowbreak{}
\item
  {[}EIP-2929{]}
  V.
  Buterin,
  M.
  Swende.
  EIP-2929:
  Gas cost
  increases
  for
  state
  access
  opcodes.
  eips.\allowbreak{}ethereum.\allowbreak{}org/\allowbreak{}EIPS/\allowbreak{}eip-\allowbreak{}2929.\allowbreak{}
\item
  {[}EIP-2935{]}
  V.
  Buterin,
  T.
  Stanczak,
  G.
  Ballet,
  G.
  Singh,
  T.
  Jasoria,
  I.
  Hagopian,
  J.
  Brouwer,
  S.
  Mahmoodi.
  EIP-2935:
  Serve
  historical
  block
  hashes
  from
  state.
  eips.\allowbreak{}ethereum.\allowbreak{}org/\allowbreak{}EIPS/\allowbreak{}eip-\allowbreak{}2935.\allowbreak{}
\item
  {[}EIP-3529{]}
  V.
  Buterin,
  M.
  Swende.
  EIP-3529:
  Reduction
  in
  refunds.
  eips.\allowbreak{}ethereum.\allowbreak{}org/\allowbreak{}EIPS/\allowbreak{}eip-\allowbreak{}3529.\allowbreak{}
\item
  {[}EIP-4788{]}
  A.
  Stokes,
  A.
  Dietrichs,
  D. Ryan,
  M. H.
  Swende,
  lightclient.
  EIP-4788:
  Beacon
  block
  root in
  the EVM.
  eips.\allowbreak{}ethereum.\allowbreak{}org/\allowbreak{}EIPS/\allowbreak{}eip-\allowbreak{}4788.\allowbreak{}
\item
  {[}EIP-6110{]}
  M.
  Kalinin,
  D. Ryan,
  P.
  Davies.
  EIP-6110:
  Supply
  validator
  deposits
  on
  chain.
  eips.\allowbreak{}ethereum.\allowbreak{}org/\allowbreak{}EIPS/\allowbreak{}eip-\allowbreak{}6110.\allowbreak{}
\item
  {[}EIP-7002{]}
  D. Ryan,
  M.
  Kalinin,
  A.
  Dietrichs,
  H.-W.
  Wang,
  lightclient,
  F.
  Lange.
  EIP-7002:
  Execution
  layer
  triggerable
  withdrawals.
  eips.\allowbreak{}ethereum.\allowbreak{}org/\allowbreak{}EIPS/\allowbreak{}eip-\allowbreak{}7002.\allowbreak{}
\item
  {[}EIP-7251{]}
  mike,
  Francesco,
  dapplion,
  Mikhail,
  Aditya,
  Justin,
  lightclient,
  F. Lange
  (the
  EIP's
  header
  gives no
  full
  name for
  the
  others).
  EIP-7251:
  Increase
  the
  MAX\_EFFECTIVE\_BALANCE.
  eips.\allowbreak{}ethereum.\allowbreak{}org/\allowbreak{}EIPS/\allowbreak{}eip-\allowbreak{}7251.\allowbreak{}
\item
  {[}EIP-7600{]}
  T.
  Beiko.
  EIP-7600:
  Hardfork
  Meta -
  Pectra.
  eips.\allowbreak{}ethereum.\allowbreak{}org/\allowbreak{}EIPS/\allowbreak{}eip-\allowbreak{}7600.\allowbreak{}
\item
  {[}EIP-7607{]}
  T.
  Beiko,
  A.
  Stokes,
  A.
  Dietrichs.
  EIP-7607:
  Hardfork
  Meta -
  Fusaka.
  eips.\allowbreak{}ethereum.\allowbreak{}org/\allowbreak{}EIPS/\allowbreak{}eip-\allowbreak{}7607.\allowbreak{}
\item
  {[}EIP-7623{]}
  T.
  Wahrstätter,
  V.
  Buterin.
  EIP-7623:
  Increase
  calldata
  cost.
  eips.\allowbreak{}ethereum.\allowbreak{}org/\allowbreak{}EIPS/\allowbreak{}eip-\allowbreak{}7623.\allowbreak{}
\item
  {[}EIP-7685{]}
  lightclient,
  F.
  Lange.
  EIP-7685:
  General
  purpose
  execution
  layer
  requests.
  eips.\allowbreak{}ethereum.\allowbreak{}org/\allowbreak{}EIPS/\allowbreak{}eip-\allowbreak{}7685.\allowbreak{}
\item
  {[}EIP-7702{]}
  V.
  Buterin,
  S.
  Wilson,
  A.
  Dietrichs,
  lightclient.
  EIP-7702:
  Set code
  for
  EOAs.
  eips.\allowbreak{}ethereum.\allowbreak{}org/\allowbreak{}EIPS/\allowbreak{}eip-\allowbreak{}7702.\allowbreak{}
\item
  {[}EIP-7825{]}
  G.
  Rebuffo,
  T.
  Wahrstätter.
  EIP-7825:
  Transaction
  gas
  limit
  cap.
  eips.\allowbreak{}ethereum.\allowbreak{}org/\allowbreak{}EIPS/\allowbreak{}eip-\allowbreak{}7825.\allowbreak{}
\item
  {[}Elipmoc{]}
  N.
  Grech,
  S.
  Lagouvardos,
  I.
  Tsatiris,
  Y.
  Smaragdakis.
  Elipmoc:
  advanced
  decompilation
  of
  Ethereum
  smart
  contracts.
  Proc.
  ACM
  Program.
  Lang.
  6(OOPSLA1),
  Article
  77,
  2022.
  DOI
  10.1145/3527321.
\item
  {[}Elle{]}
  M. M.
  Alvarez.
  Elle, a
  work-in-progress,
  foundationally
  verified
  compiler
  to the
  EVM in
  Isabelle/HOL.
  Documentation,
  2019,
  elle.readthedocs.io;
  github.\allowbreak{}com/\allowbreak{}mmalvarez/\allowbreak{}eth-\allowbreak{}isabelle.\allowbreak{}
\item
  {[}EquiVM{]}
  L.
  Lazaropoulos,
  Z.
  Paraskevopoulou.
  Foundational
  refinement
  proofs
  for
  deployed
  bytecode,
  at the
  price of
  tokens.
  arXiv:2607.26306v1,
  28 July
  2026.
  Repository
  github.\allowbreak{}com/\allowbreak{}argotorg/\allowbreak{}EquiVM,
  commit
  7230c24bff6c
  (7
  October
  2026);
  semantics
  github.\allowbreak{}com/\allowbreak{}lefterislazar/\allowbreak{}EVMLean,
  commit
  63f61339dd17.
\item
  {[}ERC-1167{]}
  P.
  Murray,
  N.
  Welch,
  J.
  Messerman.
  ERC-1167:
  Minimal
  proxy
  contract.
  ercs.\allowbreak{}ethereum.\allowbreak{}org/\allowbreak{}ERCS/\allowbreak{}erc-\allowbreak{}1167.\allowbreak{}
\item
  {[}EtherSolve{]}
  F.
  Contro,
  M.
  Crosara,
  M.
  Ceccato,
  M. Dalla
  Preda.
  EtherSolve:
  computing
  an
  accurate
  control-flow
  graph
  from
  Ethereum
  bytecode.
  ICPC
  2021,
  pp.~127--137.
  DOI
  10.\allowbreak{}1109/\allowbreak{}ICPC52881.\allowbreak{}2021.\allowbreak{}00021.\allowbreak{}
\item
  {[}eThor{]}
  C.
  Schneidewind,
  I.
  Grishchenko,
  M.
  Scherer,
  M.
  Maffei.
  eThor:
  practical
  and
  provably
  sound
  static
  analysis
  of
  Ethereum
  smart
  contracts.
  CCS
  2020,
  pp.~621--640.
  DOI
  10.\allowbreak{}1145/\allowbreak{}3372297.\allowbreak{}3417250.\allowbreak{}
\item
  {[}evm-smith{]}
  L. Alt.
  evm-smith.
  github.\allowbreak{}com/\allowbreak{}leonardoalt/\allowbreak{}evm-\allowbreak{}smith,
  commit
  14711eefeaef
  (24 June
  2026).
\item
  {[}EVMYulLean{]}
  Nethermind.
  EVMYulLean.
  github.\allowbreak{}com/\allowbreak{}NethermindEth/\allowbreak{}EVMYulLean,
  commit
  047f63070309
  (23
  September
  2025);
  ``How we
  formalized
  Ethereum
  execution:
  a
  trustworthy
  semantics
  of the
  EVM and
  Yul in
  Lean for
  Cancun'',
  6
  February
  2026,
  nethermind.\allowbreak{}io/\allowbreak{}blog/\allowbreak{}a-\allowbreak{}trustworthy-\allowbreak{}formal-\allowbreak{}model-\allowbreak{}of-\allowbreak{}evm-\allowbreak{}yul-\allowbreak{}in-\allowbreak{}lean.\allowbreak{}
\item
  {[}FIPS180-4{]}
  NIST.
  FIPS PUB
  180-4:
  Secure
  Hash
  Standard
  (SHS),
  2015.
  DOI
  10.\allowbreak{}6028/\allowbreak{}NIST.\allowbreak{}FIPS.\allowbreak{}180-\allowbreak{}4.\allowbreak{}
\item
  {[}GASOL{]}
  E.
  Albert,
  J.
  Correas,
  P.
  Gordillo,
  G.
  Román-Díez,
  A.
  Rubio.
  GASOL:
  gas
  analysis
  and
  optimization
  for
  Ethereum
  smart
  contracts.
  TACAS
  2020,
  Part II,
  LNCS
  12079,
  pp.~118--125.
  DOI
  10.\allowbreak{}1007/\allowbreak{}978-\allowbreak{}3-\allowbreak{}030-\allowbreak{}45237-\allowbreak{}7\_\allowbreak{}7.\allowbreak{}
\item
  {[}GASTAP{]}
  E.
  Albert,
  P.
  Gordillo,
  A.
  Rubio,
  I.
  Sergey.
  Running
  on
  fumes:
  preventing
  out-of-gas
  vulnerabilities
  in
  Ethereum
  smart
  contracts
  using
  static
  resource
  analysis.
  VECoS
  2019,
  LNCS
  11847,
  pp.~63--78.
  DOI
  10.\allowbreak{}1007/\allowbreak{}978-\allowbreak{}3-\allowbreak{}030-\allowbreak{}35092-\allowbreak{}5\_\allowbreak{}5.\allowbreak{}
  First
  circulated
  as
  ``GASTAP:
  a gas
  analyzer
  for
  smart
  contracts'',
  arXiv:1811.10403.
\item
  {[}Gigahorse{]}
  N.
  Grech,
  L.
  Brent,
  B.
  Scholz,
  Y.
  Smaragdakis.
  Gigahorse:
  thorough,
  declarative
  decompilation
  of smart
  contracts.
  ICSE
  2019,
  pp.~1176--1186.
  DOI
  10.\allowbreak{}1109/\allowbreak{}ICSE.\allowbreak{}2019.\allowbreak{}00120.\allowbreak{}
\item
  {[}Glamsterdam{]}
  Ethereum
  Foundation.
  Glamsterdam
  testnet
  announcement,
  28
  September
  2026.
  blog.\allowbreak{}ethereum.\allowbreak{}org/\allowbreak{}2026/\allowbreak{}09/\allowbreak{}17/\allowbreak{}glamsterdam-\allowbreak{}testnet-\allowbreak{}announcement;
  with
  EIP-7773,
  eips.\allowbreak{}ethereum.\allowbreak{}org/\allowbreak{}EIPS/\allowbreak{}eip-\allowbreak{}7773.\allowbreak{}
\item
  {[}Grishchenko18{]}
  I.
  Grishchenko,
  M.
  Maffei,
  C.
  Schneidewind.
  A
  semantic
  framework
  for the
  security
  analysis
  of
  Ethereum
  smart
  contracts.
  POST
  2018,
  LNCS
  10804,
  pp.~243--269.
  DOI
  10.\allowbreak{}1007/\allowbreak{}978-\allowbreak{}3-\allowbreak{}319-\allowbreak{}89722-\allowbreak{}6\_\allowbreak{}10.\allowbreak{}
\item
  {[}Grossman18{]}
  S.
  Grossman,
  I.
  Abraham,
  G.
  Golan-Gueta,
  Y.
  Michalevsky,
  N.
  Rinetzky,
  M.
  Sagiv,
  Y.
  Zohar.
  Online
  detection
  of
  effectively
  callback
  free
  objects
  with
  applications
  to smart
  contracts.
  Proc.
  ACM
  Program.
  Lang.
  2(POPL),
  Article
  48,
  2018.
  DOI
  10.1145/3158136.
\item
  {[}halmos{]}
  a16z.
  halmos:
  a
  symbolic
  testing
  tool for
  EVM
  smart
  contracts.
  github.\allowbreak{}com/\allowbreak{}a16z/\allowbreak{}halmos,
  commit
  079bb4241d1b
  (6
  August
  2025);
  hashing
  in
  src/halmos/sevm.py.
\item
  {[}Halmos-Pectra{]}
  D. Park.
  Formal
  verification
  of
  Pectra
  system
  contracts
  with
  Halmos.
  a16z
  crypto,
  30
  January
  2025.
  a16zcrypto.\allowbreak{}com/\allowbreak{}posts/\allowbreak{}article/\allowbreak{}formal-\allowbreak{}verification-\allowbreak{}of-\allowbreak{}pectra-\allowbreak{}system-\allowbreak{}contracts-\allowbreak{}with-\allowbreak{}halmos;
  github.\allowbreak{}com/\allowbreak{}daejunpark/\allowbreak{}sys-\allowbreak{}asm-\allowbreak{}halmos.\allowbreak{}
\item
  {[}hevm{]}
  dxo, M.
  Soos, Z.
  Paraskevopoulou,
  M.
  Lundfall,
  M.
  Brockman.
  Hevm, a
  fast
  symbolic
  execution
  framework
  for EVM
  bytecode.
  CAV
  2024,
  LNCS
  14681,
  pp.~453--465.
  DOI
  10.\allowbreak{}1007/\allowbreak{}978-\allowbreak{}3-\allowbreak{}031-\allowbreak{}65627-\allowbreak{}9\_\allowbreak{}22.\allowbreak{}
  Documentation:
  hevm.dev.
\item
  {[}Hirai17{]}
  Y.
  Hirai.
  Defining
  the
  Ethereum
  Virtual
  Machine
  for
  interactive
  theorem
  provers.
  WTSC
  2017 (FC
  2017
  Workshops),
  LNCS
  10323,
  pp.~520--535.
  DOI
  10.\allowbreak{}1007/\allowbreak{}978-\allowbreak{}3-\allowbreak{}319-\allowbreak{}70278-\allowbreak{}0\_\allowbreak{}33.\allowbreak{}
\item
  {[}IsabelleEVM{]}
  A. Card,
  D.
  Marmsoler.
  Isabelle/EVM:
  a novel
  formalization
  of the
  EVM in
  Isabelle/HOL.
  FMBC
  2026,
  OASIcs
  142,
  3:1--3:16.
  DOI
  10.\allowbreak{}4230/\allowbreak{}OASIcs.\allowbreak{}FMBC.\allowbreak{}2026.\allowbreak{}3.\allowbreak{}
\item
  {[}Jaune{]}
  Jaune.
  github.\allowbreak{}com/\allowbreak{}skbaek/\allowbreak{}jaune,
  commit
  2737c8eb
  (README,
  TRUSTED.md
  and
  scripts/GATES.md
  at that
  commit);
  archived
  by
  Software
  Heritage
  as
  swh:\allowbreak{}1:\allowbreak{}rev:\allowbreak{}2737c8ebcaec2f3a\allowbreak{}9a701415\allowbreak{}6403300b\allowbreak{}c664daab\allowbreak{}.\allowbreak{}
\item
  {[}KEVM{]}
  E.
  Hildenbrandt
  et
  al.~KEVM:
  a
  complete
  formal
  semantics
  of the
  Ethereum
  Virtual
  Machine.
  IEEE CSF
  2018,
  pp.~204--217.
  DOI
  10.\allowbreak{}1109/\allowbreak{}CSF.\allowbreak{}2018.\allowbreak{}00022.\allowbreak{}
  Repository
  github.\allowbreak{}com/\allowbreak{}runtimeverificat\allowbreak{}ion/\allowbreak{}evm-\allowbreak{}semantics,
  commit
  fbb2edbb8e84
  (5
  October
  2026).
\item
  {[}KEVM-Deposit{]}
  D. Park,
  Y.
  Zhang,
  G. Roșu.
  End-to-end
  formal
  verification
  of
  Ethereum
  2.0
  deposit
  smart
  contract.
  CAV
  2020,
  LNCS
  12224,
  pp.~151--164.
  DOI
  10.\allowbreak{}1007/\allowbreak{}978-\allowbreak{}3-\allowbreak{}030-\allowbreak{}53288-\allowbreak{}8\_\allowbreak{}8.\allowbreak{}
\item
  {[}Kontrol{]}
  Runtime
  Verification.
  Kontrol
  documentation:
  \texttt{kontrol prove}
  flags,
  \texttt{--use-gas}.
  docs.\allowbreak{}runtimeverificat\allowbreak{}ion.\allowbreak{}com/\allowbreak{}kontrol/\allowbreak{}glossary/\allowbreak{}kontrol-\allowbreak{}prove-\allowbreak{}flags.\allowbreak{}
\item
  {[}Lean4{]}
  L. de
  Moura,
  S.
  Ullrich.
  The Lean
  4
  theorem
  prover
  and
  programming
  language.
  CADE-28,
  LNCS
  12699,
  pp.~625--635,
  2021.
  DOI
  10.\allowbreak{}1007/\allowbreak{}978-\allowbreak{}3-\allowbreak{}030-\allowbreak{}79876-\allowbreak{}5\_\allowbreak{}37.\allowbreak{}
\item
  {[}lean4\#15226{]}
  Lean 4
  issue
  15226:
  \texttt{\#print axioms}
  /
  \texttt{collectAxioms}
  under-reports
  axioms
  of
  imported
  inductives
  (in-progress
  sentinel
  cached
  by
  \texttt{exportedAxio\penalty200{}msEx\penalty200{}t}).
  github.\allowbreak{}com/\allowbreak{}leanprover/\allowbreak{}lean4/\allowbreak{}issues/\allowbreak{}15226.\allowbreak{}
  Reported
  by the
  author
  of this
  paper.
\item
  {[}leanchecker{]}
  \texttt{leanchecker},
  the
  environment-replay
  checker
  distributed
  with
  Lean 4
  toolchains
  from
  v4.28.0
  (lean4
  pull
  request
  11887);
  formerly
  lean4checker,
  github.\allowbreak{}com/\allowbreak{}leanprover/\allowbreak{}lean4checker,
  now
  archived.
\item
  {[}LIP-34{]}
  A.
  Serikov.
  LIP-34:
  CircuitBreaker
  ---
  Programmable
  panic
  layer,
  2026.
  github.\allowbreak{}com/\allowbreak{}lidofinance/\allowbreak{}lido-\allowbreak{}improvement-\allowbreak{}proposals/\allowbreak{}blob/\allowbreak{}develop/\allowbreak{}LIPS/\allowbreak{}lip-\allowbreak{}34.\allowbreak{}md.\allowbreak{}
\item
  {[}LlamaRisk{]}
  LlamaRisk.
  Curve
  pool
  reentrancy
  exploit
  postmortem,
  July
  30th,
  2023.
  HackMD,
  30 July
  2023.
  hackmd.io/@LlamaRisk/BJzSKHNjn.
\item
  {[}MadMax{]}
  N.
  Grech,
  M. Kong,
  A.
  Jurisevic,
  L.
  Brent,
  B.
  Scholz,
  Y.
  Smaragdakis.
  MadMax:
  surviving
  out-of-gas
  conditions
  in
  Ethereum
  smart
  contracts.
  Proc.
  ACM
  Program.
  Lang.
  2(OOPSLA),
  Article
  116,
  2018.
  DOI
  10.1145/3276486.
\item
  {[}Marescotti18{]}
  M.
  Marescotti,
  M.
  Blicha,
  A. E. J.
  Hyvärinen,
  S.
  Asadi,
  N.
  Sharygina.
  Computing
  exact
  worst-case
  gas
  consumption
  for
  smart
  contracts.
  ISoLA
  2018,
  LNCS
  11247,
  pp.~450--465.
  DOI
  10.\allowbreak{}1007/\allowbreak{}978-\allowbreak{}3-\allowbreak{}030-\allowbreak{}03427-\allowbreak{}6\_\allowbreak{}33.\allowbreak{}
\item
  {[}Myreen08{]}
  M. O.
  Myreen,
  M. J. C.
  Gordon,
  K.
  Slind.
  Machine-code
  verification
  for
  multiple
  architectures
  --- an
  application
  of
  decompilation
  into
  logic.
  FMCAD
  2008,
  pp.~1--8.
  DOI
  10.\allowbreak{}1109/\allowbreak{}FMCAD.\allowbreak{}2008.\allowbreak{}ECP.\allowbreak{}24.\allowbreak{}
\item
  {[}Pectra-Audits{]}
  Ethereum
  Foundation,
  Protocol
  Security
  Research
  Team.
  Audit
  results
  for the
  Pectra
  system
  contracts,
  28
  February
  2025.
  blog.\allowbreak{}ethereum.\allowbreak{}org/\allowbreak{}2025/\allowbreak{}02/\allowbreak{}28/\allowbreak{}pectra-\allowbreak{}audit-\allowbreak{}results;
  reports
  at
  github.\allowbreak{}com/\allowbreak{}ethereum/\allowbreak{}audits/\allowbreak{}tree/\allowbreak{}master/\allowbreak{}Pectra.\allowbreak{}
\item
  {[}Pnueli98{]}
  A.
  Pnueli,
  M.
  Siegel,
  E.
  Singerman.
  Translation
  validation.
  TACAS
  1998,
  LNCS
  1384,
  pp.~151--166.
  DOI
  10.1007/BFb0054170.
\item
  {[}powdr{]}
  powdr
  labs.
  yul-compiler:
  a
  formally
  verified
  Yul to
  EVM
  compiler,
  and the
  verified
  stack
  underneath
  it.
  powdr.\allowbreak{}org/\allowbreak{}blog/\allowbreak{}yul-\allowbreak{}compiler,
  3 July
  2026;
  github.\allowbreak{}com/\allowbreak{}powdr-\allowbreak{}labs/\allowbreak{}yul-\allowbreak{}compiler,
  commit
  330923e0be35
  (4
  September
  2026),
  and
  github.\allowbreak{}com/\allowbreak{}powdr-\allowbreak{}labs/\allowbreak{}evm-\allowbreak{}semantics,
  commit
  2f8714d6ba96
  (18
  August
  2026).
\item
  {[}Pusceddu24{]}
  D.
  Pusceddu,
  M.
  Bartoletti.
  Formalizing
  automated
  market
  makers
  in the
  Lean 4
  theorem
  prover.
  FMBC
  2024,
  OASIcs
  118,
  5:1--5:13.
  DOI
  10.\allowbreak{}4230/\allowbreak{}OASIcs.\allowbreak{}FMBC.\allowbreak{}2024.\allowbreak{}5.\allowbreak{}
\item
  {[}RV-DepositSol{]}
  D. Park,
  Y.
  Zhang,
  G. Roșu.
  End-to-end
  formal
  verification
  of
  Ethereum
  2.0
  deposit
  smart
  contract
  (Solidity
  reimplementation).
  Runtime
  Verification,
  2020.
  github.\allowbreak{}com/\allowbreak{}runtimeverificat\allowbreak{}ion/\allowbreak{}deposit-\allowbreak{}contract-\allowbreak{}verification.\allowbreak{}
\item
  {[}RV-UniV1{]}
  Runtime
  Verification.
  Lightweight
  formal
  verification
  of
  Uniswap
  smart
  contract
  (the
  Vyper
  Uniswap
  V1
  exchange),
  12
  October
  2018.
  github.\allowbreak{}com/\allowbreak{}runtimeverificat\allowbreak{}ion/\allowbreak{}verified-\allowbreak{}smart-\allowbreak{}contracts/\allowbreak{}tree/\allowbreak{}master/\allowbreak{}uniswap.\allowbreak{}
\item
  {[}SigmaPrime-7002{]}
  Sigma
  Prime.
  Ethereum
  Foundation
  Pectra
  system
  contracts
  bytecode:
  security
  assessment
  report,
  January
  2025.
  Finding
  PEC-03,
  ``Fee
  Calculation
  Can
  Overflow''.
  github.\allowbreak{}com/\allowbreak{}ethereum/\allowbreak{}audits,
  Pectra/\allowbreak{}Sigma\_\allowbreak{}Prime\_\allowbreak{}Ethereum\_\allowbreak{}Foundation\_\allowbreak{}Pectra\_\allowbreak{}System\_\allowbreak{}Contracts\_\allowbreak{}Bytecode.\allowbreak{}pdf.\allowbreak{}
\item
  {[}Solidus{]}
  Paradigm.
  Solidus.
  github.\allowbreak{}com/\allowbreak{}paradigmxyz/\allowbreak{}solidus,
  commit
  32c7cdbc0b46
  (24 July
  2026);
  D.
  Robinson,
  ``Formally
  verifying
  a
  compiler
  using
  automated
  research'',
  24 July
  2026,
  paradigm.\allowbreak{}xyz/\allowbreak{}writing/\allowbreak{}solidus.\allowbreak{}
\item
  {[}Sourcify{]}
  Sourcify
  verification
  records
  for the
  seven
  compiled
  implementations.
  sourcify.dev.
\item
  {[}sys-asm{]}
  Ethereum.
  sys-asm:
  geas
  implementations
  of
  Ethereum's
  system
  contracts.
  github.\allowbreak{}com/\allowbreak{}ethereum/\allowbreak{}sys-\allowbreak{}asm
  (formerly
  lightclient/sys-asm).
\item
  {[}ultrasound{]}
  ultrasound.money.
  ETH
  supply
  data:
  execution-layer
  balances,
  beacon-chain
  balances
  and
  deposits,
  read 7
  October
  2026.
  ultrasound.money.
\item
  {[}UniswapV2{]}
  H.
  Adams,
  N.
  Zinsmeister,
  D.
  Robinson.
  Uniswap
  v2 Core,
  March
  2020.
  app.\allowbreak{}uniswap.\allowbreak{}org/\allowbreak{}whitepaper.\allowbreak{}pdf;
  sources
  at
  github.\allowbreak{}com/\allowbreak{}Uniswap/\allowbreak{}v2-\allowbreak{}core,
  tag
  v1.0.0
  (commit
  8160750),
  whose
  Solidity
  files
  tag
  v1.0.1
  leaves
  unchanged.
\item
  {[}Vandal{]}
  L.
  Brent,
  A.
  Jurisevic,
  M. Kong,
  E. Liu,
  F.
  Gauthier,
  V.
  Gramoli,
  R. Holz,
  B.
  Scholz.
  Vandal:
  a
  scalable
  security
  analysis
  framework
  for
  smart
  contracts.
  arXiv:1809.03981,
  2018.
\item
  {[}Verifereum{]}
  Verifereum.
  github.\allowbreak{}com/\allowbreak{}verifereum/\allowbreak{}verifereum,
  commit
  f5e172ccbf45
  (1
  October
  2026);
  conformance
  table
  verifereum.\allowbreak{}org/\allowbreak{}table.\allowbreak{}html,
  as
  modified
  25
  September
  2026.
\item
  {[}Verity{]}
  T.
  Marchand.
  Verity:
  a
  verified
  compiler
  for a
  core
  fragment
  of EVM
  smart
  contracts
  in Lean
  4.
  ETHReS:
  Ethereum
  Research
  Symposium
  @
  ETHPrague,
  May
  2026.
  veritylabs.\allowbreak{}dev/\allowbreak{}papers/\allowbreak{}verity.\allowbreak{}pdf;
  github.\allowbreak{}com/\allowbreak{}lfglabs-\allowbreak{}dev/\allowbreak{}verity,
  commit
  d42019149188
  (6
  October
  2026),
  with the
  axiom-dependency
  report
  of its
  CI run
  37448155682.
\item
  {[}VerX{]}
  A.
  Permenev,
  D.
  Dimitrov,
  P.
  Tsankov,
  D.
  Drachsler-Cohen,
  M.
  Vechev.
  VerX:
  safety
  verification
  of smart
  contracts.
  IEEE
  S\&P
  2020,
  pp.~1661--1677.
  DOI
  10.\allowbreak{}1109/\allowbreak{}SP40000.\allowbreak{}2020.\allowbreak{}00024.\allowbreak{}
\item
  {[}Vyper-0.2.15{]}
  Vyper
  v0.2.15
  release,
  published
  24 July
  2021.
  github.\allowbreak{}com/\allowbreak{}vyperlang/\allowbreak{}vyper/\allowbreak{}releases/\allowbreak{}tag/\allowbreak{}v0.\allowbreak{}2.\allowbreak{}15.\allowbreak{}
\item
  {[}VyperPM{]}
  Vyper
  team,
  with
  Omniscia.
  Vyper
  nonreentrancy
  lock
  vulnerability
  technical
  post-mortem
  report.
  HackMD,
  5 August
  2023.
  hackmd.io/@vyperlang/HJUgNMhs2.
\item
  {[}YellowPaper{]}
  G. Wood.
  Ethereum:
  a secure
  decentralised
  generalised
  transaction
  ledger
  (the
  Yellow
  Paper).
  ethereum.\allowbreak{}github.\allowbreak{}io/\allowbreak{}yellowpaper/\allowbreak{}paper.\allowbreak{}pdf;
  latest
  version
  Shanghai,
  efc5f9a,
  4
  February
  2025.
\end{itemize}

\end{document}